\documentclass[prd,aps,preprintnumbers,floats,floatfix,superscriptaddress,preprintnumbers,showpacs,eqsecnum,nofootinbib,notitlepage,twocolumn]{revtex4} 

\usepackage[final]{graphicx}
\usepackage{dcolumn}
\usepackage{bm}
\usepackage{amsmath}	
\usepackage{amssymb}	
\usepackage{hyperref}
\usepackage{cleveref}
\usepackage{comment}
\usepackage{url}
\usepackage{float}
\newcommand{\beq}{\begin{equation}}
\newcommand{\eeq}{\end{equation}}
\usepackage{hyperref}
\usepackage{placeins}

\newcommand{\bea}{\begin{eqnarray}}
\newcommand{\ena}{\end{eqnarray}}
\newcommand{\beann}{\begin{eqnarray*}}
\newcommand{\enann}{\end{eqnarray*}}

\usepackage[dvipsnames]{xcolor}

\begin{document}


\title{Modeling transient resonances in extreme-mass-ratio inspirals}


\author{Priti \sc{Gupta}} 
\email{priti.gupta@tap.scphys.kyoto-u.ac.jp}
\affiliation{
Department of Physics, Kyoto University, Kyoto 606-8502, Japan
}
\author{Lorenzo \sc{Speri}} 
\affiliation{
Max Planck Institute for Gravitational Physics (Albert Einstein Institute), Am M\"uhlenberg 1, Potsdam 14476, Germany
}
\author{Be\'atrice \sc{Bonga}} 
\affiliation{
Institute for Mathematics, Astrophysics and Particle Physics, Radboud University, 6525 AJ Nijmegen, The Netherlands
}
\author{Alvin \sc{J. K. Chua}} 
\affiliation{
Theoretical Astrophysics Group, California Institute of Technology, Pasadena, CA 91125, United States
}
\author{Takahiro \sc{Tanaka}} 
\affiliation{
Department of Physics, Kyoto University, Kyoto 606-8502, Japan
}
\affiliation{Center for Gravitational Physics, Yukawa Institute for Theoretical Physics,
Kyoto University, Kyoto 606-8502, Japan
}

\date{\today}

\begin{abstract}
Extreme-mass-ratio inspirals are one of the most exciting and promising target sources for space-based interferometers (such as LISA, Taiji, and TianQin). The observation of their emitted gravitational waves will offer stringent tests on general theory of relativity, and provide a wealth of information about the dense environment in galactic centers. To unlock such potential, it is necessary to correctly characterize EMRI signals. However, resonances are a phenomena that occurs in EMRI systems and can impact parameter inference, and therefore the science outcome, if not properly modeled. Here, we explore how to model resonances and develop an efficient implementation. Our previous work \cite{PaperI} has demonstrated that tidal resonances induced by the tidal field of a nearby astrophysical object alters the orbital evolution, leading to a significant dephasing across observable parameter space. Here, we extensively explore a more generic model for the tidal perturber with additional resonance combinations, to study the dependence of resonance strength on the intrinsic orbital and tidal parameters. To analyze the resonant signals, accurate templates that correctly incorporate the effects of the tidal field are required. The evolution through resonances is obtained using a step function, whose amplitude is calculated using an analytic interpolation of the resonance jumps. We benchmark this procedure by comparing our approximate method to a numerical evolution. We find that there is no significant error caused by this simplified prescription, as far as the astronomically reasonable range in the parameter space is concerned. Further, we use Fisher matrices to study both the measurement precision of parameters and the systematic bias due to inaccurate modeling. Modeling of self-force resonances can also be carried out using the implementation presented in this study, which will be crucial for EMRI waveform modeling.
\end{abstract}
\maketitle

\section{Introduction}
\label{sec:intro}
The detection of the first gravitational wave (GW) signal in 2015 by LIGO observatories commenced a new era of astronomy. Since then, ground-based LIGO-VIRGO networks have observed about a hundred GW signals in the 10 Hz to 1 kHz frequency band \cite{Abbott_2020,Abbott_2021,ligo2021population,theligo2020tests}. In the near future, planned space-based interferometric detectors such as LISA (Laser Interferometer Space Antenna), Taiji and Tianquin will observe GW in the 1-100 mHz frequency band. Extreme mass ratio inspiral (EMRI) is one of the most exciting possible sources and also one of the most challenging to model emitting gravitational radiation in the mHz range ~\cite{amaroseoane2017laser,berry2019unique,Mei_2020}. During such an inspiral, a stellar-mass compact object spirals into a massive black hole (MBH) at the center of a galaxy. EMRIs are characterized by a small mass ratio, typically between $10^{-4}$ and $10^{-7}$, in contrast to comparable mass binaries observed by ground-based interferometers. An EMRI can stay in the LISA bandwidth for years before it plunges, orbiting many cycles near the innermost stable circular orbit (ISCO), Thus, offering a very accurate mapping of spacetime around MBHs. EMRIs provide a chance to probe the environment of (dense) galactic centers and tests for deviations from the predictions of General Relativity (GR) \cite{berry2019unique,amaroseoane2017laser}

In the test particle limit, the small object with a mass $\mu$ follows a geodesic around the spinning MBH. In the framework of black hole perturbation theory (BHPT), the small mass ratio $\eta = \mu/M \sim 10^{-4} -10^{-7} $ is used as an expansion parameter to account for the finite mass of $\mu$. Consequently, the forcing term known as the ``self-force" moves the body away from its geodesic and is responsible for the inspiral motion. Relativistic bounded orbits around massive BHs have three frequencies --- the radial $\omega_{r}$, polar $\omega_\theta$, and azimuthal $\omega_\phi$ frequencies. These frequencies smoothly evolve as the small object gets closer to the massive one due to the self-force. Flanagan and Hinderer \cite{PhysRevLett.109.071102} highlighted an interesting phenomenon that occurs during the EMRI evolution: \textit{self-force resonances}\footnote{There is a common term in the literature for these resonances: ``transient" since the frequencies are continually evolving and the orbit does not stay at a resonance. To distinguish them from tidal resonances, which are also transient in nature, we call them ``self-force resonances", here.}. During such a resonance, radial and polar frequencies become commensurate such that $n\,\omega_{r}+k\,\omega_\theta=0,$ where $n,k$ are integers.  Recent studies have shown the impact of self-force resonances on detection and parameter estimation \cite{2016Berry,speri2021assessing}, although the precise evaluation of 
self-force resonance effects is still to be performed. 

The event rate of EMRIs depends on highly uncertain parameters such as the stellar density profile around each galactic center, the population of compact objects, and rates of dynamical processes that can lead to the capture of the stellar-mass body in the gravitational potential of a MBH \cite{Amaro_Seoane_2019,amaroseoane2020gravitational,Emami_2020,emami2020detectability,pan2021formation,pan2021formation2}. Therefore, the expected range varies from a few to a few hundred EMRI signals over a four-year mission duration for LISA \cite{Gair_2017,pan2021formation2}. To take the full advantage of the scientific potential of such astrophysical sources, data analysis methods rely on theoretical waveform templates to compare against the data. Thus, we must have waveforms for generic orbits that are modelled accurately within a fraction of a radian, even after hundreds of thousands of orbital cycles. Another necessity is that the templates should cover the high dimensional parameter space of possible EMRI configurations and their generation must be fast enough to be able to deal with templates in large numbers.  Significant efforts by the scientific community focusing on the computation of the self-force, together with LISA working groups and mock data challenges, are concentrated on realizing the goal of accurate and fast waveform modeling  \cite{Fujita_2020,hughes2021adiabatic,Chua_2021,2021Katz,Wardell:2021fyy,lynch2021eccentric}.  

Environmental effects will introduce systematic parameter estimation errors, potentially spoiling the efforts of the community towards accurate waveform models and precision gravitational wave astrophysics. This can lead to the erroneous conclusion that the data conflicts with GR \cite{byh}. Thus, quantifying and modeling resonances resulting from self-force and external tidal fields on inspirals is another challenge to overcome, if we want to perform precision tests of GR \cite{byh,Amaro-Seoane:2022rxf}. Our paper is motivated by this issue, and we investigate the modeling of resonances induced by an external tidal field. We developed for the first time the implementation of a realistic EMRI waveform passing through a resonance. This is essential for the scientific success of LISA. In particular, full waveforms will be essential for the search \cite{2016Berry} and parameter estimation of EMRIs \cite{speri2021assessing}.
The insights gained from this paper will be also relevant to self-force resonances, which we do not model in this paper as there are no precise jump size estimations available at present, but we hope they will be available in the near future \cite{Flanagan:2012kg,Isoyama:2013yor,Isoyama:2021jjd, Zachary}.

Most of the current models are focused on isolated EMRI systems. However, EMRIs may exist within noisy astrophysical environments, and their evolution can therefore deviate from the pure vacuum predictions of GR. For instance, studies based on a Fokker-Planck simulation suggest that a population of 40$M_\odot$ BHs can be close to Sagittarius $\,$A$\!^\star$, with a median distance $\sim$ 5$\,$AU~\cite{Amaro_Seoane_2011,emami2020detectability,byh}. According to~\cite{Amaro_Seoane_2019,Gourgoulhon_2019}, brown dwarfs can be at an approximate distance of $\sim$ 30$\,$AU for Sgr$\,$A$\!^\star$. The focus of our work is to study the influence on EMRI evolution by a nearby stellar-mass compact object with mass $M_\star$, although our results apply to any kind of external sources whose main contribution can be modelled by a quadrupole tidal field.
The tidal perturbation (the external force), can modify the orbital dynamics, and hence the GWs radiated from that EMRI. In particular, a new type of resonance is induced in EMRIs by the tidal force of a nearby object \cite{byh}, named \textit{tidal resonances}, when the condition $n\,\omega_{r}+k\,\omega_\theta+m\,\omega_\phi=0$ is satisfied. 
During the resonance crossing, a ``jump" is induced in the constants of motion which alters the subsequent orbital evolution. Unlike self-force resonances, tidal resonances are caused by the tidal force of the tertiary. Although the magnitude of the tidal field depends on the situation, here we assume that the magnitude is not excessively large, and hence the resonances are transient, {\it i.e.},  the evolution of orbital frequencies is dominated by the radiation reaction due to gravitational self-force.

Our recent paper \cite{PaperI} (hereafter Paper I) surveyed how common and vital tidal resonance encounters are over a large part of the relevant parameter space of the orbital evolution tracks. The results showed that an EMRI typically crosses multiple resonances 
during an observationally important regime leading to a significant dephasing of waveforms. We also provided analytic fits for tidal resonant jumps for an efficient generation of EMRI waveform models taking into account these features, which are at the foundation of the present work. The analytic fits also provide insight into the dependence of the resonance strength on the orbital parameters such as the spin of the massive BH $a$, the orbital eccentricity $e$, and the inclination $I$. In Paper I, the position of the perturber was restricted to the equatorial plane, and its tidal influence on the EMRI was implemented taking only the $m=2$ quadrupole tidal deformation into account.

This paper aims to generalize the results of Paper I in two important directions. First, the position of the tertiary is generalized. Namely, we include the perturber's inclination as a parameter, while maintaining the stationary perturber approximation. This additional inclination parameter introduces new non-vanishing resonances with $m = \pm 1$ and thus, enhancing the allowed resonances. We also take into account the $m=0$ mode, which was neglected in our previous work. 
Treating the tertiary as a perturber, we can obtain the metric perturbation using black hole perturbation theory \cite{Yunes2006}. From the tidally perturbed metric, we calculate the tidal force on the EMRI, and the resonant jumps in the constants of motion are determined semi-analytically. 

Second, we go beyond semi-analytic fits to resonant jumps by proposing a new waveform model taking the resonances into account. 
To detect and analyze GW signals, the phase evolution of our waveform models need to be accurate enough because detections rely on matched filtering techniques, which are extremely sensitive to the errors in the phase evolution of the template waveforms. 
If the resonance effects are large enough, post-resonance waveform evolution can become totally out of phase compared with the template neglecting resonances. 
It requires a practical, {\it i.e.}, fast and accurate, model to efficiently detect EMRIs and correctly estimate the parameters of EMRI and the perturber. A recent work \cite{speri2021assessing} presented a partially phenomenological Effective Resonance Model (ERM) with additional free parameters for the resonance jumps. We use techniques from this model to incorporate tidal resonances that are constrained by physics, and hence our model is no longer ``effective'' in the above sense.

A consistency check confirms that the obtained fitting formulae accurately estimate the jump size by comparing it with the slow evolution forced osculating elements trajectory \cite{osculating-kerr}. Hence, these fittings allow incorporating resonances at inexpensive computational costs. To model the jump, we use a step function approach rather than a `smooth' impulse function \cite{speri2021assessing}, and show that this simplified treatment is enough to maintain the accuracy required for data analysis. The accuracy of post-resonance evolution depends far more on the fitting formulae than the profile of the jump. For a small tidal perturbation (examined in this paper), the phase accumulated during the passage of the resonance is negligible, which makes the step function approach suitable. In case of large tidal perturbations (sustained resonances), the impulse function must be carefully selected. However, this occurs in a less astrophysically relevant region of the parameter space, and is beyond the scope of this paper.

With our model, we explore the parameter measurement precision when tidal resonances are present and study the parameter bias induced by ignoring them \cite{2007Curt}. Based on the studied EMRI configurations, we find that biases are larger than noise-induced statistical errors. As a result of our findings, parameter estimates of resonant EMRIs will likely be biased if resonances are not taken into consideration in parameter estimation models. The Fisher matrices are also used to discuss the threshold magnitude of tidal perturbation below which the observed signal cannot be interpreted as indicative of tidal perturbation.

The outline of the paper is as follows. In Sec.~\ref{sec:2}, we recall the evolution equations for Kerr geodesic motion and the framework to compute jumps due to tidal resonances. In Sec.~\ref{sec:3}, we present the first part of our results and show the dependence of tidal resonances and accumulated phase shift on orbital and tidal parameters. In Sec.~\ref{sec:4}, we review gravitational wave data analysis concepts and the key concepts of the Resonance Model (RM). In Sec.~\ref{sec:5}, we analyze the agreement between the RM and forced osculating evolution. We examine the bias in parameter estimation using Fisher matrices and present our results. We conclude our paper with a discussion and future outlook in Sec.~\ref{sec:6}.
In App.~\ref{appex:A}, we discuss the combination of resonances that are suppressed and do not contribute to dephasing the waveform. Throughout this paper, we use geometrical units with $c = G = 1$ where $c$ is the speed of light and $G$ is the gravitational constant.

\section{Background}
\label{sec:2}
In this section, we first describe the motion of a non-spinning compact object of mass $\mu$ moving in the Kerr spacetime and set up the notation and conventions that we use. Next, we introduce the tidal resonance condition and briefly describe the tidally perturbed metric used to model the tidal force and calculate the jump in conserved quantities due to a tidal resonance. For an in-depth discussion, we refer the reader to Paper I.
\subsection{Overview of Kerr geodesic}
Consider a small body of mass $\mu$ moving in the spacetime of a large black hole described by mass $M$ and spin parameter $a$. We use Boyer-Lindquist coordinates $\{r$,$\theta$,$\phi\}$ and Carter-Mino time $\lambda$ to describe the geodesic equations \cite{1972ApJBardeen,Schmidt_2002,Mino_2003,Fujita1_2009}:

\begin{subequations}
	\begin{align}
	\bigg(\frac{dr}{d\lambda}\bigg)^{2}
            &= \big[E(r^{2}+a^{2}) - a L_{z}\big]^2\nonumber\\
        &\qquad    - \Delta \big[r^2+(L_{z} - a E)^2 +Q \big]
            \nonumber\\
             &\hskip 0.06cm\equiv R(r)\,, \label{eq:geo1}\\
    \bigg(\frac{d\theta}{d\lambda}\bigg)^{2}
            &= Q -  {\rm cot}^{2} \theta L_{z}^{2}-a^{2} {\rm cos}^{2}\theta (1 -E^2)
            \nonumber \\   
             &\hskip 0.06cm \equiv \Theta(\theta)\,, \label{eq:geo2}\\
    \frac{d\phi}{d\lambda}
            &=\Phi_{r}(r)+ \Phi_{\theta}({\rm cos}\, \theta)-a\,L_{z}\,, \label{eq:geo3}\\
 	\frac{dt}{d\lambda} 
 	        &=T_{r}(r)+ T_{\theta}({\rm cos}\, \theta)-a\,E\,, \label{eq:geo4}
	\end{align}
\end{subequations}
The quantities $E, L_{z}$, and $Q$ correspond to the orbit’s energy (in unit $\mu$), axial angular momentum (in unit $\mu M$), and Carter constant (in unit $\mu^2 M^2$). Here, $\Delta = r^2 - 2Mr + a^2$, and the Carter-Mino time parameter $\lambda$ is related to the proper time $\tau$ through $d\lambda = d\tau/\Sigma$, where $\Sigma = r^2 + a^2 {\rm cos^2} \theta$. The explicit forms of the functions, $\Phi_r(r), \Phi_\theta(\cos\theta), T_r(r)$ and $T_\theta(\cos\theta)$ in Eqs. (\ref{eq:geo3}) and (\ref{eq:geo4}) can be found in Ref~\cite{Fujita1_2009}.

The Kerr geodesic orbit can be also parameterized by another set of parameters: the semi-latus rectum $p$, the orbital eccentricity $e$, and orbital inclination angle $I$, instead of $E, L_{z}$, and $Q$. These parameters are defined by
\begin{eqnarray}
       &&p := \frac{2 r_\mathrm{p}r_\mathrm{a}}{M (r_\mathrm{p}+r_\mathrm{a})}\,,\\
       &&e := \frac{r_\mathrm{a}-r_\mathrm{p}}{r_\mathrm{a}+r_\mathrm{p}}\,,\\
       &&I := \pi/2 - {\rm sgn}(L_z) \, \theta_{\rm min} \, .
\label{eq:pex}   
\end{eqnarray}
where $r_\mathrm{a}$, $r_\mathrm{p}$ are, respectively,  the values of $r$ at the apoapsis and periapsis, and $\theta_{\rm min}$ is the minimum value of $\theta$ (measured from the black hole’s spin axis). For later convenience, we also introduce  $\rm{x}$ $= \cos I$.

\subsection{Framework to study tidal resonances}
We consider an EMRI within the influence of an external tidal field. The tidal environment created by a stellar-mass object near the EMRI is treated in a relativistic framework by computing the complete linear metric perturbation to the Kerr spacetime \cite{poisson2015tidal,Yunes2006}. 

We use a set of action-angle variables to study the orbital evolution, such that the angle variables $q_i$ parameterize a torus and the conjugate action variables $J_i$ are functions of the constants of motion $\{E, L_{z},Q\}$. This method offers a simple formulation to incorporate and study deviations from the geodesic motion due to different forces~\cite{MTW_2017}:
\begin{eqnarray}
        &&\frac{dq_{i}}{d\tau}
            = \omega_{i} (\bold{J}) +\epsilon g_{i,\rm td}^{(1)}(q_\phi,q_\theta,q_r,\bold{J}) +
            \eta g_{i,\rm sf}^{(1)}(q_\theta,q_r,\bold{J})
         \nonumber \\   
         &&\hskip 0.7cm                                         
        + \hskip 0.1cm O(\eta^2,\epsilon^2,\eta\epsilon)\,,
\label{eq:EOM1}\\
        &&\frac{dJ_{i}}{d\tau}
            = \epsilon G_{i,\rm td}^{(1)}(q_\phi,q_\theta,q_r,\bold{J}) +
            \eta G_{i,\rm sf}^{(1)}(q_\theta,q_r,\bold{J})
         \nonumber \\   
         &&\hskip 0.7cm                                                         
        + \hskip 0.1cm O(\eta^2,\epsilon^2,\eta\epsilon)\,,
\label{eq:EOM2}
\end{eqnarray}
where the terms with subscript ``td” are from the tidal force, and the terms with subscript ``sf” are from the self-force. Here, the parameter 
\beq
\epsilon = M_{\star} M^2 \,{\rm{x}}_{\star}/R^3
\eeq
characterizes the strength of the tidal field produced by the perturber $M_{\star}$ at an inclination $I_{\star}$. Here, ${\rm x}_{\star}$ is a sinusoidal function of $I_{\star}$ depending on mode $m$ of the quadrupole ($l=2$) tidal perturbation. The distance of the tidal perturber from the central MBH is denoted by $R$. As mentioned in the introduction, the frequencies of EMRI orbital evolution associated with distant observer time are $\omega_r$ (oscillations in the radial direction), $\omega_\theta$ (oscillations in the polar direction),  and $\omega_\phi$ (rotations around the central BH spin axis). 

From the expressions above, we see that at the zeroth order (neglecting the terms with the superscript $(1)$ and hither order), the action variables are conserved whereas the angle variables increase at a fixed rate in time, which are denoted by $\omega_i$. At leading order in $\eta$, the EMRI orbit deviates from the geodesic motion due to the particle’s self-force ($g_{i,\rm sf}$,$G_{i,\rm sf}$) \cite{Mino_1997,Quinn_1997,Poisson_2011,Barack_2018}.
In our model, the EMRI experiences an external tidal force  introduced in evolution equations by terms $(g_{i,\rm td}$,\,$G_{i,\rm td})$. As we proceed, we will only consider tidal resonances and hence the leading order tidal force $G_{i,\rm td}^{(1)}$, and we will drop the subscript `td', for brevity. 
The force is written in terms of its Fourier modes as
\beq
G_{i}^{(1)} (q_\phi,q_\theta,q_r,\bold{J}) = \sum_{n,k,m} G_{i,nkm}^{(1)}(\bold{J}) e^{i( nq_r+ kq_\theta+mq_\phi)} \;. 
\label{eq:FT}
\eeq
For non-resonant orbits, the exponential factor in the above equation is rapidly oscillating in time, thereby averaging to zero over many cycles. Thus, all $m,k,n$ modes, except for the one with $m=k=n=0$, do not contribute to a secular change in $\bold{J}$. 
However, the phase in Eq.~\eqref{eq:FT} will be stationary when 
\beq
\omega_{nkm} := n \omega_r +k \omega_\theta + m \omega_{\phi} =0\,,
\label{eq:TR}
\eeq
\textit{i.e.} when the tidal resonance condition is satisfied for a set of relatively small integers\footnote{When the condition is satisfied for large integers,the corresponding $G_{i,nkm}^{(1)}$ is much smaller. Hence, they tend to be irrelevant from the observational point of view, although it also depends on the magnitude of the tidal perturbation which resonances are sufficiently influential.  This holds true for self-forces resonances as well \cite{2016Berry}.} $(n,k,m)$. Thus, the exponential factor varies slowly around the resonance point, and the corresponding average of the force amplitude $G_{i,nkm}^{(1)}$ is non-vanishing, inducing a secular change in $\bold{J}$.

It is helpful to recall the relevant timescales for our physical setup. The fastest timescale is the orbital period  $\tau_{\rm orb}$ $\sim \mathcal{O}(M)$ and the slowest timescale corresponds to the radiation reaction time $\tau_{\rm rr}$ $\sim M/\eta$. The orbital period of the tidal perturber is given by $\tau_{\rm td} \sim 2\pi \sqrt{R^3/M}$.  Another key time scale is the resonance duration $\tau_{\rm res}$ \cite{PhysRevLett.109.071102,PaperI},
\beq
\tau_{\rm res} \sim \sqrt{\frac{4 \pi}{m\dot{\omega}_{\phi}+k \dot{\omega}_\theta +n \dot{\omega}_r}} \sim M\sqrt{\frac{1}{\eta}}\, .
\eeq
Overall, when the stationary perturber approximation is valid, we have 
$$\tau_{\rm orb} \ll \tau_{\rm res}\ll \tau_{\rm td}, \tau_{\rm rr}.$$

Flanagan and Hinderer~\cite{PhysRevLett.109.071102} gave an analytic expression for the `jump' in the constants of motion in the context of self-force resonances. We use a similar estimate to model the effect of the tidal resonance, and calculate the jump $\Delta J_{i}$ in conserved quantities across a resonance point. Assuming that the evolution of $\bold{J}$ and hence the orbital periods 
is dominantly determined by the gravitational radiation reaction, 
the jump $\Delta J_{i}$ is estimated as 
\begin{eqnarray}
\label{eq:Jump}
        &&\Delta J_{i}
            = \epsilon \int_{-\infty}^{\infty} G_{i}^{(1)} (q_\phi,q_\theta,q_r,\bold{J}) d\tau\nonumber\\
         &&\hskip 0.5cm                                                         
        = \epsilon \sum_{s=\pm 1} \sqrt{\frac{2 \pi}{|\Gamma s|}} {\rm exp}\bigg[{\rm sgn}(\Gamma s)\frac{i\pi}{4}+is\chi\bigg] \nonumber\\
        &&\qquad\qquad\qquad \times G_{i,sn,sk,sm}^{(1)}(\bold{J})\,, 
\end{eqnarray}
where $\chi = nq_{r0} + kq_{\theta0}+ mq_{\phi0} $ and $\Gamma = n\dot{\omega}_{r0} + k\dot{\omega}_{\theta0}+m\dot{\omega}_{\phi0}$, and the quantities $q_{i0}$ and $\dot{\omega}_{i0}$ are phases and frequency derivatives evaluated at $\tau_{\rm res,0}$ (the instant where tidal resonance condition is satisfied), respectively. Strictly speaking, higher modes with $(n,k,m)$ multiplied by an integer other than $\pm1$ are also non-vanishing, but their contribution is highly suppressed. In the estimate of $\Gamma$, the corrections due to the tidal resonance are neglected, because such corrections are higher order in $\epsilon$. 

\begin{figure}
		\includegraphics[width=5.5cm]{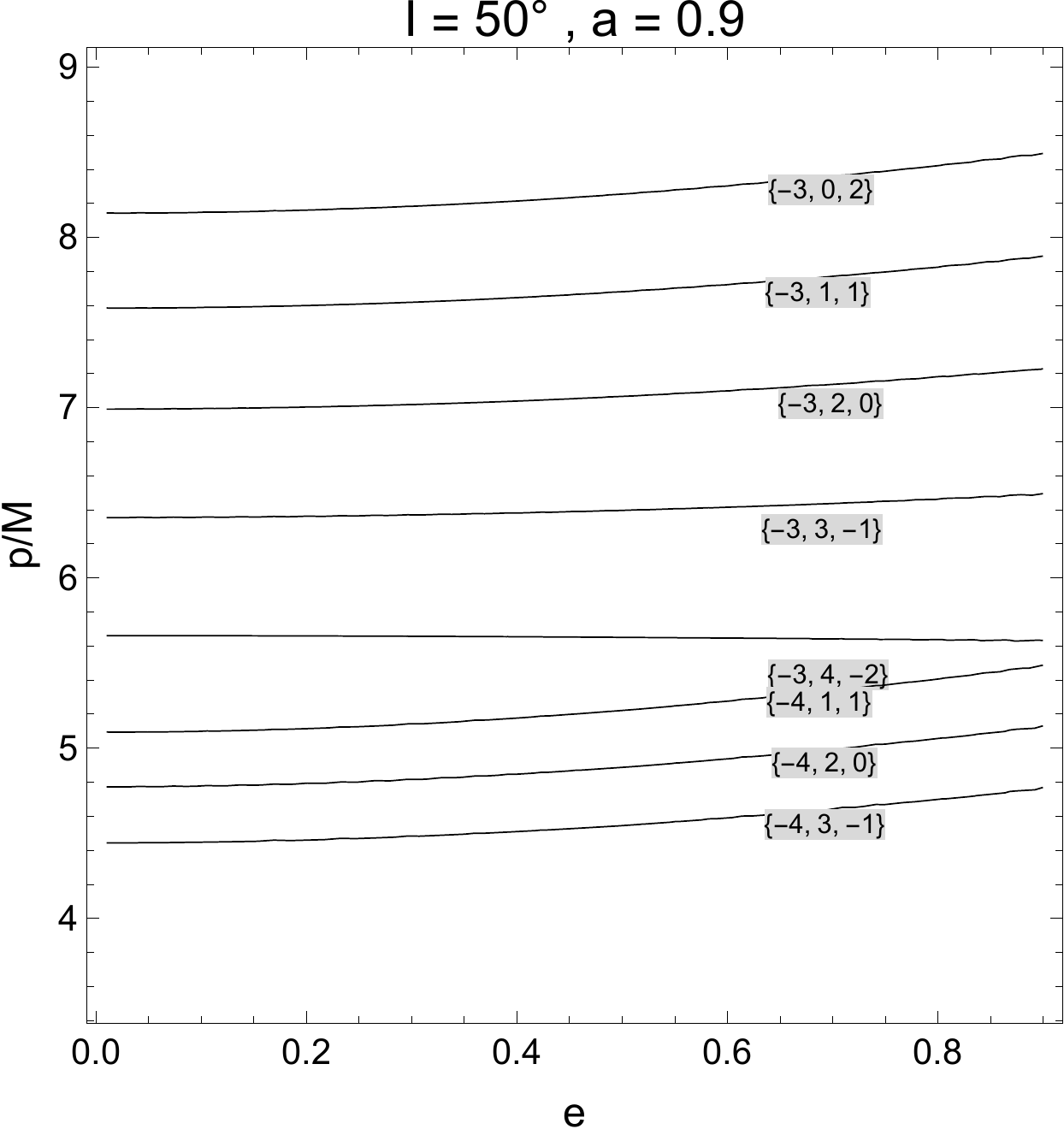}
		\includegraphics[width=5.5cm]{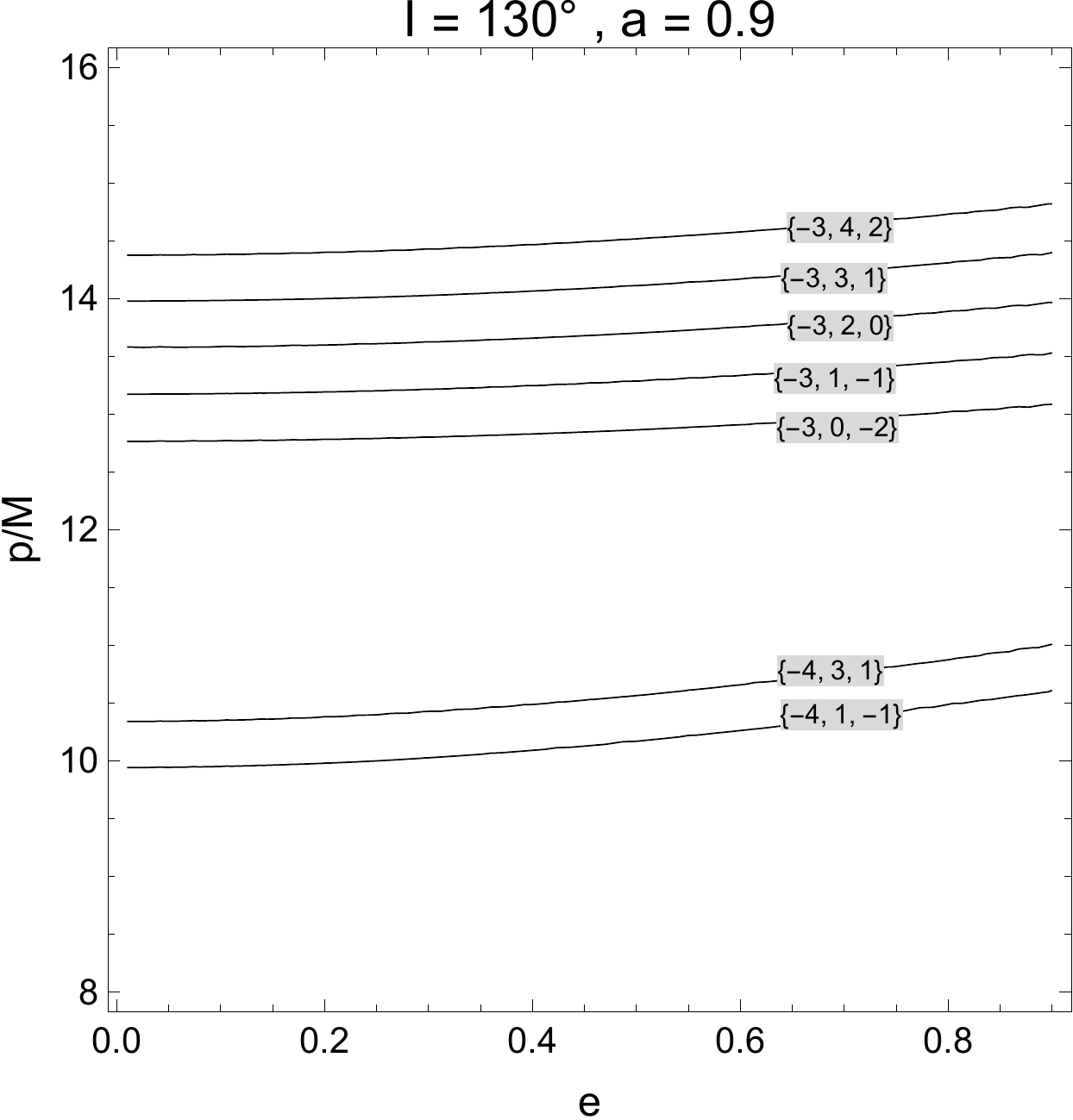}
\caption{\small The low order tidal resonance contours for a prograde orbit with inclination $50^{\circ}$ (top) and a retrograde orbit with inclination $130^{\circ}$ (bottom) in $e$ - $p$ plane. The spin parameter of the central BH is set to $a = 0.9$. The contour labels correspond to integers $\{n,k,m\}$.  We discuss the suppression of resonance combinations with $k + m = \rm{odd}$ in appendix ~\ref{appex:A}.}
		\label{fig:rescomb}
\end{figure}

In this work, we study only the leading quadrupolar $l$=2 modes, because the higher multipoles will be smaller by a power of $M/R$.
For $l$ =2, allowed values for azimuthal number $m$ are $-2$ to $2$. In Paper I only the $m = \pm 2$ modes were considered. We relax this restriction to incorporate resonances caused by $m = 0, \pm 1$ modes. In Fig~\ref{fig:rescomb}, we show the full set of low order resonance combinations investigated in our analysis. We find that resonance jumps vanish for combinations with $k + m = \rm{odd}$. This suppression is discussed in appendix ~\ref{appex:A}.

\begin{figure*}
  \centering
\includegraphics[width=0.4\linewidth]{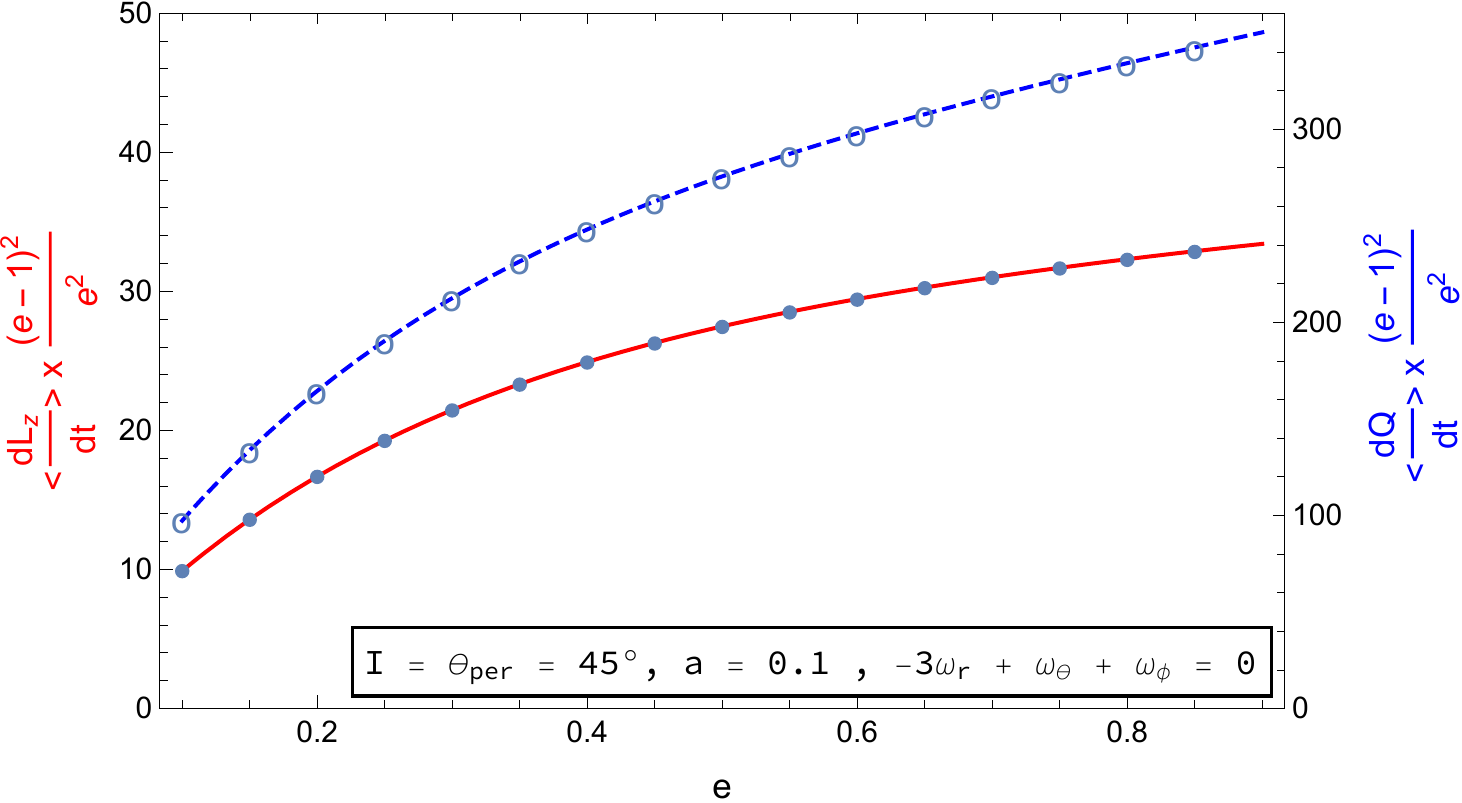}
\hspace{0.3cm}
\includegraphics[width=0.4\linewidth]{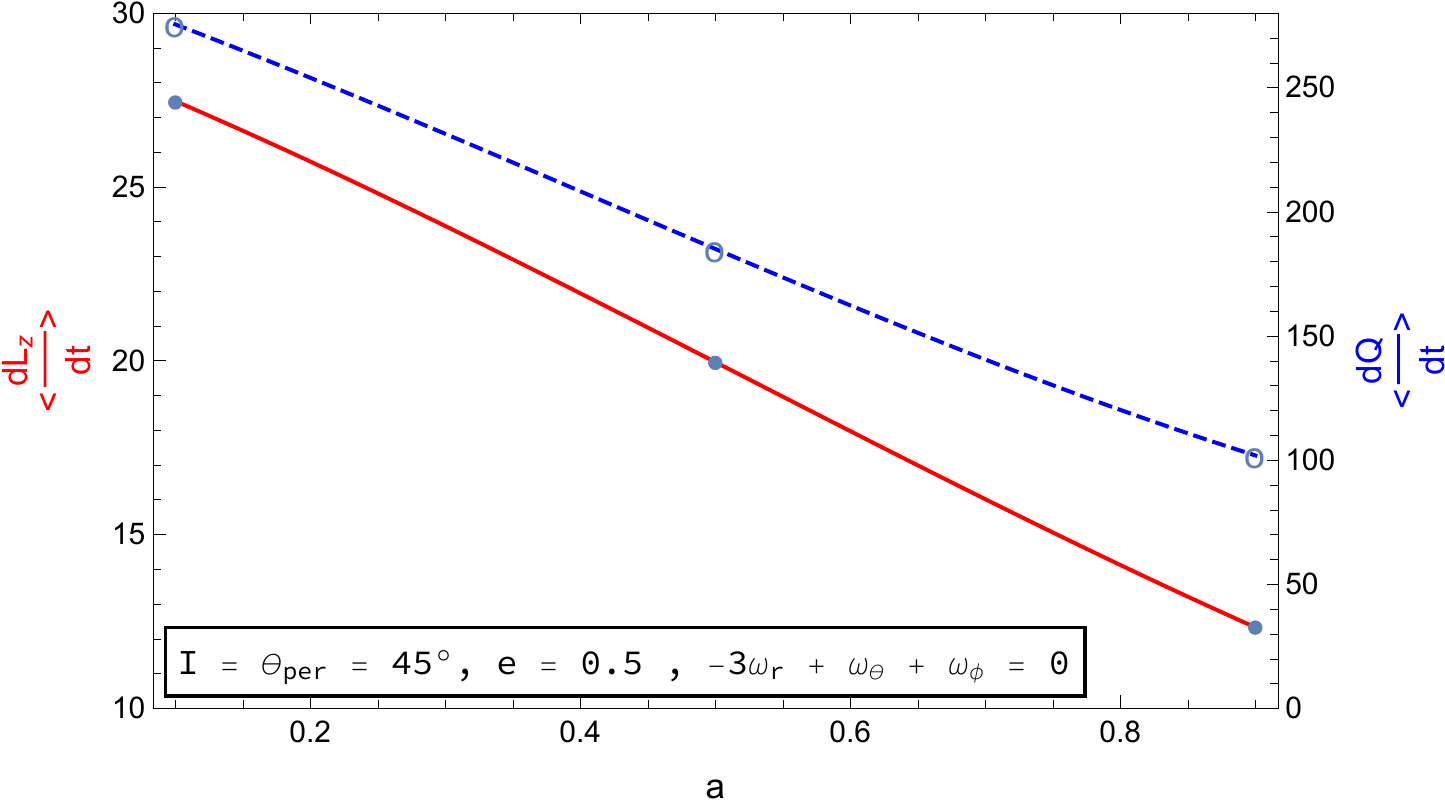}
\includegraphics[width=0.4\linewidth]{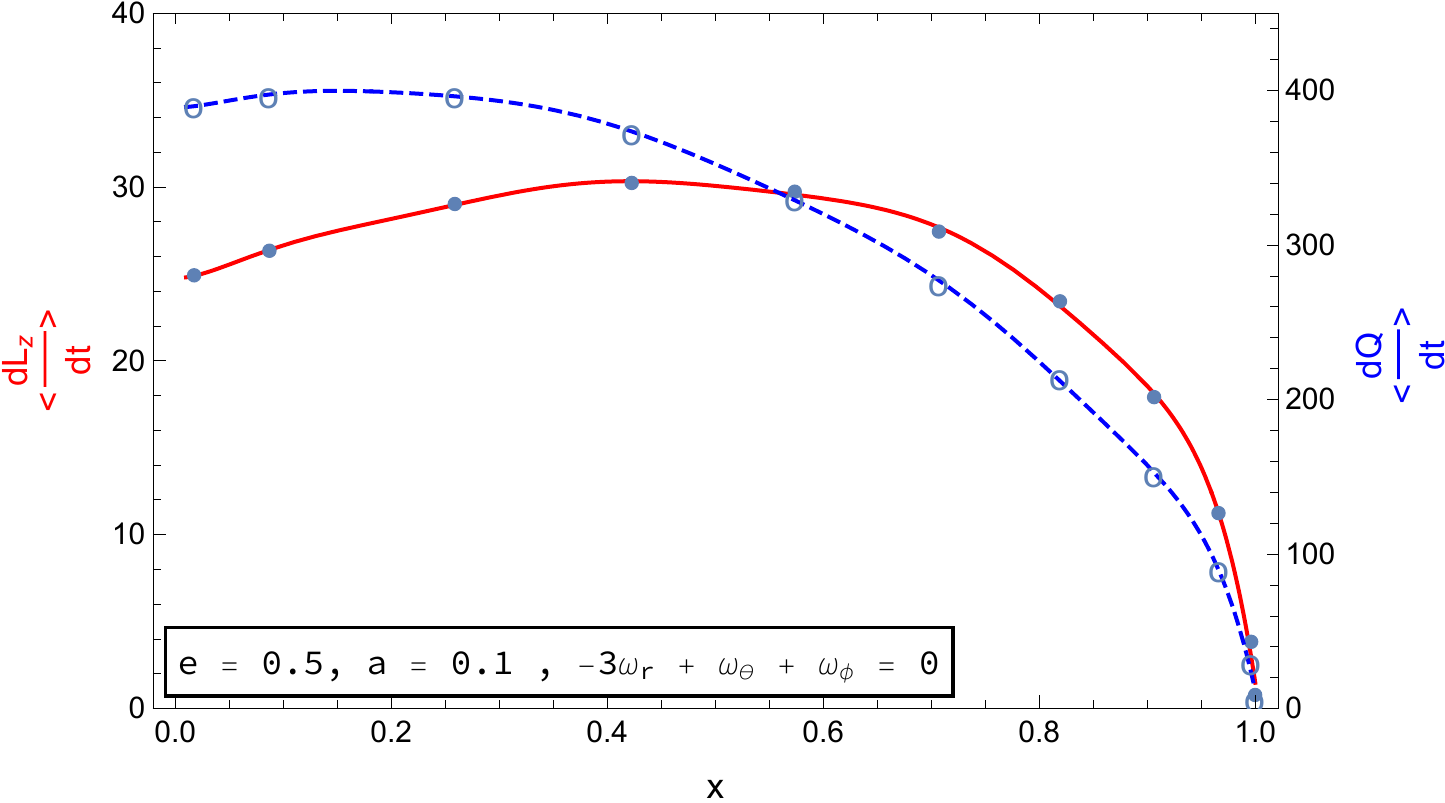}
\hspace{0.3cm}
\includegraphics[width=0.4\linewidth]{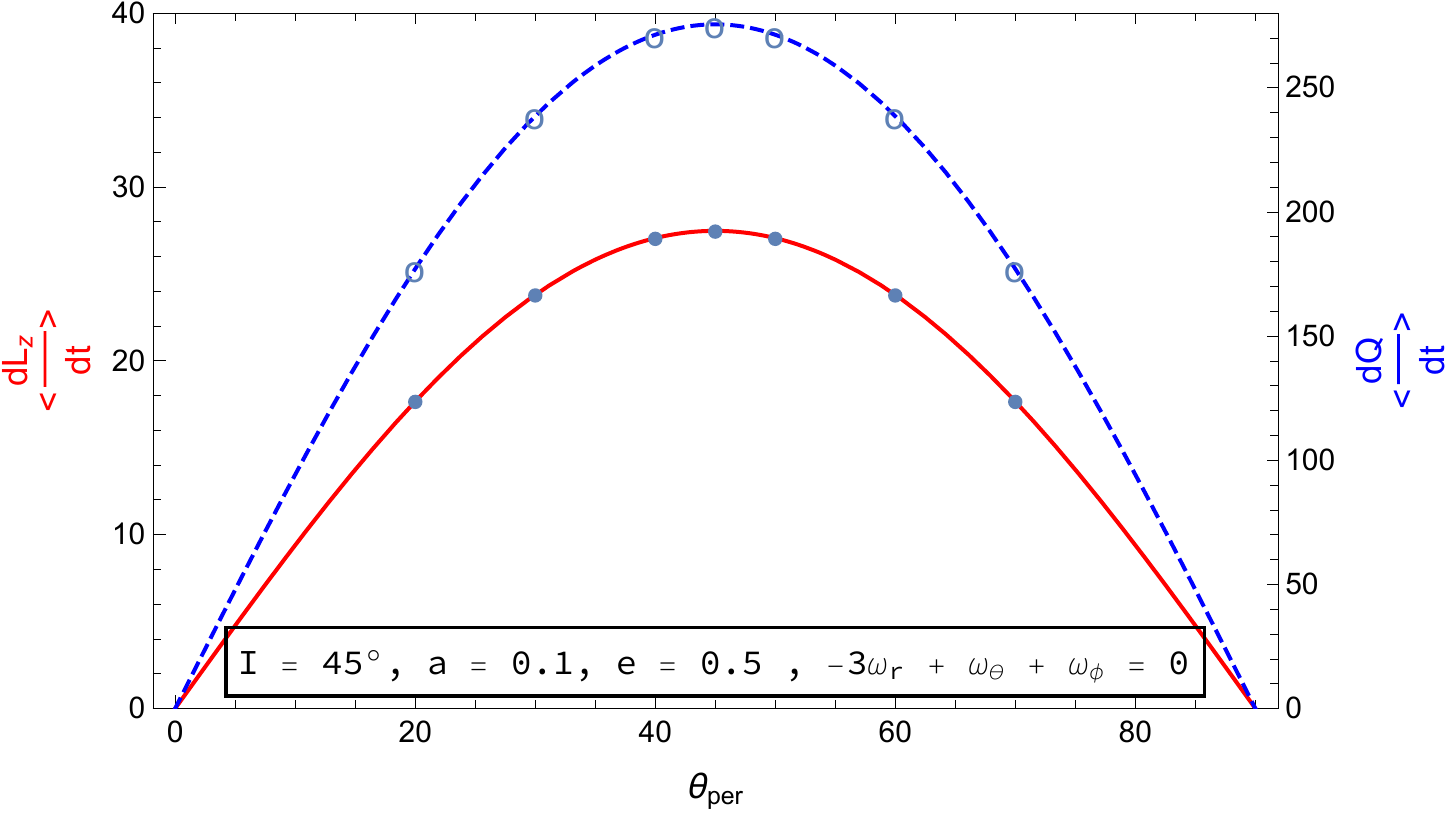}
\caption{\small Dependence of average change rate of the $z$-component of angular momentum (red-solid) and Carter constant (blue-dashed) on the orbital eccentricity (top, left), on orbital inclination (bottom, left), the spin of central BH (top, right), and perturber's inclination (bottom, right) for $n:k:m=-3:1:1$. The dots represent the values obtained from the semi-analytic calculation, and curves denote the obtained fitting. Note that both $\langle dL_{z}/dt\rangle$ and $\langle dQ/dt \rangle$ are normalised by multiplying a factor of $(\epsilon/M)^{-1}$.
}
\label{fig:meq1dependence}
\end{figure*}

\begin{figure*}
  \centering
  \includegraphics[width=0.4\linewidth]{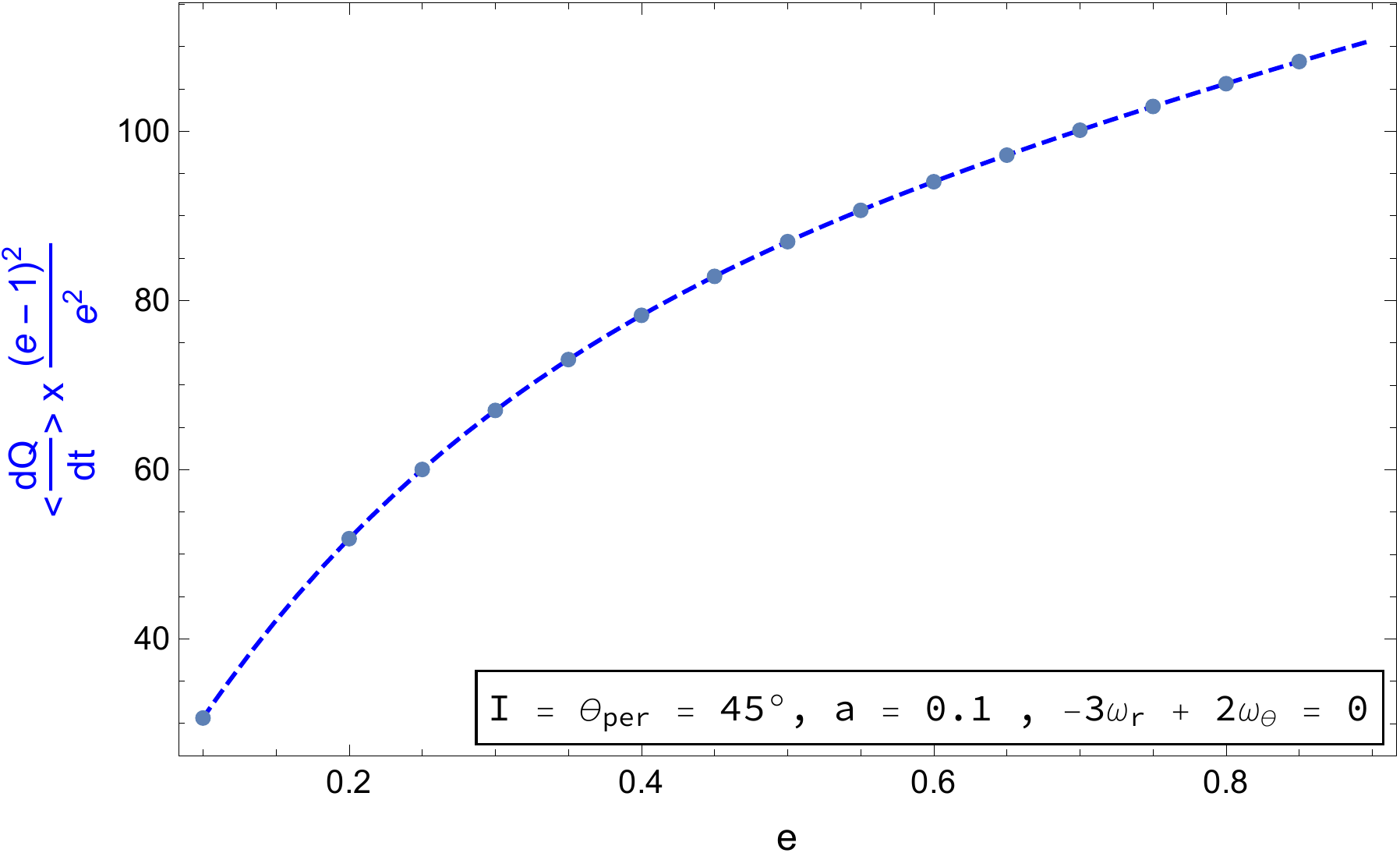}
  \hspace{0.3cm}
\includegraphics[width=0.4\linewidth]{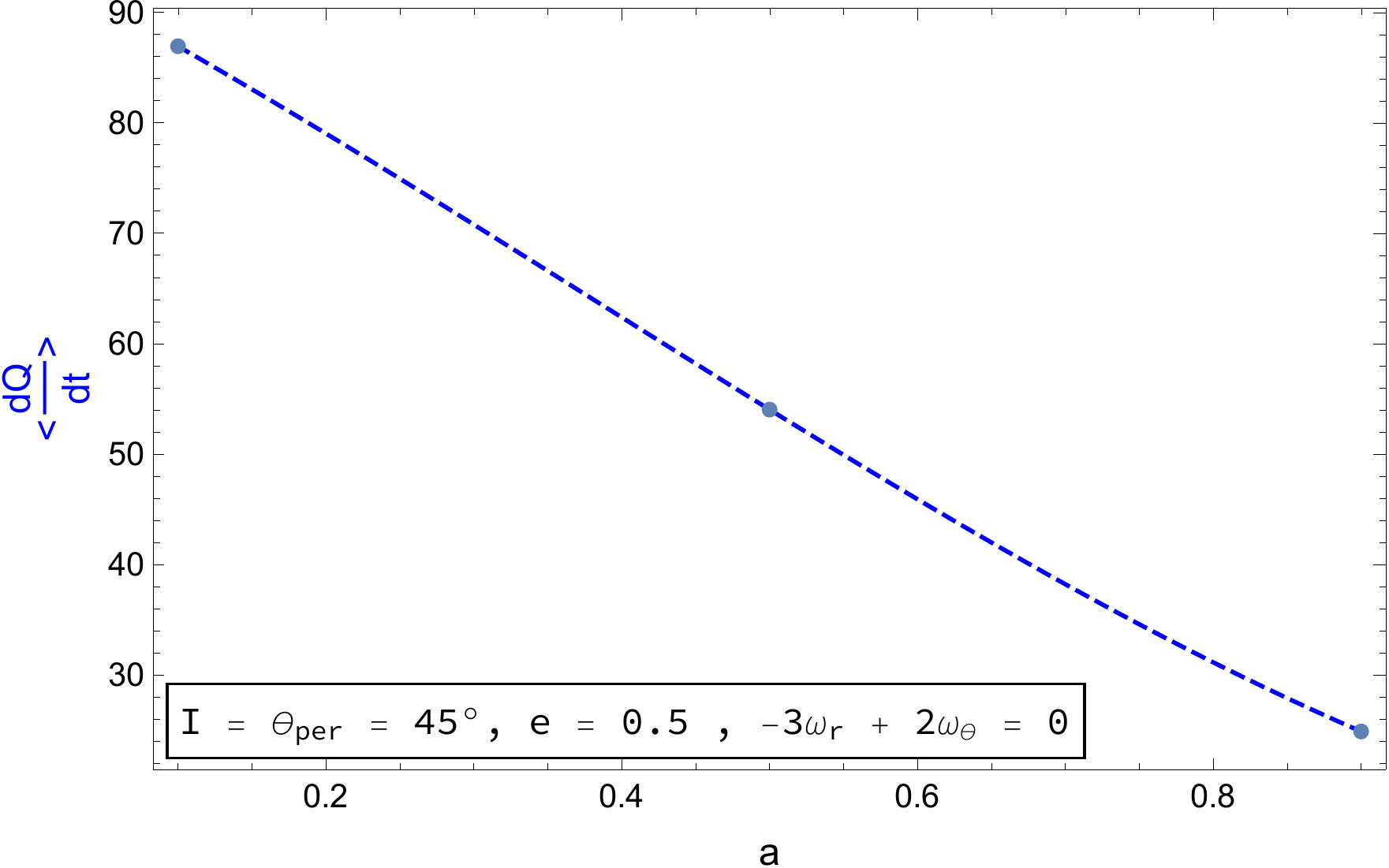}
\includegraphics[width=0.4\linewidth]{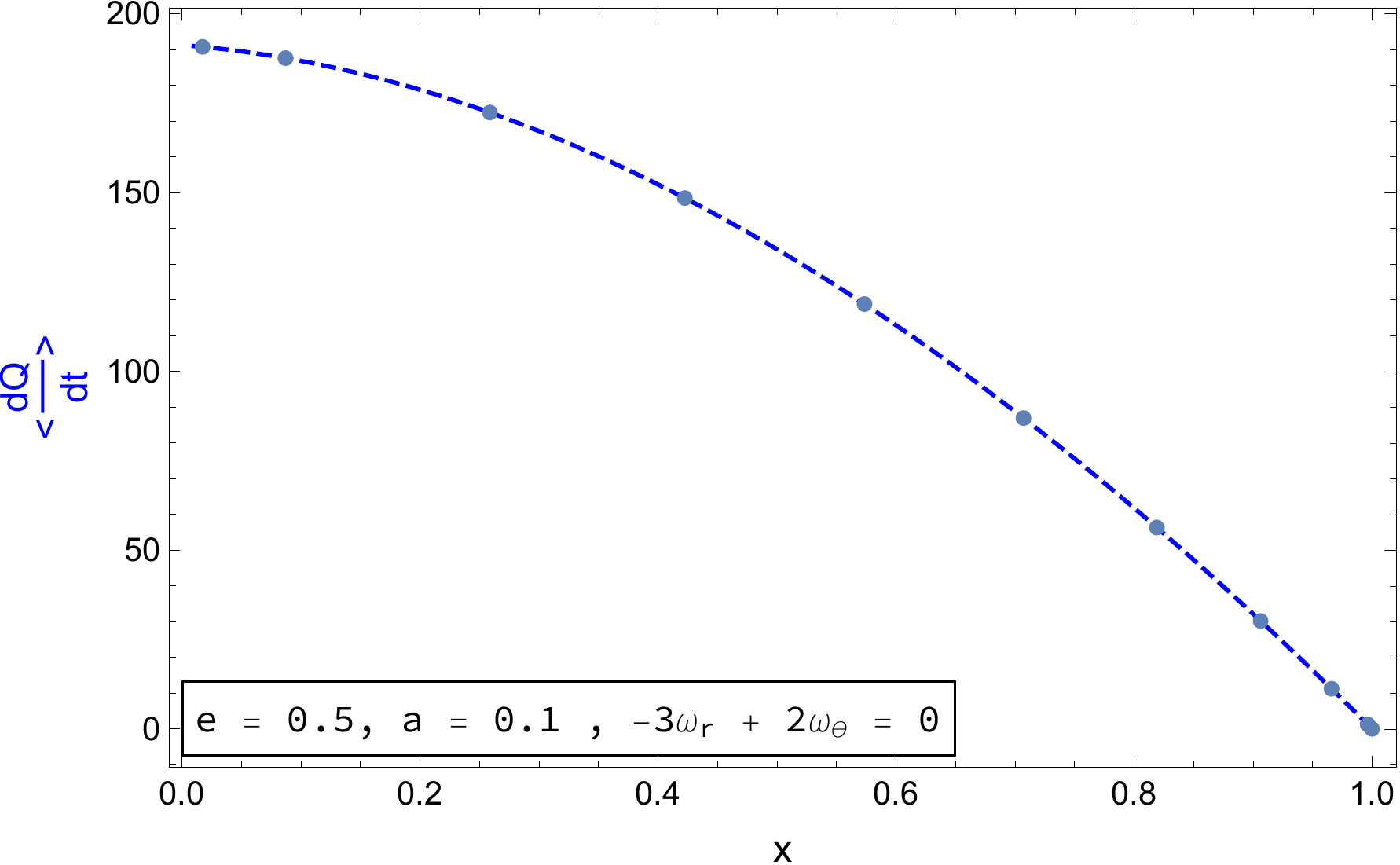}
\hspace{0.3cm}
\includegraphics[width=0.4\linewidth]{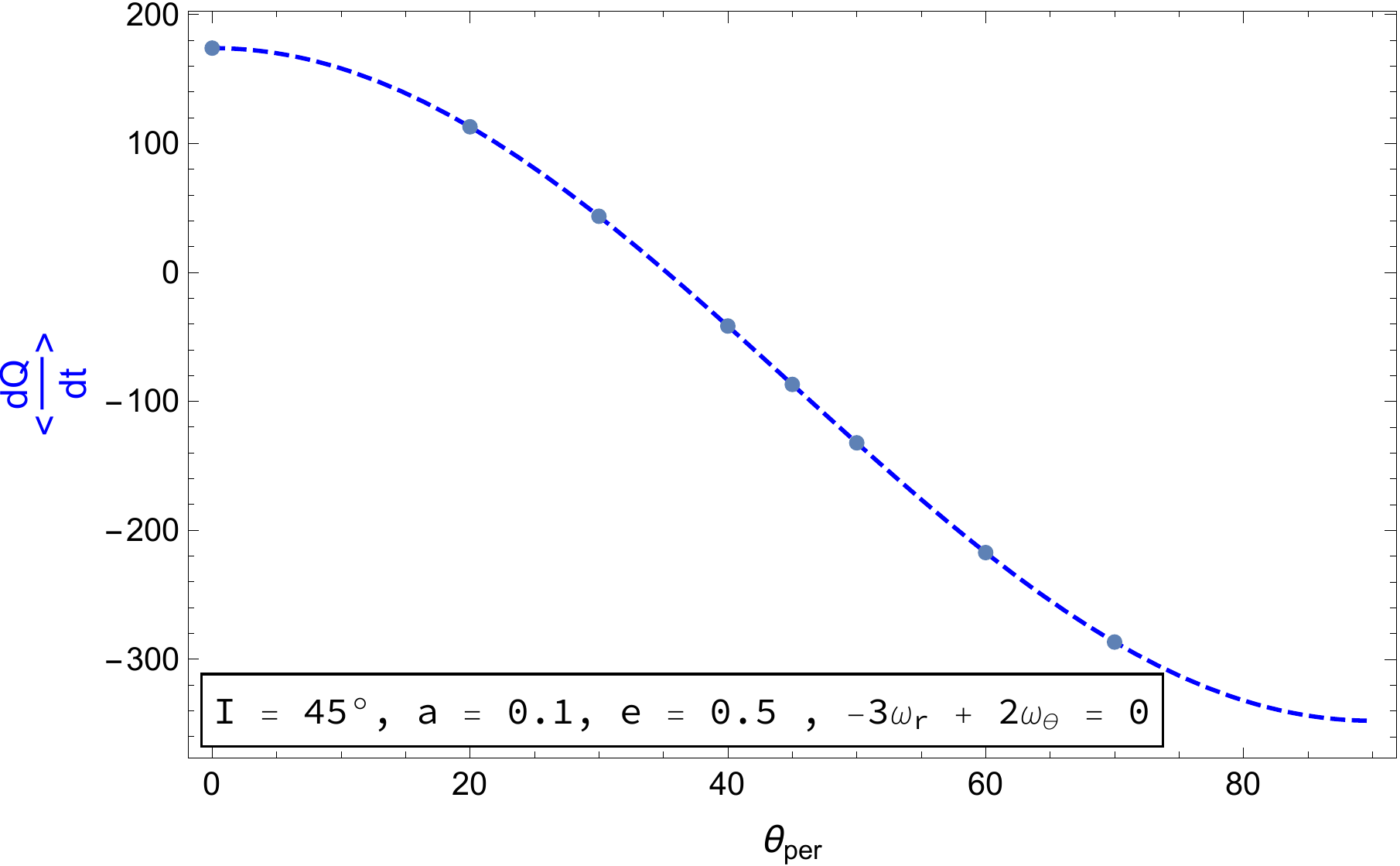}
\caption{\small Dependence of average change rate of the Carter constant (blue-dotted) on the orbital eccentricity (top, left), on orbital inclination (bottom, left), the spin of central BH (top, right), and perturber's inclination (bottom, right) for $n:k:m=-3:2:0$. The dots represent the values obtained from the semi-analytic calculation, and curves denote the obtained fitting. There is no change in the $z$-component of angular momentum given the axisymmetry of the $m = 0$ perturbation. Note that $\langle dQ/dt \rangle$ is normalised by multiplying a factor of $(\epsilon/M)^{-1}$.
}
\label{fig:meq0dependence}
\end{figure*}

To calculate the tidal force $G_i^{(1)}$, we start with the space-time metric of a rotating BH perturbed by a nearby object \cite{Yunes2006}. Given the perturbation $h_{\alpha \beta}$\footnote{An overall factor of two missing in $h_{\alpha \beta}$ in~\cite{Yunes2006}; see footnote 17 in~\cite{LeTiec:2020bos} for details.}, the induced acceleration with respect to the background Kerr spacetime is expressed as,
\begin{align}\label{eq:acc}
a^\alpha & = -\frac{1}{2} (g^{\alpha\beta}_{\rm Kerr}+u^\alpha u^\beta)(2h_{\beta \lambda;\rho}-h_{\lambda \rho; \beta}) u^\lambda u^\rho\;,
\end{align}
with $u^\alpha$ the unit vector tangent to the worldline of the inspiralling object with a small mass $\mu$.
The instantaneous change rates of the constants of motion are~\cite{PhysRevD.96.083015}
\begin{align}
    \frac{dL_{z}}{d\tau}& = a_\phi\,,\label{eq:Ldot}\\
    \frac{dQ}{d\tau} &= 2 u_\theta a_\theta - 2 a^2 {\rm cos}^2 \theta u_t a_t + 2 {\rm cot}^2 \theta u_\phi a_\phi\,.
    \label{eq:Qdot}
\end{align}

As our perturber is treated to be stationary, the change in orbital energy $E$ due to a tidal resonance is zero. Moreover, $m = 0$ mode drives an axisymmetric perturbation, leaving  $L_z$ unchanged. Nonetheless, this mode can cause a significant non-zero change in $Q$. 

\section{Resonance strength and dephasing}
\label{sec:3}
In the following, we first explore the strength of $ m=0, \pm 1$ mode resonances as a function of orbital parameters $(a,p,e,\rm{x})$ and inclination $\theta_{per}$ of the tidal perturber. Next, similar to the analysis in Paper I, we show accumulated phase shift for typical cases for $m=0, \pm1$ resonances and discuss the impacted orbital parameter space of EMRIs due to a tidal resonance encounter. Fitting formulae are constructed for each resonance combination shown in Fig.~\ref{fig:rescomb} for both prograde and retrograde cases.

\subsection{Dependence on orbital and tidal parameters}
When we introduce a tidal perturber, the  spacetime describing the central black hole and the tidal perturber is no longer axisymmetric. As shown in Eq.~\eqref{eq:EOM2}, the tidal force depends on the axial position of the small body. Hence, the changes in conserved quantities are sensitive to EMRI's orbital phases on encountering the resonance, {\it i.e.}, $q_{\phi0}, q_{\theta0}, q_{r0}$. To demonstrate this dependence, we first compute $d{L}_{z}/dt$ and $dQ/dt$ for some resonances. 
After orbit averaging at the resonance point, the right hand side in Eq.~\eqref{eq:FT} is well approximated by,
\begin{align}
&\big<G_{i}^{(1)} (q_\phi,q_\theta,q_r,\bold{J})\big> \cr 
 &\qquad \approx G_{i,mkn}^{(1)} (\bold{J})e^{ i(m q_{\phi0}+k q_{\theta0}+n q_{r0})} + \{\rm c.c.\}\,.
\label{eq:phasedependence}
\end{align}
The resonant phase is defined as $q_{\rm {res}}:=m q_{\phi0}+k q_{\theta0}+n q_{r0}$ and from Eq.~\eqref{eq:phasedependence}, it is clear that the jump size due to the tidal resonance has a sinusoidal dependence on the resonant phase \cite{PaperI,byh}. Therefore, depending on this phase, an orbit may cross the tidal resonance with a negligible jump in $L_z$ and $Q$, even if the magnitude of the tidal perturbation itself is sufficiently large. To analyze the strength of resonance on orbital and tidal parameters, we will adopt the fine-tuned value of $q_{\rm {res}}$ that maximizes the changes in ${L}_z$ and $Q$. Hence, our results show the upper limit of influence caused by these resonances.

Using \eqref{eq:Jump}, we compute the change in $L_z$ and $Q$ for different resonances and note some interesting trends for $m= 0, \pm 1$ modes.  In Fig.~\ref{fig:meq1dependence}, we show dependence of a sample resonance $-3:1:1$ (prograde orbit) on $a, e, \rm x$ and $\theta_{per}$.
\begin{itemize}
 	\item We find that, irrespective of the resonance combinations, \textit{i.e}. $m= 0, \pm 1,  \pm 2$, and the direction of the orbit (prograde or retrograde), both $d{L}_{z}/dt$ and $dQ/dt$ increase with increasing orbital eccentricity $e$. The prefactor $e^2/(e-1)^2$ ensures that $d{L}_{z}/dt$ and $dQ/dt$ are zero for circular orbits ($e=0$) since the amplitude of radial oscillations is zero for this case. 
	\item Another pattern is observed for variation in the spin parameter of MBH. Similar to $m =  \pm 2$ modes analysed in Paper I, for prograde orbits, $m =  \pm 1$ mode resonances show a decrease in both $d{L}_{z}/dt$ and $dQ/dt$ as $a$ increases whereas for retrograde orbits both quantities increase as $a$ increases. The difference between prograde and retrograde orbits is expected because the resonance occurs at smaller (larger) $p$ values for prograde (retrograde) orbits for larger values of $a$ (see vertical scale of lower panel in Fig.~\ref{fig:rescomb}) for which the acting tidal force is greater.
	\item As for orbital inclination parameter ${\rm {x}} = \cos \,I$, we find that, as $\rm x$ increases, both $d{L}_{z}/dt$ and $dQ/dt$ decreases regardless of the orbit's direction. This feature is again qualitatively similar to the trend found for $m =  \pm 2$ in Paper I. 
	\item Next, we note the dependence of resonance strength on inclination of the tidal parameter $\theta_{per}$. For the sample resonance $-3:1:1$ and other resonance combinations with $m = \pm 1$, the change in $d{L}_{z}/dt$ and $dQ/dt$ is maximum for the perturber at an inclination of $\theta_{per} = 45^\circ$. This behaviour can be qualitatively explained for $L_z$ using Newtonian arguments --- the spherical harmonic decomposition of ($l = 2, m = \pm 2$) mode of the tidal force and hence the torque turns out to be proportional to $\rm{sin} \theta_{per} \, \rm{cos} \theta_{per}$ \cite{poisson_will_2014}. This dependence also clarifies that $m = \pm 1$ resonance gives no contribution for an equatorial perturber ($\theta_{per} = 0^\circ$).
\end{itemize}

In Fig~\ref{fig:meq0dependence}, we show the dependence on orbital and tidal parameter for a $m = 0$ mode focusing on $-3:2:0$ resonance. For this mode, the axisymmetry of the background Kerr spacetime remains intact. Therefore there is no jump induced in $L_z$. Nonetheless, we find that such resonances can still drive a jump in $Q$ as shown in Fig~\ref{fig:meq0dependence}. The dependency on $e, a,\rm x$ are qualitatively similar to $m = 1$ resonances discussed above. In contrast, for $m = 0$ resonances, we find that the absolute jump size is largest when the perturber is aligned with the rotation axis of the MBH. This finding is important because $m = 0$ modes can cause a jump in $Q$, implying that other axisymmetric sources such as accretion disks can also induce a jump and impact waveforms through tidal resonances. Furthermore, tidal resonances with $m=0$ modes are degenerate with self-force resonances, for which only the radial and polar integers ($n$ and $k$) determine the resonance combination due to the axisymmetry of the Kerr space-time. In order to dissociate such resonances, waveforms need to be accurately modeled.  If multiple tidal resonances due to the same perturber are encountered by an EMRI, they might be sufficient to break the degeneracy.

\begin{figure}
  \centering
  \includegraphics[width=7.5cm]{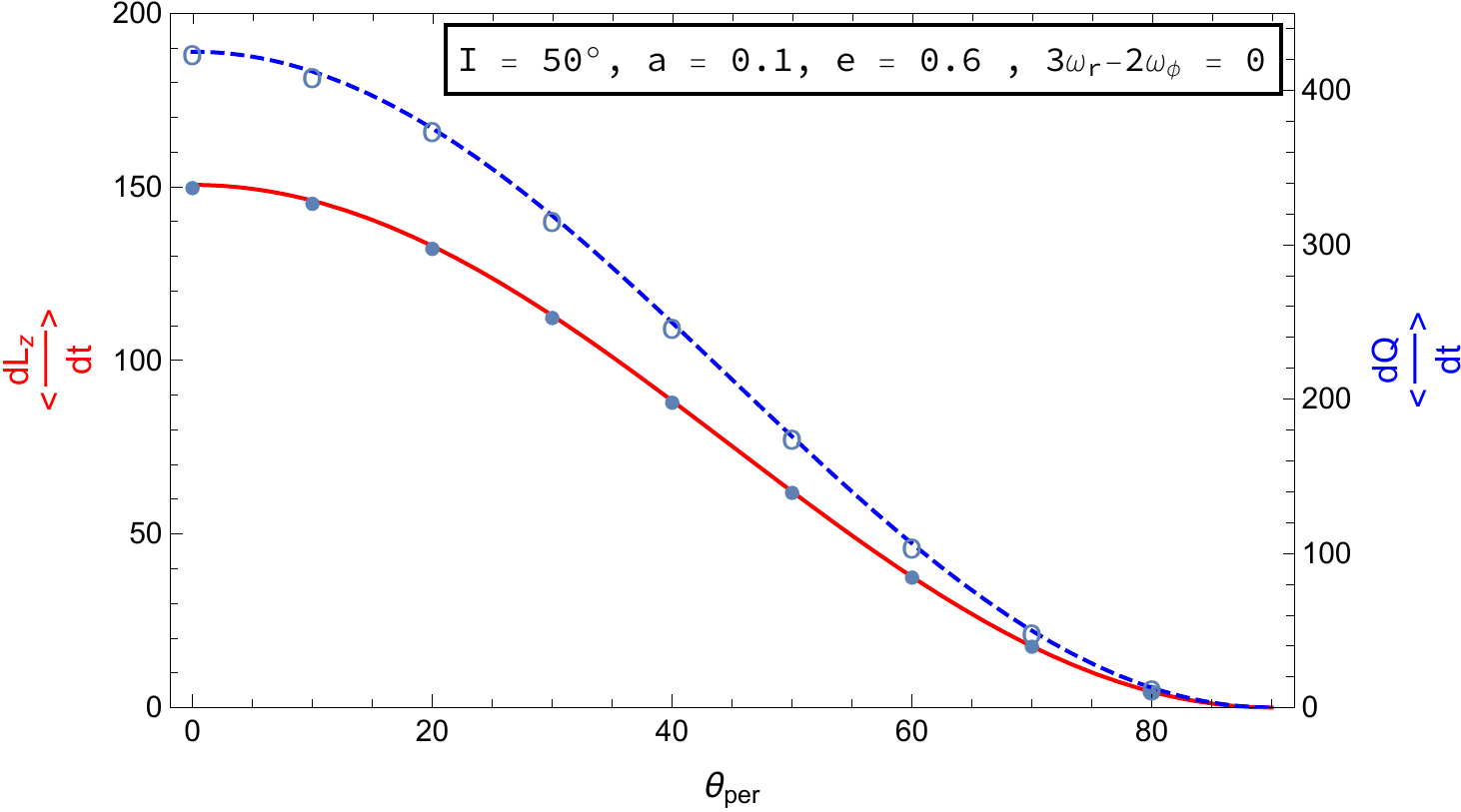}
\caption{\small Dependence of average change rate of the Carter constant (blue-dotted) on the perturber's inclination for a prograde orbit crossing $n:k:m=3:0:-2$. The dots represent the values obtained from the semi-analytic calculation, and curves denote the obtained fitting. Note that both $\langle dL_{z}/dt\rangle$ and $\langle dQ/dt \rangle$ are normalised by multiplying a factor of $(\epsilon/M)^{-1}$.}
\label{fig:meq2dependence}
\end{figure}

For the completeness, in Fig~\ref{fig:meq2dependence} we show the dependence of the $m =2$ mode on $\theta_{per}$. The $\rm{cos}^2 \theta_{per}$ like dependence highlights that the jump size from $m =2$ modes is maximum when the perturber is on the equatorial plane. This holds true irrespective of the orbit's direction.

Note that in Fig.~\ref{fig:meq1dependence}-\ref{fig:meq2dependence} $\langle dL_{z}/dt\rangle$ and $\langle dQ/dt \rangle$ are normalised by multiplying a factor of $(\epsilon/M)^{-1}$. The dots represent the values obtained from the semi-analytic calculations, and curves denote the obtained fitting (see Paper I for discussion on the construction of fitting formulae). The agreement between the semi-analytic evaluation and fitting agrees remarkably well with the error always less than $1\%$. The Mathematica notebook with fittings for all significant resonances is made available on \cite{BHPC}.

\begin{figure*}
  \centering
  \includegraphics[width=0.3\linewidth]{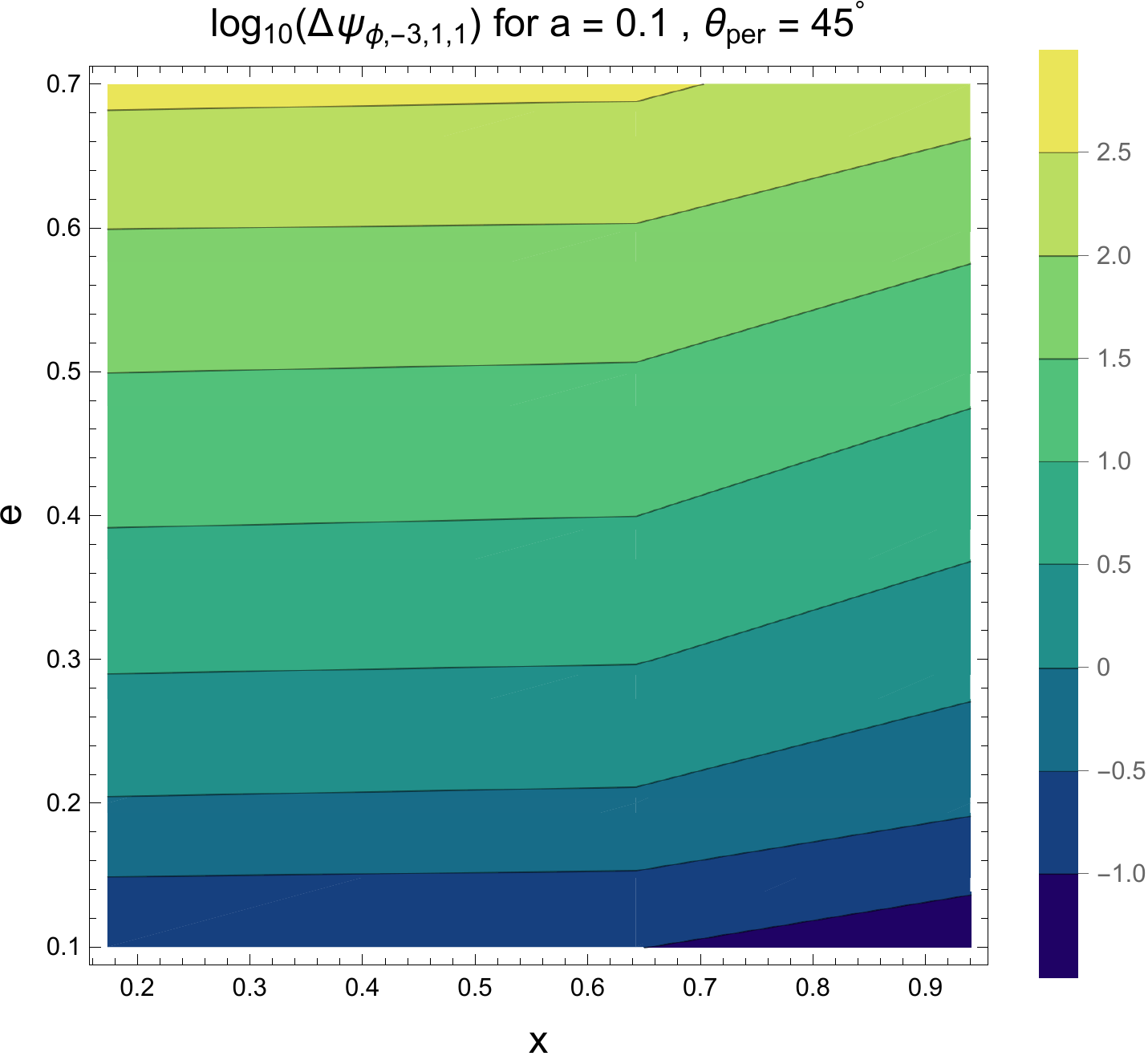}
    \hskip 0.5cm
\includegraphics[width=0.3\linewidth]{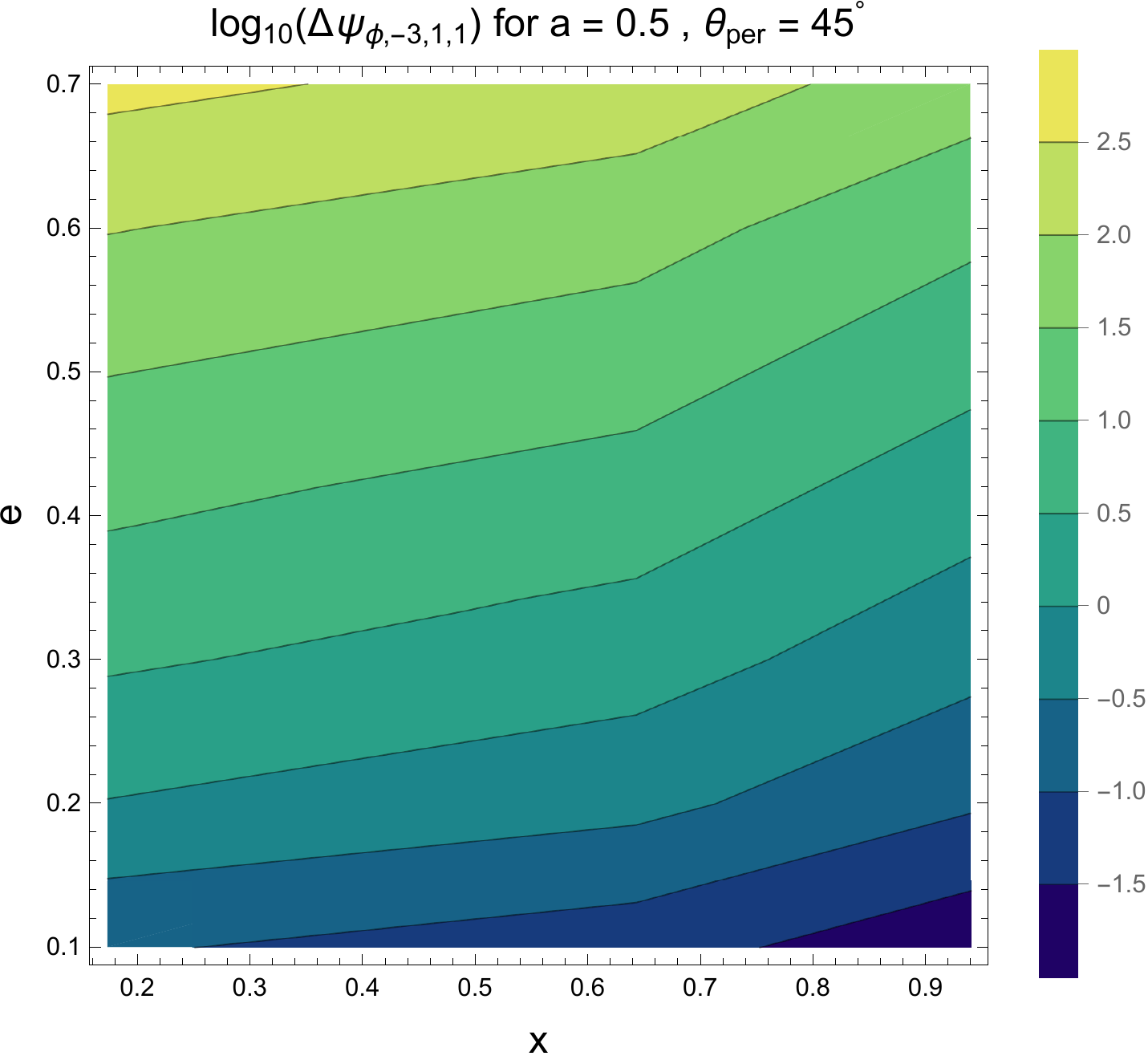}
\hskip 0.5cm
\includegraphics[width=0.3\linewidth]{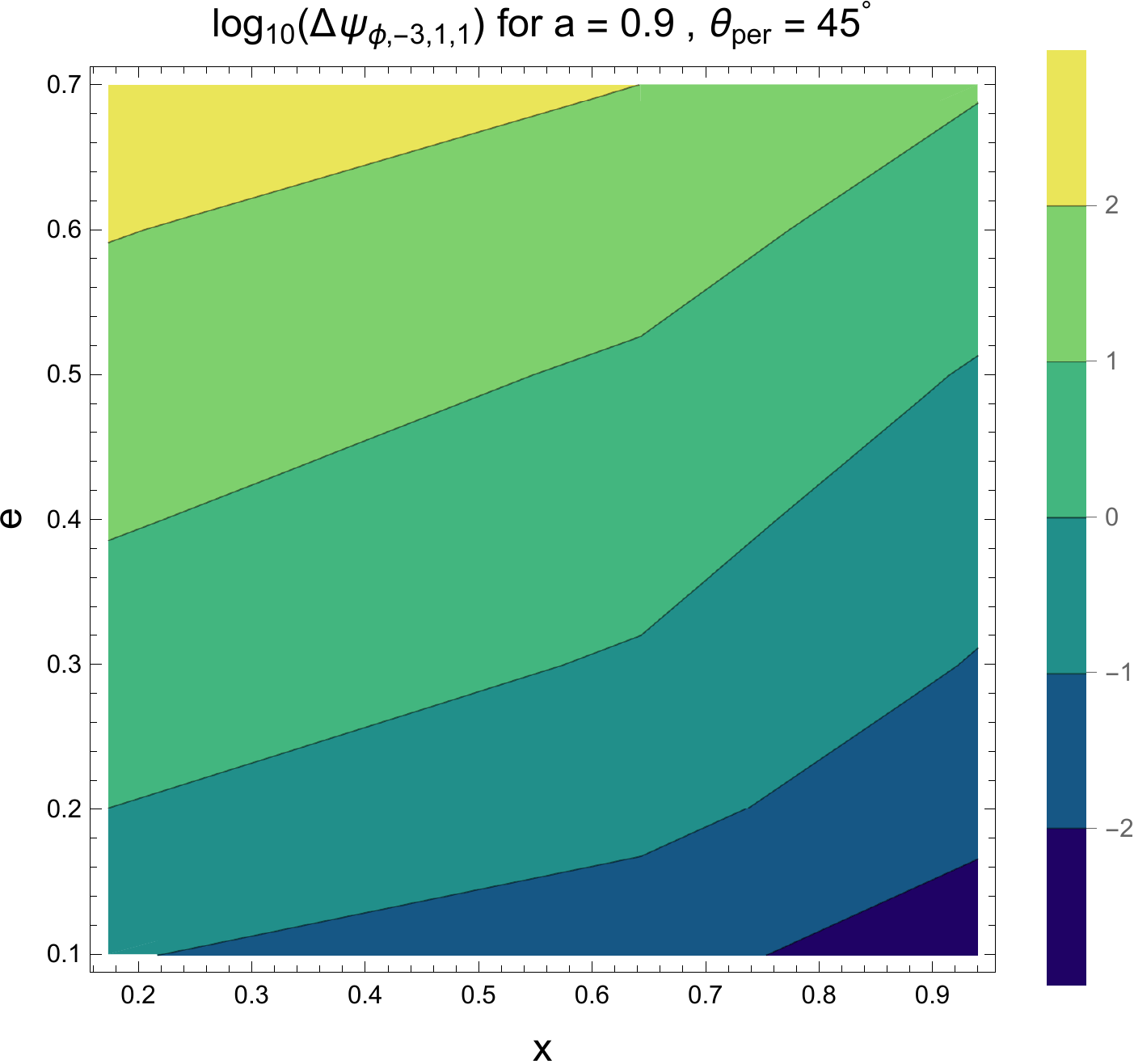}

\includegraphics[width=0.3\linewidth]{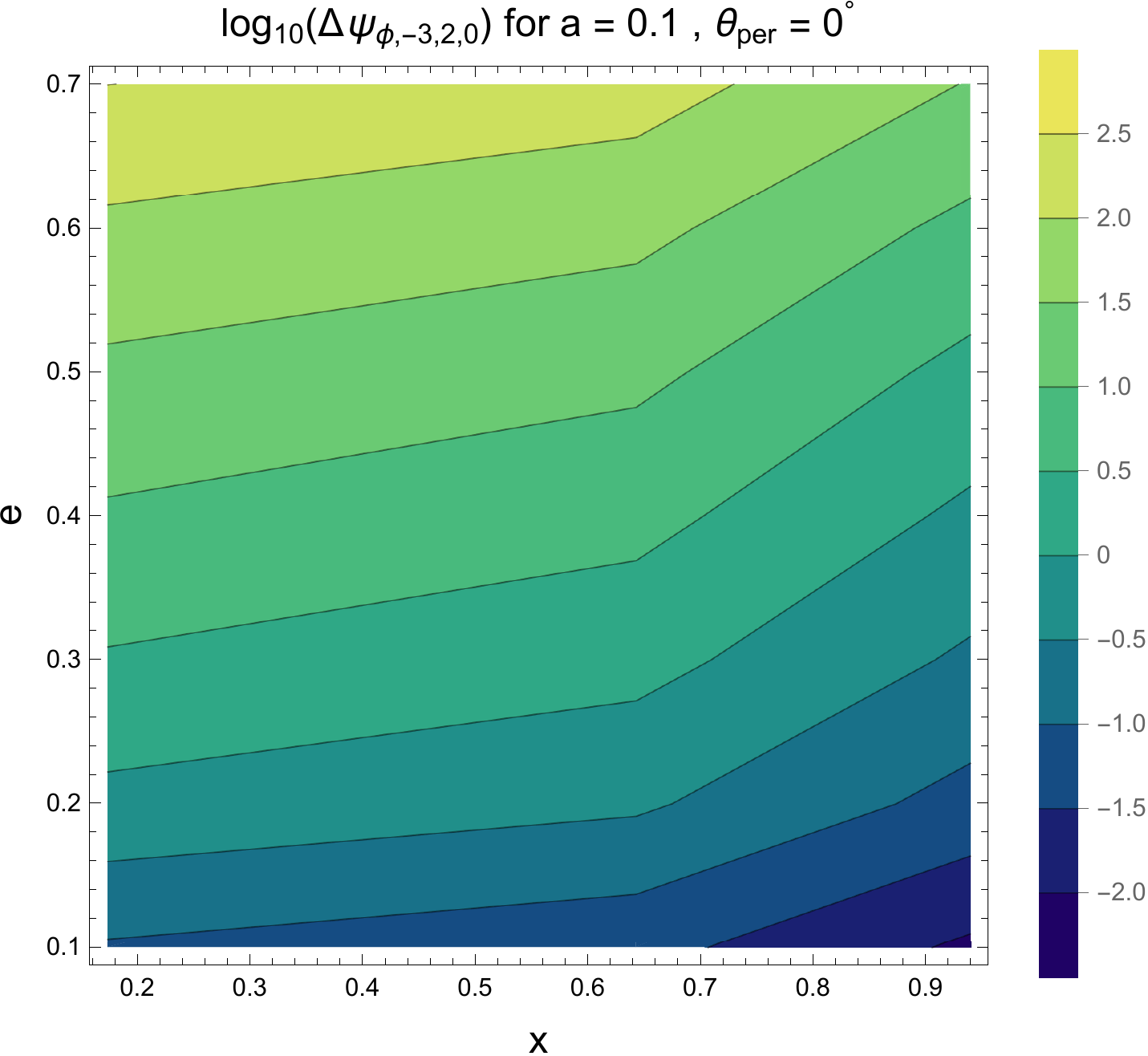}
    \hskip 0.5cm
\includegraphics[width=0.3\linewidth]{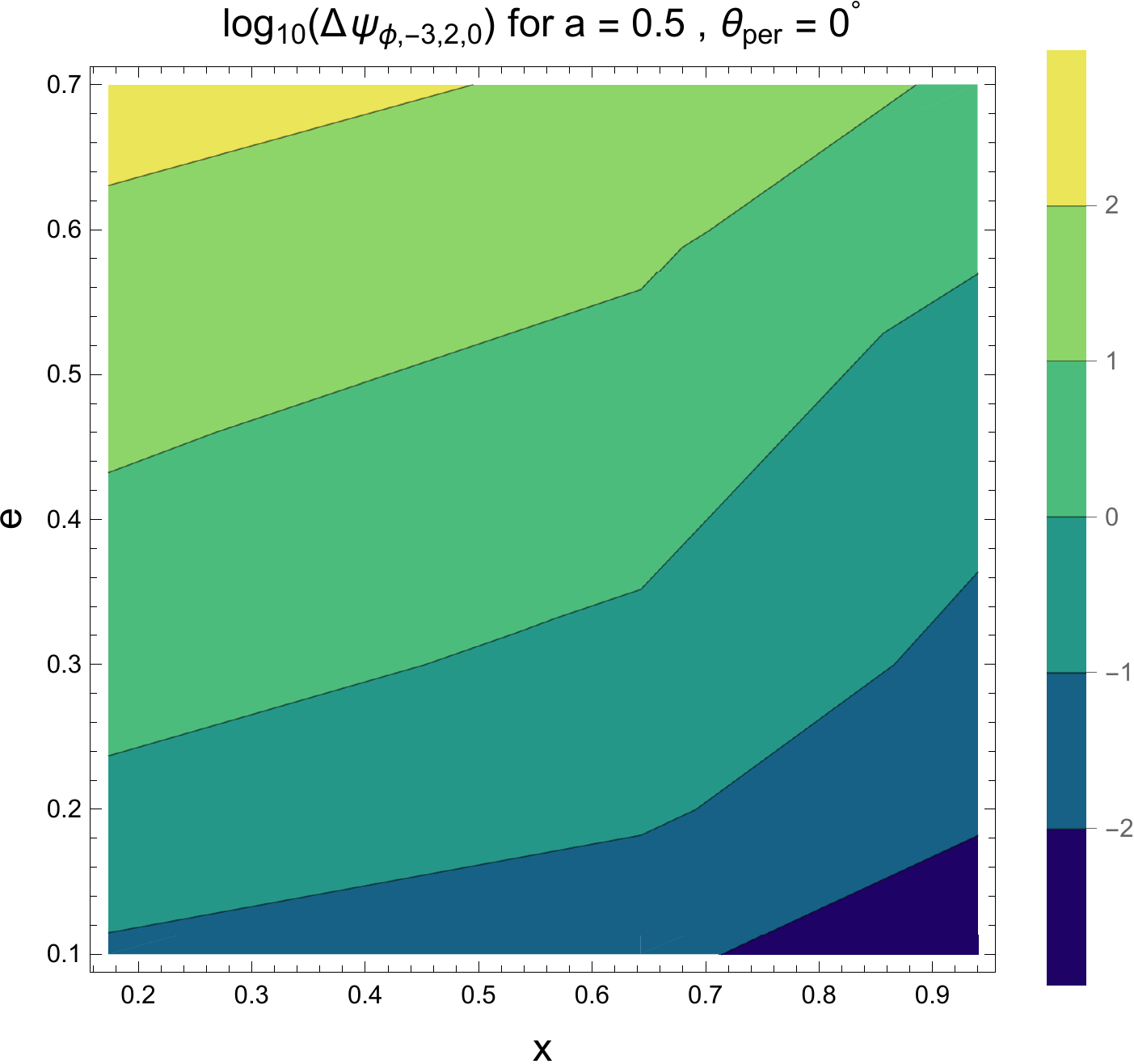}
\hskip 0.5cm
\includegraphics[width=0.3\linewidth]{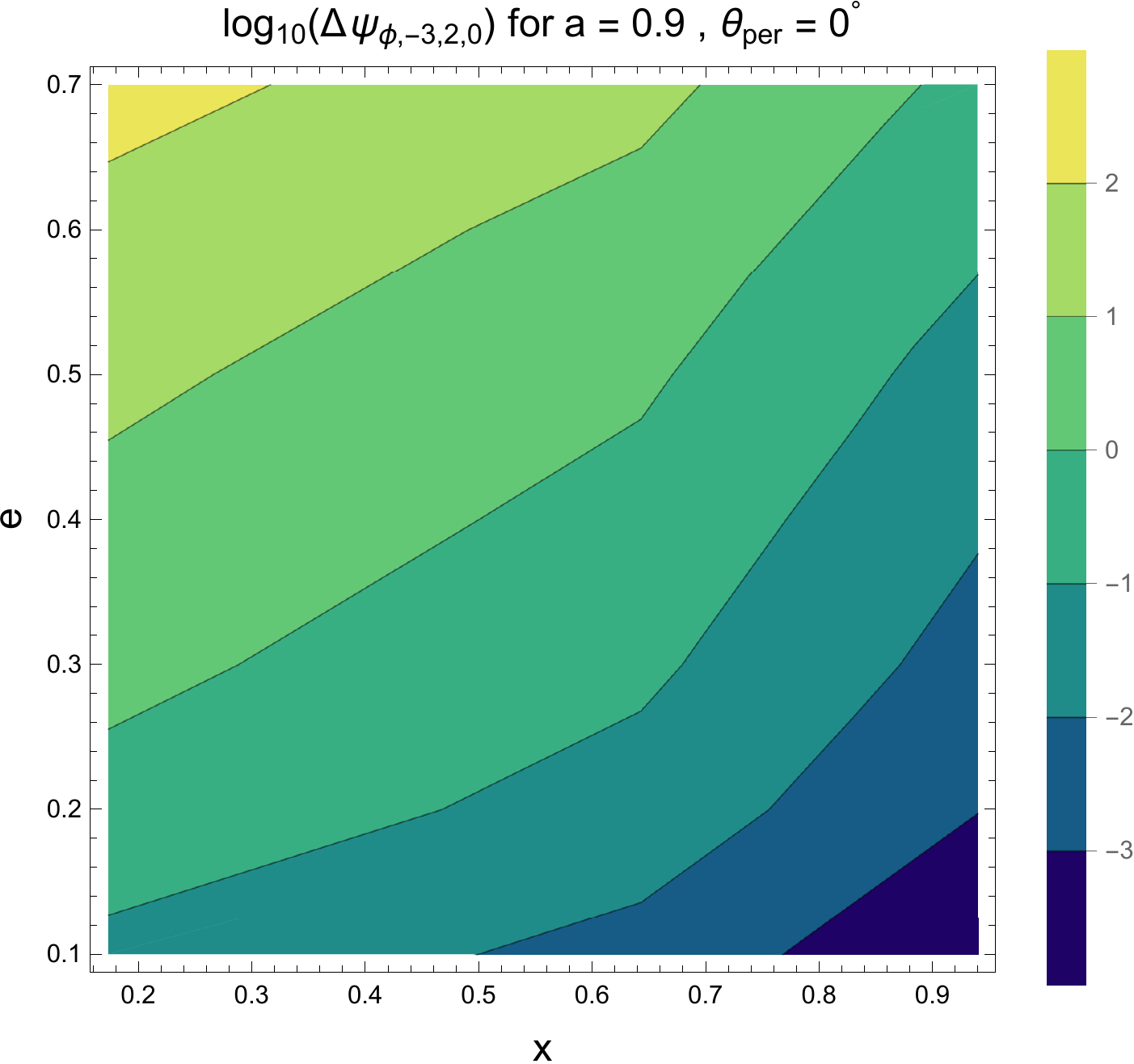}

\caption{Log of accumulated phase $\Delta \Psi_{\phi}$ for spin parameter $a = 0.1, 0.5, 0.9$ for a prograde orbit crossing the $-3:1:1$ (top panel) and $-3:2:0$ (bottom panel) resonance in the $\rm x$ - $e$ plane. The phase shift is computed for an EMRI with $M=4 \times 10^6 M_\odot, \mu =30 M_\odot$ under the influence of a tidal perturber with mass $M_\star=30 M_\odot$ at a distance of 10$\,$AU from the central MBH. Results for different sets of parameters can be estimated from the scaling relation given in Eq.~\eqref{eq:scale}.}
\label{fig:PSpro311-320}
\end{figure*}
\subsection{Dephasing due to tidal resonance}
Low order resonances encountered by EMRI orbits lie within the LISA frequency band for a typical mass ratio of $10^{-4}-10^{-7}$. As discussed in previous sections, an orbit crossing resonance experiences a jump in the constants of motion. Thus, resonances cause the orbit and hence, the phases to depart from the standard adiabatic evolution. Given a high SNR ($\sim 30$) of the waveform, LISA may resolve the phase in $\phi$ with an approximate sensitivity of $\Delta \Psi_{\phi}\sim 0.1$~\cite{Gair_2017,byh}. To quantify the dephasing, we compute the deviation in the GW phase using 
\beq
\Delta \Psi_{\phi} = \int_{0}^{T_{\rm plunge}} 2 \Delta \omega_{\phi} dt\,.
\label{eq:phase}
\eeq
The accumulation in phase is integrated from the onset of resonance (when the resonance condition is satisfied) up to the plunge time $T_{\rm plunge}$. The method of determining the phase evolution during the subsequent inspiral is discussed in detail in Paper I (Sec III-B). In short, for the implementation of the analytic expressions of fundamental frequencies~\cite{Fujita1_2009, Schmidt_2002}, our code employs the `Kerr Geodesic' Package from the Black Hole Perturbation Toolkit~\cite{BHPToolkit}. We evolve two orbits--- one with and without $\Delta J_{i}$ included. At each time $\omega_{\phi}$ for both the orbits are compared, and the difference in frequencies for these two evolutions is given by $\Delta \omega_{\phi}$. The factor of 2 in Eq.~\eqref{eq:phase} appears because the strongest harmonic in GWs (for quasi-circular equatorial EMRIs) is the quadrupolar mode ($l=2,m=2$). For eccentric orbits such as the one we have here, higher harmonics dominate, which can increase the amplitude of mismatch due to dephasing.

We set $M = 4 \times 10^6 M_\odot$, $\mu=M_\star=30 M_\odot$ and $R=10 \rm{AU}$. This distance is the same as in Paper I, but twice as far compared to \cite{byh} to give a more conservative estimate. In Fig~\ref{fig:PSpro311-320}, $\Delta \Psi_{\phi}$ is shown for prograde orbits crossing the $-3:1:1$ (top panel) and $-3:2:0$ (bottom panel) resonances in the ${\rm x}$ - $e$ plane for different spin parameters of the MBH.  The whole parameter space except for low eccentricity orbits and/or for a large spin is measurably affected by the $-3:1:1$ resonance. In a similar way, the $-3:2:0$ resonance impacts a large parameter space.
The dephasing increases with increasing eccentricity. Since both sample resonances are encountered early in the inspiral phase (see the upper panel of Fig~\ref{fig:rescomb}), the dephasing accumulates over hundreds of thousands of cycles before the plunge, and therefore affects most of the parameter range.

The accumulated phase shown for the sample resonances is calculated for fixed masses of the MBH, EMRI and the tidal perturber.  The accumulated phase $\Delta \Psi'_{nkm}$ for a different set of parameters $M',\mu',M'_\star,R'\,,\rm{x'}_\star$ simply scales as 
\begin{align}
    \Delta \Psi'= \Delta \Psi \bigg(\frac{M'}{M}\bigg)^{\!\!7/2} \bigg(\frac{\mu'}{\mu}\bigg)^{\!\!-3/2} \bigg(\frac{M'_\star}{M_\star}\bigg) 
 \bigg(\frac{\rm{x'}_\star}{\rm{x}_\star}\bigg)\bigg(\frac{R'}{R}\bigg)^{\!\!-3}\,.
\label{eq:scale}
\end{align}

So far, our results suggest that resonance jumps are sensitive to intrinsic orbital parameters, especially the orbital phases at resonance as discussed below Eq.~\eqref{eq:phasedependence}. Also, dephasing due to low-order tidal resonances can strongly impact the EMRIs detectable by LISA, assuming such tidal perturbers exist. Consequently, the waveform evolution becomes out of phase, compared to a template neglecting resonances --- reducing the detection rate because the signal-to-noise ratio falls as the phase error accumulates. It calls for careful modeling of waveforms that correctly detect EMRIs and estimate the parameters of EMRI and perturber. This serves as our motivation for the rest of the paper.

\section{Modeling Tidal Resonances}
\label{sec:4}
In this section, we first review how to evaluate the expected accuracy and systematic bias in parameter estimation, based on Fisher analysis. 
Next, we introduce the structure of the resonance model (RM), which is used to incorporate tidal resonances in waveforms and investigate the loss of signal and the systematic bias due to inaccurate modeling.

\subsection{Gravitational wave data analysis}
\label{subsec:data-analysis}
The output data $s(t)$ of a gravitational detector consists of random noise, $n(t)$ and possibly a gravitational wave signal $h(t;\bm{\lambda})$ characterized by a set of parameters $\bm{\lambda} =[\lambda_1 \dots \lambda_n]$ in $n$-dimensional parameter space.
\beq
s(t) = h(t ;\bm{\lambda} ) + n(t). 
\eeq
We assume that noise is given by a weakly stationary, Gaussian random process with zero mean. Under these assumptions, the Likelihood for the parameters $\bm{\lambda}$ is given by \cite{dataanalysis},
\beq
\label{eq:likelihood}
p(s|\bm{\lambda}) \propto \exp \left(-\frac{1}{2} \langle{s - h(\bm{\lambda})}|{s- h(\bm{\lambda})\rangle}\right)\, ,
\eeq
where $\langle{\cdot}|{\cdot}\rangle$ is a noise-weighted inner product defined as,
\beq
\label{eq:innerproduct}
\langle{a (t)}|{b (t)}\rangle :=4 \,\operatorname{Re} \int _{0} ^\infty \frac{\tilde{a} ^* (f) \tilde{b} (f) }{S_n (f)} \, df \, .
\eeq
$S_{n}(f)$ is the power spectral density (PSD) of the noise and the variable with tilde indicates the Fourier transform of the corresponding time series data.
Additionally, it is customary to define the signal-to-noise ratio (SNR),
\beq
\label{SNR}
\rho = \sqrt{\langle h|h\rangle},
\eeq
which characterizes the detectability of a signal by a detector with a given noise power spectrum.

We define two other quantities which serve as a measure of similarity between two template waveforms $h_a = h(t;\bm{\lambda}_a)$ and $h_b = h(t;\bm{\lambda}_b)$, the Overlap $\mathcal{O}(h_a,h_b)$ and Mismatch $\mathcal{M}(h_a,h_b)$, by 
\begin{align}
    \mathcal{O}\langle h_{a},h_{b}\rangle &= \frac{\langle h_{a}|h_{b}\rangle}{\sqrt{\langle h_{a}|h_{a}\rangle \langle h_{b}|h_{b}\rangle}} \label{eq:overlap}\\
    \mathcal{M}(h_{a},h_{b}) &= 1 - \mathcal{O}(h_{a},h_{b}) \label{eq:mismatch}.
\end{align}
If $\mathcal{O}(h_{a},h_{b}) =  1$, the two waveforms are identical. Waveforms with $\mathcal{O}(h_{1},h_{2}) = 0$ are mutually orthogonal. 
In contrast, by definition, the smaller $\mathcal{M}(h_{a},h_{b})$, the better the match is.

If we want to estimate how accurately parameters are measured, it is helpful to calculate the Fisher Information matrix $\Gamma_{ij}$.
When a strong signal with parameters $\bm{\lambda}$ is present in the detector output, the likelihood is strongly peaked in the parameter space at the best-fit (BF) parameter set close to the true values. Namely, the measurement error 
\beq
\Delta\bm{\lambda} = \bm{\lambda}_{\rm{BF}} - \bm{\lambda}\,,
\eeq
is small.
Then, we expand $h(\bm{\lambda})$ up to linear order in $\Delta \bm{\lambda}$ (truncating higher orders terms given the smallness of $\Delta \bm{\lambda}$ in the strong signal limit) and substitute it into~\eqref{eq:likelihood}. On substitution, the Likelihood function becomes
\beq
\label{FisherProbDens}
p(s|\bm{\lambda})  \propto \exp\left(-\frac{1}{2}\sum_{i,j}\Gamma_{ij}\Delta\lambda^{i}\Delta\lambda^{j}\right).
\eeq
where
\beq
\label{eq:Fisher_Matrix}
\Gamma_{ij} = \left(\frac{\partial h}{\partial \lambda^{i}} \bigg\rvert \frac{\partial h}{\partial \lambda^{j}}\right).
\eeq
The waveform derivatives $\partial_j h$ are computed numerically using the five-point stencil formula such that the numerical error scales at fourth order in the derivative spacing. The probability function shows that the inverse of $\Gamma_{ij}$, known as the covariance matrix, contains information about variances of parameter measurement error (diagonal elements) as well as correlations of errors among different parameters (off-diagonal elements). In particular, the statistical error in the estimate of the parameter $\lambda_i$ can be evaluated by
\beq
\label{eq:stat_err}
\Delta \lambda_i^{stat} = \sqrt{(\Gamma^{-1})^{ii}}\,.
\eeq
From~\eqref{eq:Fisher_Matrix}, the Fisher Matrix scales as $\sim \rho^{2}$, therefore $\Delta \boldsymbol{\lambda}$ scales as $\sim \rho^{-1}$.

Besides the errors induced by noise, there can be a possible systematic uncertainty that is not parameterized in our waveform models.  For instance, if we use an inaccurate waveform model $h_m(\bm{\lambda})$ to estimate the parameters $\bm{\lambda}_0 $ of a signal actually described by a model $h_t(\bm{\lambda})$, the recovered parameters will be affected by systematic errors given by \cite{2007Curt},
\beq
\label{eq:sys_err}
    \Delta \lambda_i^{sys}  =  
        (\Gamma ^{-1})^{ki} \langle{\partial_k h(\bm{\lambda}_0)}|{
        h_t(\bm{\lambda}_0) - h_m(\bm{\lambda}_\text{BF}) \rangle}      
     \, .
\eeq
This error is independent of the strength of the signal. Therefore, if exists, it will dominate over noise-induced error, 
whenever the data quality is sufficiently good.

In this work, we evaluate the above mentioned errors, by comparing two kinds of waveforms: resonant waveforms which are produced using the AAK module \cite{Chua:2017ujo,michael_l_katz_2020_4005001}  implemented in the RM (discussed in Sec.~\ref{subsec:ERM}), and non-resonant waveforms where we ``turn-off" the jumps induced by tidal resonances. For our Fisher analysis, we assume that from the data we determine the following ten parameters:
\beq
\bm{\lambda} = \left( \,\log_{10} \frac{M}{M_\odot},\log_{10} \frac{\mu}{M_\odot}, a, p, e, {\rm x} , q_{r},q_{\theta},q_{\phi},\widetilde{\epsilon} \,\right) \, ,
\label{eq:fisherparam}
\eeq

where $q_{r},q_{\theta},q_{\phi}$ are the initial phases of an EMRI orbit and  $\widetilde{\epsilon}$ is the normalized (by the fiducial value of $\epsilon$) tidal parameter. These intrinsic parameters govern the detailed dynamical evolution of a system, regardless of where or how an observer observes it. For computational convenience, we are not including extrinsic parameters such as the sky location angles ($\theta_S, \phi_S$) and the angles pointing to the direction of the MBH’s spin ($\theta_K, \phi_K$) in this list, since they are not strongly correlated with the intrinsic parameters. The luminosity distance $D_L$ of the source is rescaled for each waveform to fix the SNR to 30. Our fiducial values for the masses of the EMRI system are $M = 10^6 M_\odot$ and $\mu = 30 M_\odot$. The perturber of mass $M_{\star} = 30 M_\odot$ is placed at a distance of $5AU$ on the equatorial plane, resulting in the following fiducial tidal parameter $\epsilon \sim 2.3 \times 10^{-13}$ for ${\rm x} = 1$. 

\begin{figure}
  \centering
  \includegraphics[width=7.5cm]{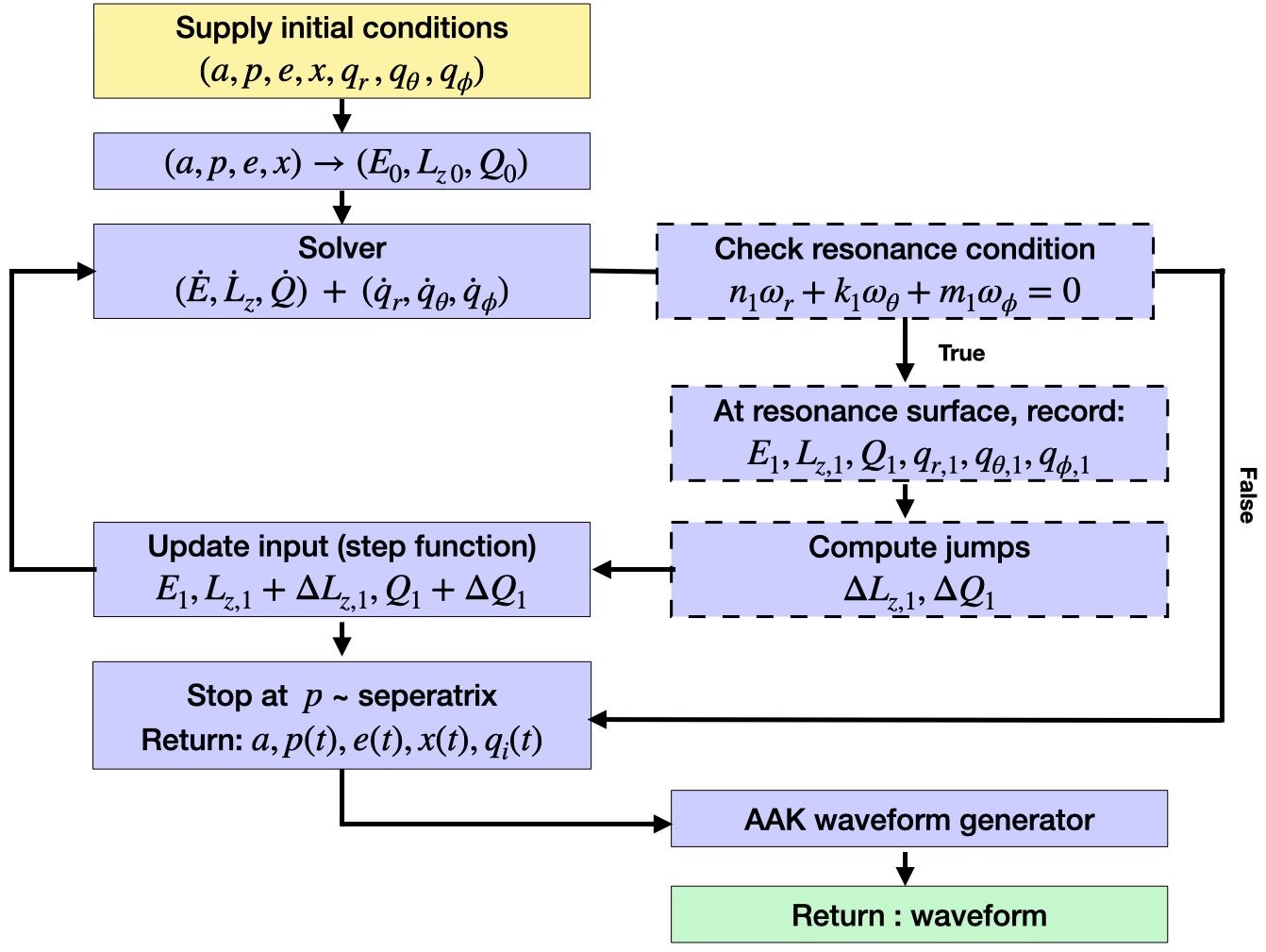}
\caption{\small Workflow of Resonance Model.}
\label{fig:flowchart}
\end{figure}
\subsection{Resonance model}
\label{subsec:ERM}
\begin{figure*}
  \centering
  \includegraphics[width=0.4\linewidth]{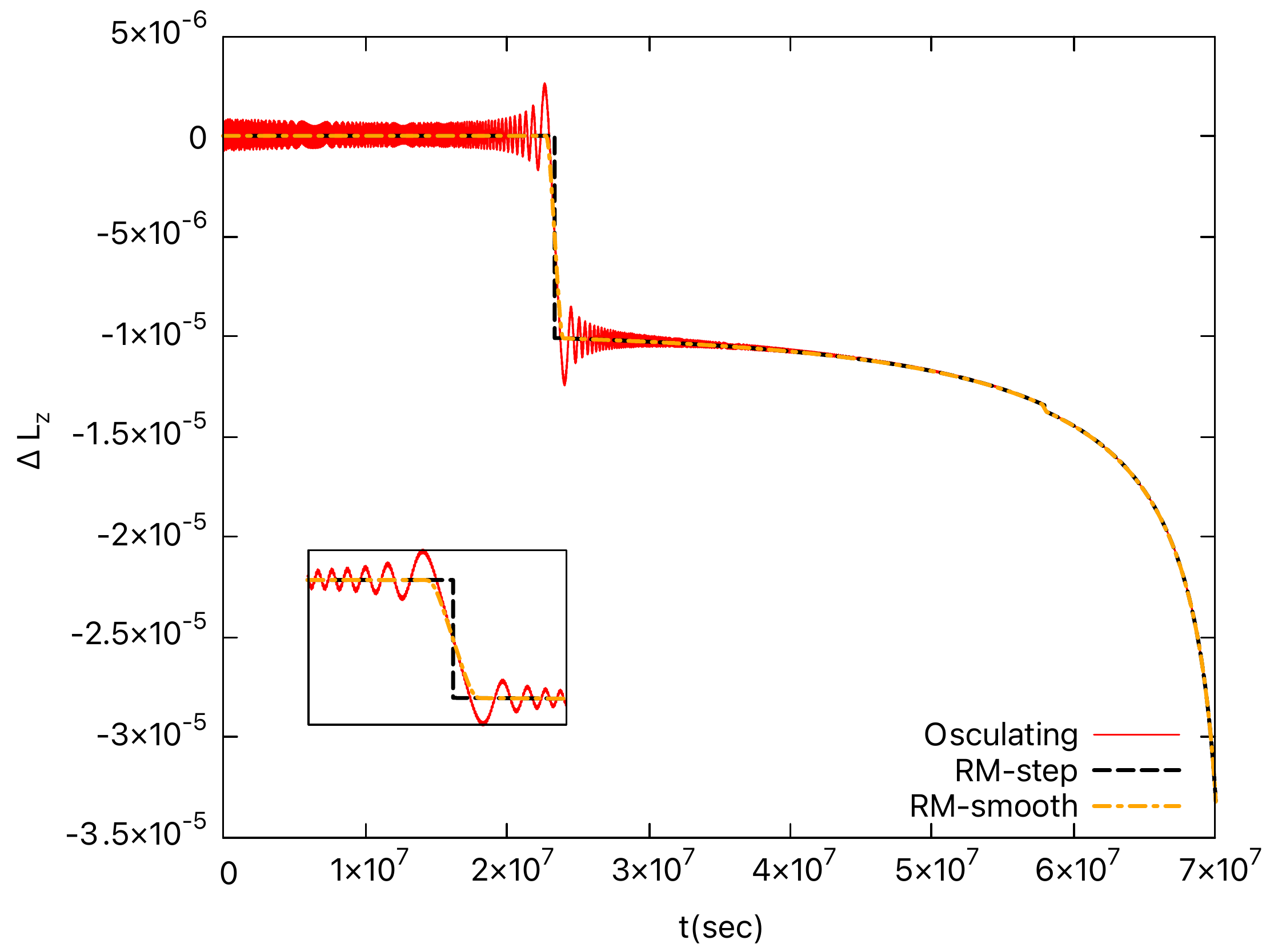}
    \hskip 0.5cm
\includegraphics[width=0.4\linewidth]{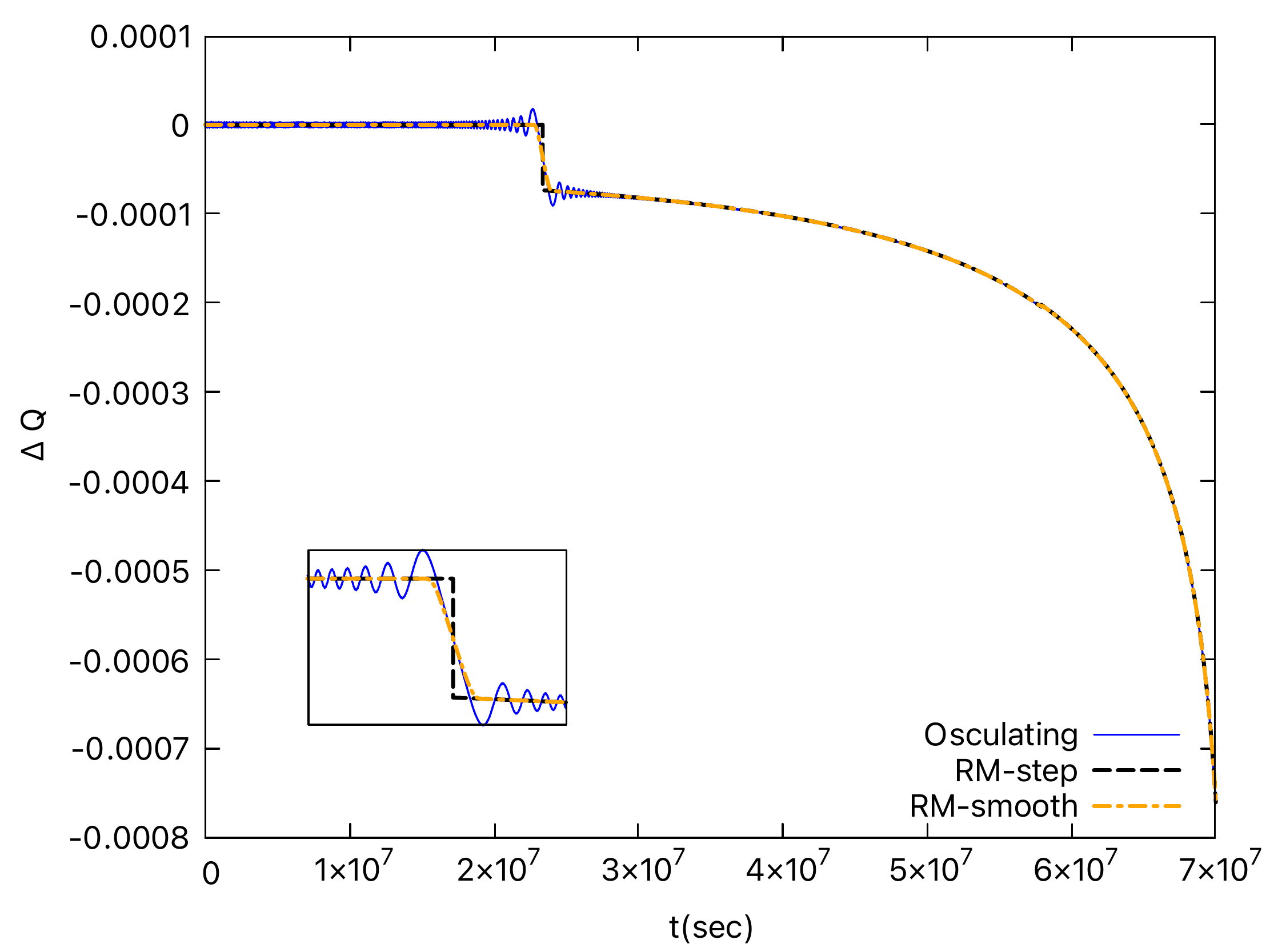}
\caption{\small The left figure shows the difference in $L_z$ between the orbits evolved with and without tidal resonance effect. When the orbit undergoes a resonance, there is a jump in the action variables {\bf J}. Black dashed lines illustrate the evolution of $\Delta L_z$ using a step impulse function in the RM, whereas orange (dashed-dotted) lines represent evolution tracks using a `smooth' impulse function. Similarly, the right figure shows the evolution of the Carter constant $Q$. The initial conditions for this orbit are $(a,p,e,\rm{x}) =(0.9,11.8,0.8,0.0187)$, and the trajectory crosses two resonances, $n:k:m = 3:0:-2$ and $n:k:m = 3:-4:-2$ around $t \sim 2.2 \times 10^{7} \text{and} \sim 5.8 \times 10^{7}$, respectively. The fast oscillations in both figures correspond to timescales of the orbital motion. The inset plot shows zoomed-in evolution near the $3:0:-2$ resonance.}
\label{fig:deltaLQ-osc-ERM-stepvssmooth}
\end{figure*}

The Effective Resonance Model (ERM) is a phenomenological model developed recently to study EMRI resonances. It was constructed using the resonance jumps as free parameters and applied to the case of self-force resonances \cite{speri2021assessing}. Following the implementation of \cite{speri2021assessing}, we extend the ERM to incorporate tidal resonances.
We refer to our working code as the resonance model (RM); the word ``Effective" has been discarded since we are not using the resonance jumps as free parameters. The flowchart is shown in Fig~\ref{fig:flowchart}. The solver employs flux and phase evolution equations to obtain the trajectory, given some initial condition ($E_0, L_{z0}, Q_0$). Our calculations use the fifth order post-Newtonian (5PN) fluxes generated by the post-Newtonian (PN) approximation in first-order black hole perturbation theory \cite{Fujita_2020}. The right-hand side of the phase evolution equations are corresponding Kerr orbital frequencies \cite{Schmidt_2002}. The resonance condition is checked at each time step of the solver (using the adaptive time step and event handling tool in the \texttt{Solve-ivp} ODE package in \texttt{Python}) for some low order integer $m,k, \text{and}\, n$. If the resonance condition is satisfied, we record the orbital parameters at the resonance surface and use them to estimate the jump size of the resonance due to the tidal field using the analytic fits obtained from our semi-analytic calculations  \cite{PaperI}. Once the jump sizes $ \Delta L_{z}$ and $\Delta Q$ are measured, we update the constants of motion for the next time step using a step function. In \cite{speri2021assessing}, the resonance jump is implemented using a ``smooth" impulse function. In this study, however, we find that using a smooth function instead of a step function did not affect our results (shown in Fig~\ref{fig:deltaLQ-osc-ERM-stepvssmooth}). Consequently, we choose to implement the faster and simpler step function. We stop the evolution of the trajectory once the separatrix, where $\omega_r$ vanishes, is reached. The orbital parameters and phases are then fed to the Augmented Analytic Kludge (AAK) module to obtain the waveform. Our code makes use of the modular FEW package \cite{michael_l_katz_2020_4005001}. 

\begin{figure*}
  \centering
  \includegraphics[width=0.75\linewidth]{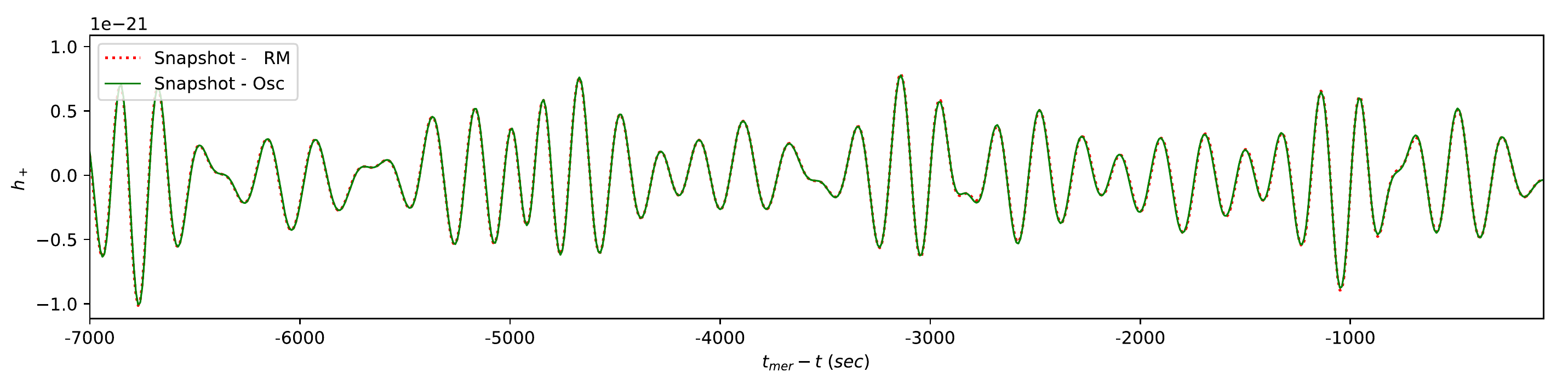}
    \vskip 0.1cm
\includegraphics[width=0.75\linewidth]{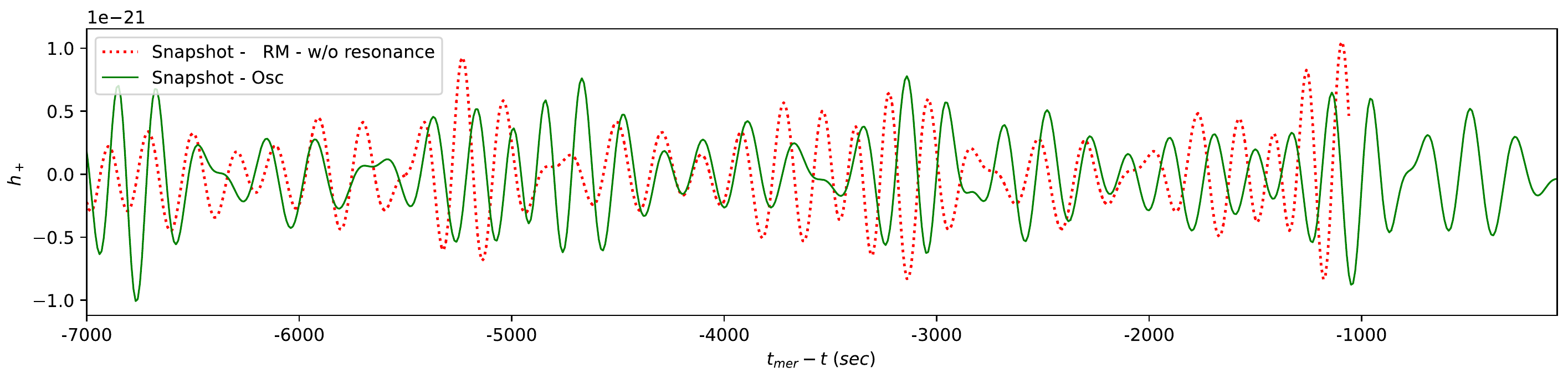}
\caption{\small Snapshot of $h_+$ waveforms obtained from the RM and osculating method a few hours before plunge. Top panel: comparison of $h+$ from RM (with resonance jump included) and 
the one from the osculating method. Lower panel: comparison of $h+$ from the RM without resonance jump and the one from the osculating method. }
\label{fig:waveform-snap}
\end{figure*}
\section{Results}
\label{sec:5}
In this section, we compare the jump obtained from analytic fits with the result obtained by the numerical osculating code, to find a good agreement between the two. Using the RM and Fisher matrices, we show mismatches for different initial conditions and assess the measurement precision of EMRI orbital parameters and tidal parameters. We also compute the systematic bias that would be induced by ignoring resonances.

\begin{table}[!]
\centering
\begin{tabular}{|c|c|c|}
\hline 
&&
\\[-.5em]
IC&$(a,p,e,\rm{x})$&$t_{3:0:-2}$ ($10^7$sec)
\\[.5em]
\hline 
1&$0.1,11.5,0.7,0.642 $&$\sim 1.64 $
\\[.4em]
2&$0.5,10.5,0.8,0.642 $&$\sim 1.85 $
\\[.4em]
3&$0.7,11.0,0.7,0.342 $&$\sim 1.71 $
\\[.4em]
4&$0.9,11.8,0.8,0.087 $&$\sim 2.24 $
\\[.4em]
\hline 
\end{tabular}
\caption{\small Initial conditions for EMRI orbit. The last column shows the time of $n:k:m = 3:0:-2$ resonance encounter.
}
\label{tab:models}
\end{table}

\subsection{Mismatch}
Dephasing induced by tidal resonances accumulates over the inspiral, resulting in a decrease in the overlap~\eqref{eq:overlap} between resonant and non-resonant waveforms after resonance encounter. In this section, we analyze the evolution of the mismatch $\mathcal{M}$~\eqref{eq:mismatch} as a function of the final time for different initial conditions listed in Table~\ref{tab:models}. These conditions were chosen since they cover a broad range of possibilities for astrophysical EMRI events that may be measured by future low-frequency GW missions. All initial conditions are subject to a 30 $M_\odot$ tidal perturber at a distance of 5 AU on the equatorial plane, and the EMRI inspiral lasts for $\sim 1-2$ years. The parameters chosen for tidal perturber are motivated by the Fokker-Planck simulation study that suggests a population of stellar-mass BHs at a median distance of $\sim 5AU$ \cite{emami2020detectability}.  We note that for the chosen set of parameters  $\tau_{\rm res} \sim \tau_{\rm td}$, thereby violating the stationary perturbation approximation. However, we leave the impact of a dynamical tidal perturber on the resonances for future work.

We first determine the consistency of the resonance model by comparing its trajectory evolution with the numerical osculating trajectory. The forced osculating orbital elements method ~\cite{osculating-kerr,PhysRevD.77.044013} uses the tidal force computed from the metric perturbation $h_{\alpha \beta}$ and for the inclusion of radiation reaction effects, 5PN fluxes ~\cite{Fujita_2020,BHPC} are employed. Using the osculating code, we ran two simulations for an inspiral orbit --- with and without the effect of the tidal force with the same initial conditions. To extract the jump size, we compute the difference ($ \Delta L_z$ and $\Delta Q$) between the full trajectory (tidal force + 5PN) and adiabatic (only 5PN) trajectory. A similar trajectory evolution is obtained by means of the resonance model, where the inspiral is derived mostly from 5PN adiabatic fluxes, and the jump is added only when the resonance condition is satisfied.

The comparison is presented in Fig~\ref{fig:deltaLQ-osc-ERM-stepvssmooth}. We show the differences $\Delta L_z$ (left, red) and $\Delta Q$ (right, blue) for IC4 crossing two resonances $3:0:-2$ and $3:-4:-2$. The apparent thickness of the lines shown in the figures is due to oscillations on the orbital timescale. In this plot, the evolution of the respective quantities obtained from the RM are overlaid for both the `step' (black, dashed) and `smooth' (orange, dashed-dotted) impulse functions that model the jump obtained from the fitting formulae. This figure shows a good agreement of jump size (and therefore resonant phase) and overall evolution between the RM and osculating method regardless of the choice for the impulse function. The difference between the evolutions from the two impulse functions is $\sim \mathcal{O}(10^{-8})$, too small to resolve on the scale in Fig~\ref{fig:deltaLQ-osc-ERM-stepvssmooth}.

Additionally, we compare the agreement between the RM and osculating methods at the waveform level. The trajectory information from both models is fed into the AAK module, and the snapshot of the waveform ($+$ polarization) a few hours just before the plunge is displayed in Fig~\ref{fig:waveform-snap}. We can see a remarkable phase match between the two in the top panel. In the lower panel, we switch off the jump in the RM waveform and compare it with the osculating waveform.  As a result of dephasing, there is a clear disagreement in the waveforms. Furthermore, we see that in the present example the merger time corresponding to the end point of the waveform is delayed for the osculating waveform, which takes the tidal jump into account, because of the positive jump in $L_z$ and $Q$.
\begin{figure}
		\includegraphics[width=8.0cm]{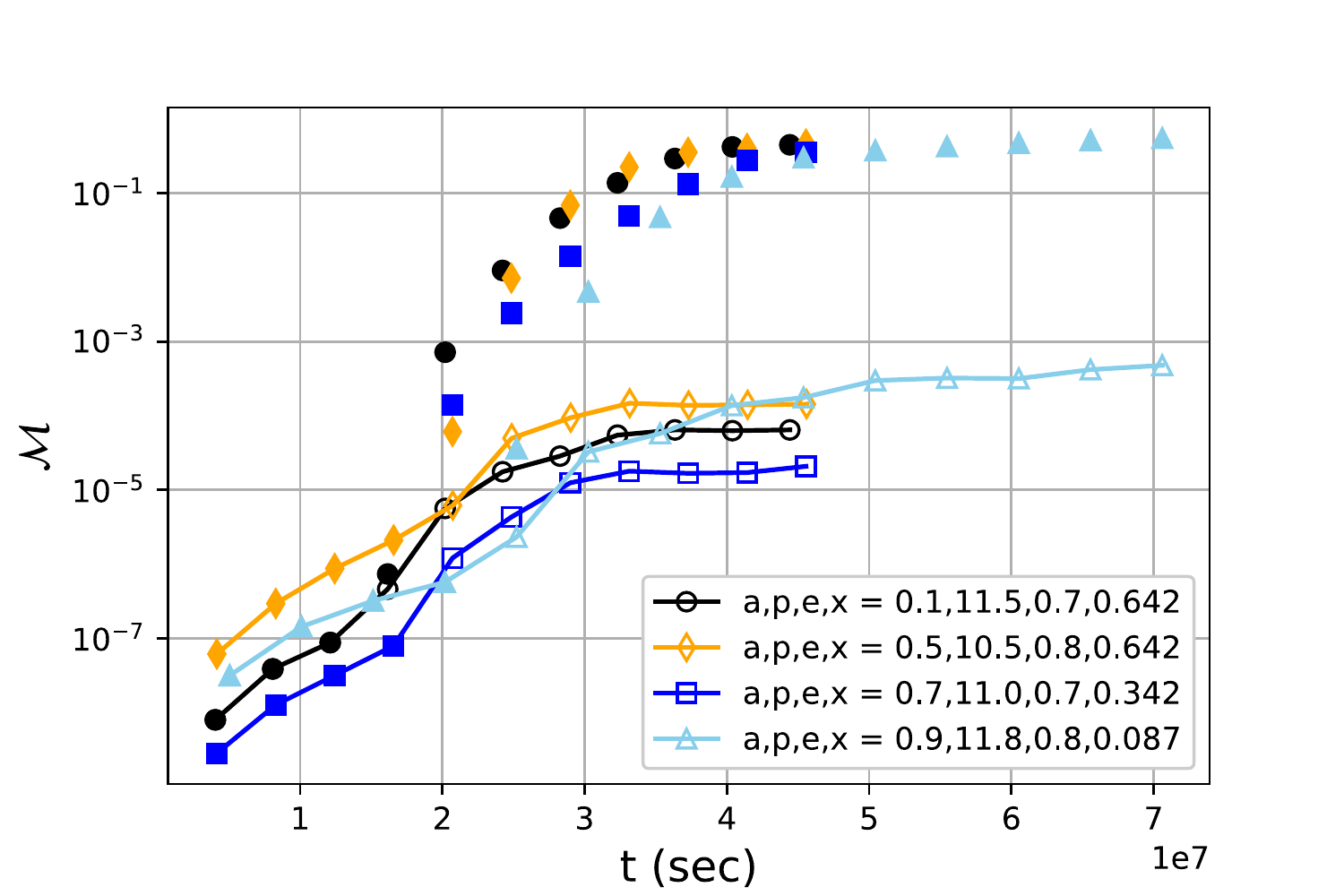}
\caption{\small The cumulative mismatch between resonant and non-resonant waveforms using the RM and osculating method. Here, the unfilled markers show the cumulative mismatch between the resonant waveforms using the RM and osculating method for different initial conditions (see Table \ref{tab:models}) crossing two resonances $n:k:m = 3:0:-2$ and $n:k:m = 3:-4:-2$ during the evolution. In contrast, the filled markers show the mismatch if resonances are neglected in the waveform model. The filled markers overlay the unfilled ones before crossing the first resonance for every initial condition. The condition with spin 0.9 has the longest inspiral time because the separatrix is close to the central BH compared to the low spin EMRIs.}
		\label{fig:Cumulative_Mismatch}
	\end{figure}
	
\begin{figure}
    \includegraphics[width=8.0cm]{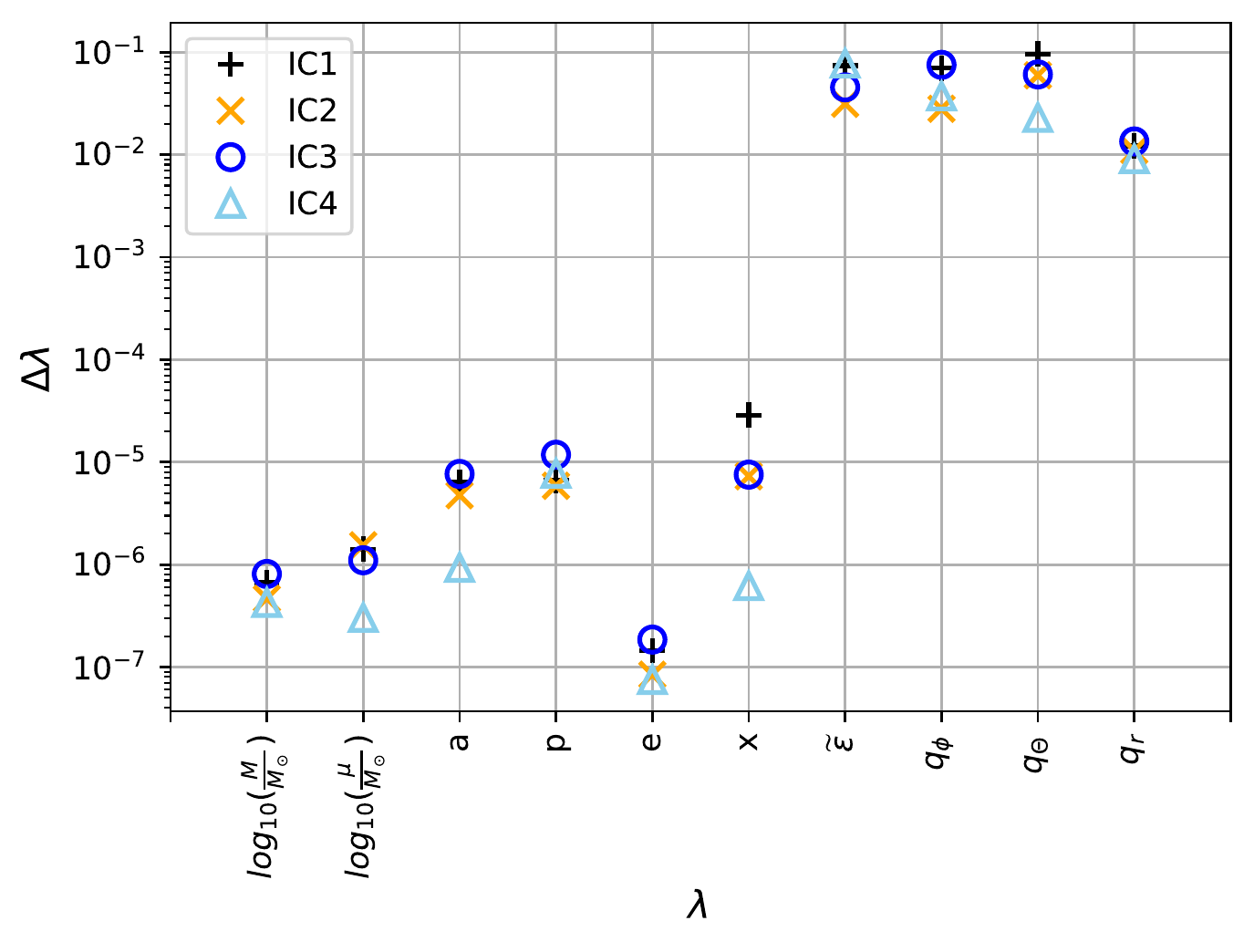}
    \caption{\small Measurement precision $\Delta \lambda$ of EMRI's intrinsic and tidal parameters for the initial conditions listed in Table~\ref{tab:models}. All the signals are normalized to SNR = 30.}
 	\label{fig:delta_params}
\end{figure}

\begin{figure}
		\includegraphics[width=8.0cm]{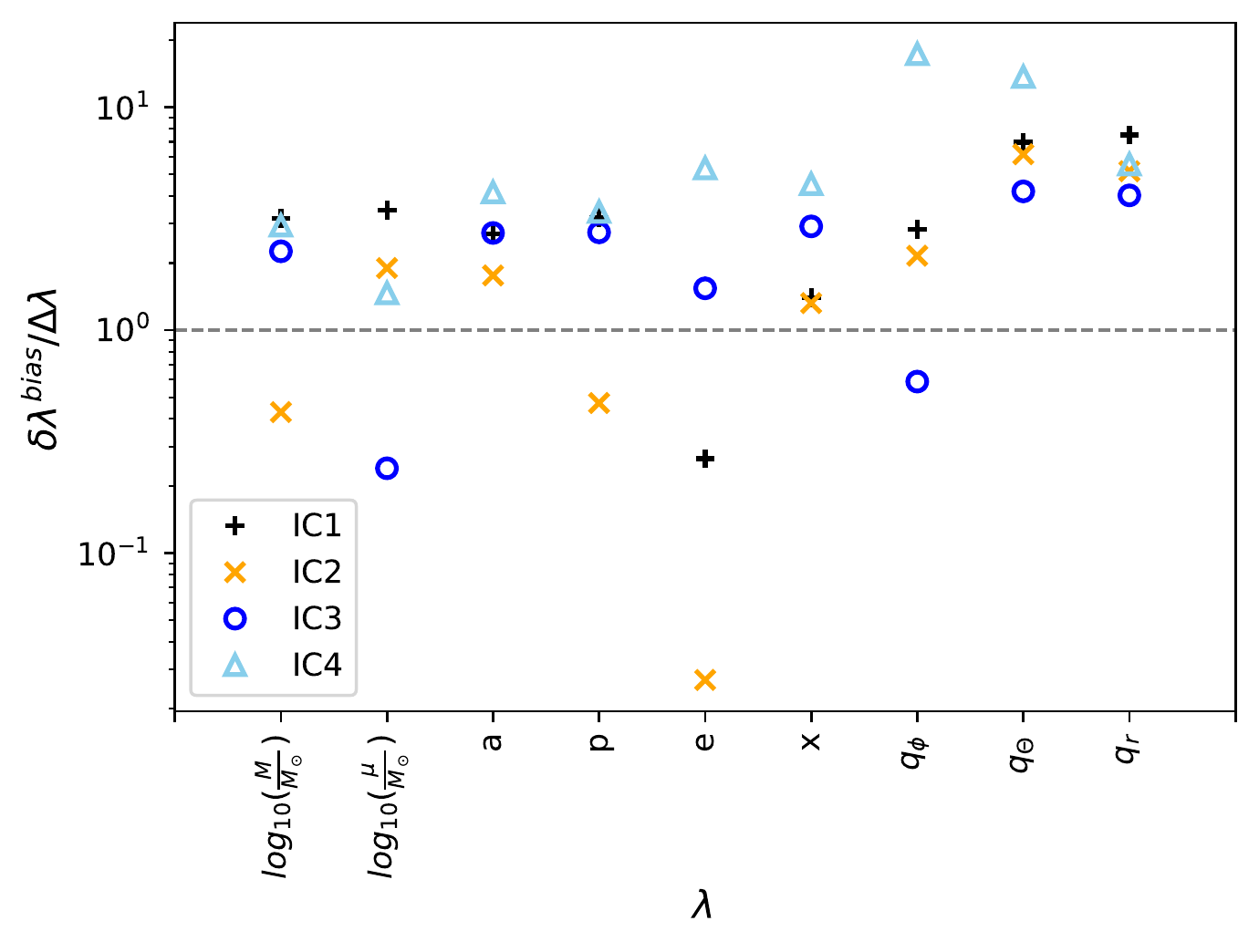}
\caption{\small The ratio $\delta\lambda^{bias}/\Delta\lambda$ between the size of the systematic and statistical errors is shown for the initial conditions listed in Table~\ref{tab:models}. The dashed grid line indicates that the ratio is equal to 1. For $\delta\,\lambda^{bias}/\Delta\lambda > 1$, the bias induced by inaccurate waveform modeling is more significant than that caused by the noise fluctuations in the detector.}
		\label{fig:sys_bias_params}
	\end{figure}	

\begin{figure*}
  \centering
  \includegraphics[width=\textwidth,height=\textheight,keepaspectratio]{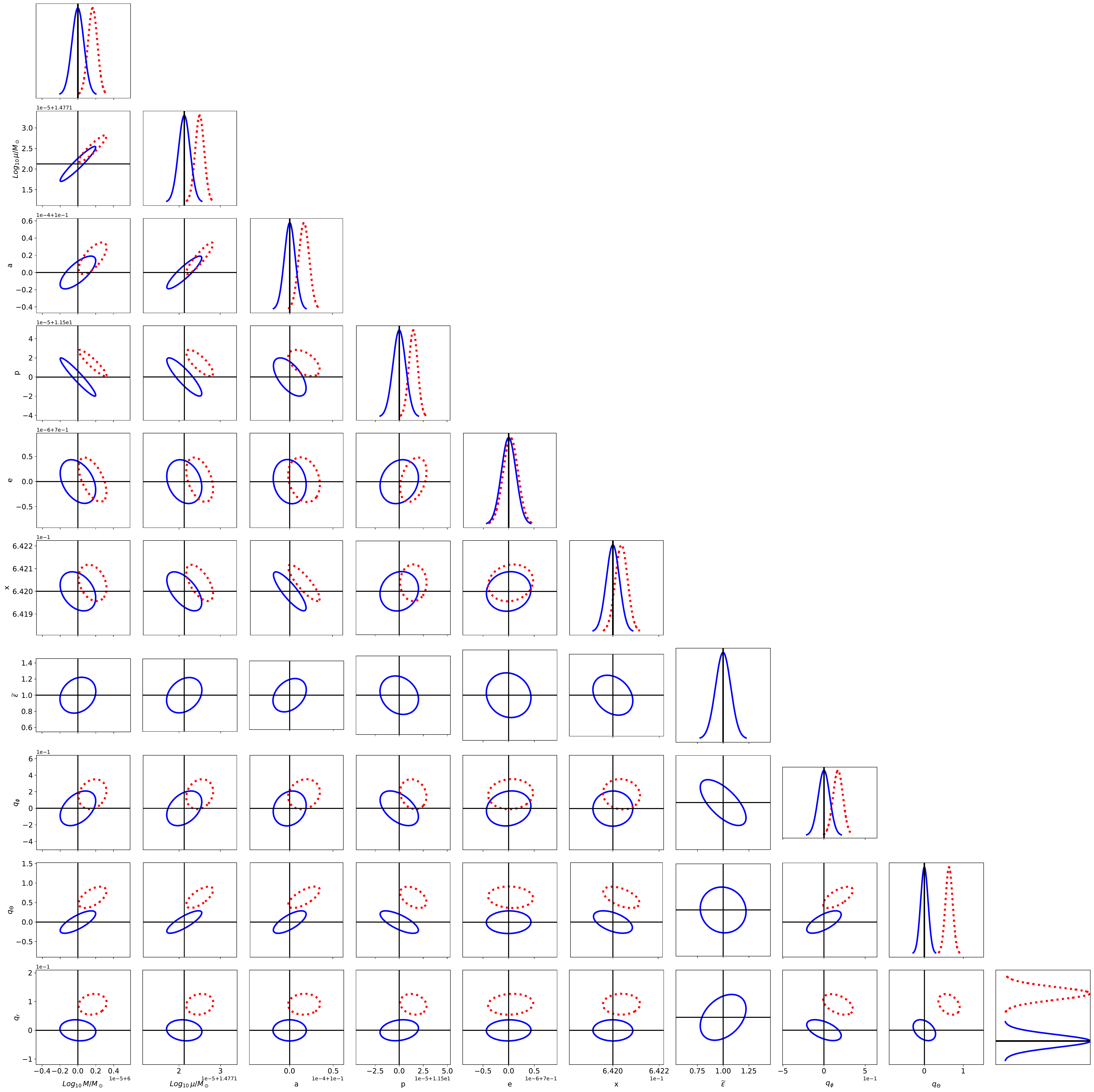}
\caption{\small The 2-dimensional posterior showing 3$\sigma$ contours for IC1 (see Table~\ref{tab:models}), where the injected signal had an SNR of 30. The solid (blue) contours are generated by the model with resonance and are centered on the true parameter values. The dotted (red) contours are derived from a model without resonance with peak shifted to parameter values estimated with induced systematic error.}
\label{fig:cornerIC1}
\end{figure*}

\begin{figure*}
  \centering
  \includegraphics[width=\textwidth,height=\textheight,keepaspectratio]{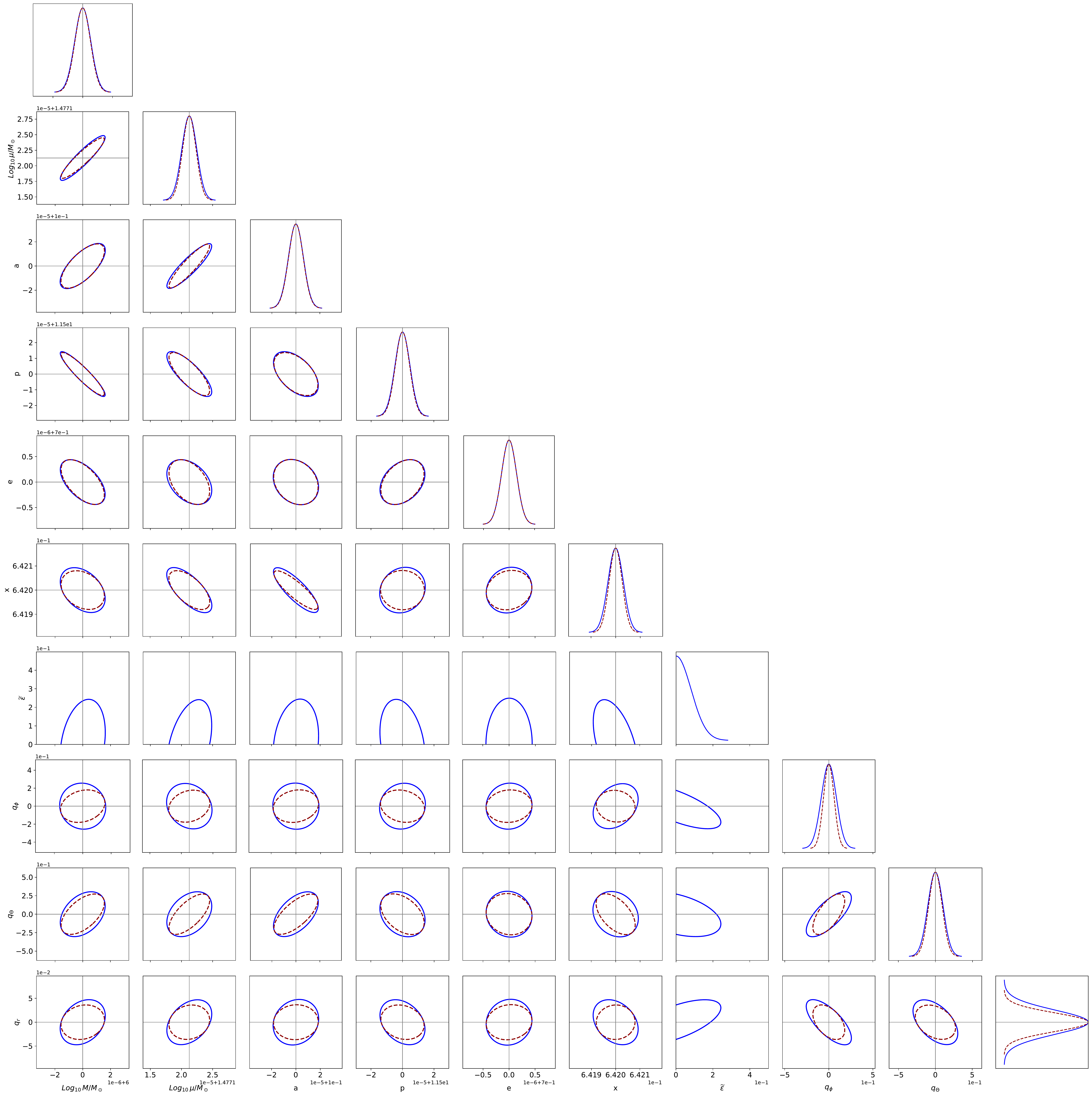}
\caption{\small The 2-dimensional posterior showing 3$\sigma$ contours for IC1 (see Table~\ref{tab:models}), where the injected signal had an SNR of 30. The blue-solid contours represent the model in which the tidal parameter is set to zero ($\widetilde{\epsilon} = 0$), and red-dotted contours represent the model in which the tidal parameter is not included in the analysis. The confidence contours are centered at the true value since both signals were unperturbed.}
\label{fig:corner-nullfullIC1}
\end{figure*}

In Fig~\ref{fig:Cumulative_Mismatch}, the cumulative mismatch between resonant and non-resonant waveforms using the RM and osculating method is shown. The unfilled markers show the cumulative mismatch between the resonant waveforms using the RM and osculating method, for four different initial conditions crossing two resonances $n:k:m = 3:0:-2$ and $n:k:m = 3:-4:-2$ during the evolution. In contrast, the filled markers show the mismatch when the resonances are neglected in the RM waveform model. Before crossing the first resonance, the filled markers overlay the unfilled ones for every initial condition. This indicates that the mismatch increases with each subsequent resonance encounter. The tiny increase in $\mathcal{M}$ before resonance is only due to numerical error arising from a `shift' in initial orbital frequencies due to tidally perturbed metric as also discussed in \cite{Bronicki:2022eqa} using a Newtonian analysis. A key point to notice is that after the resonance the mismatch between the RM and osculating resonant waveforms grows from $10^{-5}-10^{-7}$ up to $\sim 10^{-3}$.  This result is significant for the waveform modeling community, because it quantifies the mismatch induced by ignoring the resonance modeling. As long as we correctly predict the resonance jumps, it is possible to have an accurate waveform up to a mismatch of $\sim 10^{-3}$. This is expected to be sufficient for search and parameter estimation. In summary, we can model (multiple) tidal resonances by using the RM and  match the waveform until the end of the inspiral keeping $\mathcal{M}$ below $10^{-3}$.
It is worth reminding the reader that the cause of the mismatch of $\mathcal{O}(10^{-3})$ comes from a numerical error resulting from tidally perturbed metric causing a tiny `shift' in the initial conditions. If we can determine the initial conditions correctly, the mismatch would be smaller. We also find no discernible difference in mismatch between the RM with the `step' and the `smooth' impulse functions for all four initial conditions.

\subsection{Parameter estimation and systematic bias}
With the resonant waveforms derived from the RM and including only one resonance (3:0:-2), we examine the parameter measurement precision of the orbital and tidal parameters based on Fisher matrices (discussed in Sec~\ref{subsec:data-analysis}).
All the waveforms have been normalized so that their SNR $=30$ and the extrinsic parameters are set to $\{\theta_K,\phi_K,\theta_S,\phi_S\}=\{\pi/4,0,\pi/4,0\}$. The results for Fisher matrix estimates are shown in Fig.~\ref{fig:delta_params}. In this figure, we can see that the orbital parameters (except the initial phases) are well constrained, whereas the tidal parameters are less well constrained. In particular, the measurement precision for the tidal parameter $\tilde{\epsilon}$ and the phases is $\sim 10^{-1}$. In terms of the waveform, the initial phases determine the initial position of the compact object and do not affect the frequency evolution of the EMRI at adiabatic order, so their impact is weaker, which leads to the lower measurement precision. Due to the longer observation time for IC4 (see Fig.~\ref{fig:Cumulative_Mismatch}), the orbital parameters of this system are better constrained than for the other models.

In waveform modeling, using an approximate model can introduce systematic error~\eqref{eq:sys_err} into parameter estimation. We investigate the systematic error by using a non-resonant approximate waveform $h_m$, while the true waveform $h_t$ incorporates the resonance using the RM. To compare this error with the statistical error ~\eqref{eq:stat_err}, we show the ratio  $\delta\,\lambda^{bias}/\Delta\lambda$ in Fig.~\ref{fig:sys_bias_params}. With $\delta\,\lambda^{bias}/\Delta\lambda > 1$, the inaccurate waveform modeling leads to biases larger than those induced by noise fluctuations. The magnitude of systematic bias naturally depends on the magnitude of the tidal perturbation.
 For the strong but still realistic examples (motivated by \cite{Amaro_Seoane_2011,emami2020detectability,byh}) of tidal resonance that we consider, the systematic errors cannot be completely ignored.
Thus, we may need to account for the presence of tidal perturbers when performing careful inference, as also indicated by mismatch analysis in Fig.~\ref{fig:Cumulative_Mismatch}.

In addition to measurement precision, the Fisher matrix also provides the covariance relation between the parameters. To visualize this, we plot the $3\sigma$ contours in Fig.~\ref{fig:cornerIC1} for IC1. The solid (blue) contours are generated by the true model (with resonance) and are centered on the true parameter values. The dotted (red) contours are derived from the model without resonance, where the peak values are shifted by the amount of the systematic errors shown in Fig.~\ref{fig:sys_bias_params}. For the example considered in Fig.~\ref{fig:cornerIC1}, 
the bias is within the credible region for most of the EMRI parameters. However, our ability to measure the initial phases is more significantly affected if tidal effect is not modeled. The normalized tidal parameter $\widetilde{\epsilon}$ (discussed below  Eq.~\ref{eq:fisherparam}) can be constrained with an absolute precision of $0.25$.

 In the analysis above, we showed the bias induced in parameter measurement precision if tidal resonance was not modeled in the waveform. Next, we compare the same model with the one in which tidal parameter is set to zero \textit{i.e.} the signal is not tidally perturbed but the tidal parameter is included in the Fisher analysis. The goal is to check whether the error estimates are affected by the introduction of the tidal parameter.
We assume that the signal is given by a model with the tidal parameter set to zero.
In Fig.~\ref{fig:corner-nullfullIC1} blue-solid contours show the 3$\sigma$ confidence region when we use the model with 10 parameters including the tidal parameter, while the red-dotted contours corresponding to the model with 9 parameters excluding the tidal parameter. 
Because the tidal parameter is positive by definition, we show a section of ellipses in the positive range. The orbital parameters such as $M,\mu, a,p,e,\rm{x}$ are measured with approximately the same precision in both models.
Our ability to measure the EMRI's initial phasing is noticeably more degraded, but the overall impact is still fairly marginal. Thus, the tidal parameter is largely a non-degenerate degree of freedom, and its inclusion in EMRI data analysis will not pose fundamental issues in the absence of a tidal perturber at least for the magnitude of tidal perturbation considered in our work.

By combining the results from Fig.~\ref{fig:cornerIC1} and Fig.~\ref{fig:corner-nullfullIC1} for the example considered, we can infer the maximum value of tidal parameter under which the presence of a tidal resonance cannot be assessed. According to Fig.~\ref{fig:cornerIC1}, we can constrain the tidal parameter within the error bar of $\pm 0.25$ of the true value, whereas Fig.~\ref{fig:corner-nullfullIC1} says that we can rule out values larger than $0.25$ for $\widetilde{\epsilon}$. Therefore, if we choose a signal with $\widetilde{\epsilon} = 0.25$, we would likely have an ellipse centered at 0.25 and the width touching the zero (since the error bar is $\pm 0.25$). It follows that we may rule out zero for a larger $\widetilde{\epsilon}$ ($> 0.25$), thereby marking the presence of the perturber, but not for a smaller $\widetilde{\epsilon}$.

\section{Summary and Future Work}

Observations of extreme-mass-ratio inspirals may provide an excellent opportunity to test some of the key predictions of general relativity and are particularly useful for probing the stellar distribution at the galactic center. In this work, we generalized our previous study \cite{PaperI} to explore the impact of different resonance combinations caused by a stellar-mass perturber near an EMRI. We computed the accumulation in phase after a tidal resonance has been encountered by an EMRI and showed the dependence of resonance strength on orbital parameters and inclination of the perturber. Using Fisher matrices, we also analyzed how this phenomenon impacts the estimation of the intrinsic orbital and tidal parameters by using a resonance model (RM) based on a step function approach. We validated the evolution of the trajectory derived from the RM by comparing it with the forced osculating trajectory. This gives us confidence in the robustness of the fitting formulae as well as the implementation of the RM. Our study examined the systematic errors that might arise from neglecting tidal resonances in the estimation of intrinsic parameters. Our results suggest that parameter estimates are likely to be biased if resonances are not considered in waveform modeling. The analysis presented here to model tidal resonances would likely apply to self-force resonances as well.

As part of the extension of this work, we will relax the stationary perturber approximation and explore multiple resonant interactions in parameter estimation using Bayesian posterior calculations. Furthermore, once the resonances jump sizes due to the self-force is available, the ability of RM to detect and characterize EMRIs should be investigated. 
Last, the overall approach in this work, Paper I, and  modeling efforts by the EMRI community is to pursue a \emph{modeled} treatment of resonances (be it self-force or tidal) in data analysis. However, this is not the only possible approach, since phenomenological treatments such as ERM (where information on resonance jumps is recovered rather than modeled) might also prove useful; this is especially the case if sufficiently precise modeling of these jumps turns out to be unfeasible or unachievable. Thus, it is worthwhile to continue exploring both approaches in parallel, which will in turn benefit from shared techniques such as those introduced in this work.

\label{sec:6}
\begin{acknowledgments}
We thank  Soichiro Isoyama for the helpful discussions. This work makes use of the Black Hole Perturbation Toolkit~\cite{BHPToolkit}. PG is supported by JSPS fellowship and KAKENHI Grant Number 21J15826. AJKC acknowledges support from the NASA LISA Preparatory Science grant 20-LPS20-0005. TT is supported by JSPS KAKENHI Grant Number JP17H06358 (and also JP17H06357), \textit{A01: Testing gravity theories using gravitational waves}, as a part of the innovative research area, ``Gravitational wave physics and astronomy: Genesis'', and also by JP20K03928. 
\end{acknowledgments}
\begin{widetext}
\begin{figure*}
  \centering
  \includegraphics[width=0.3\linewidth]{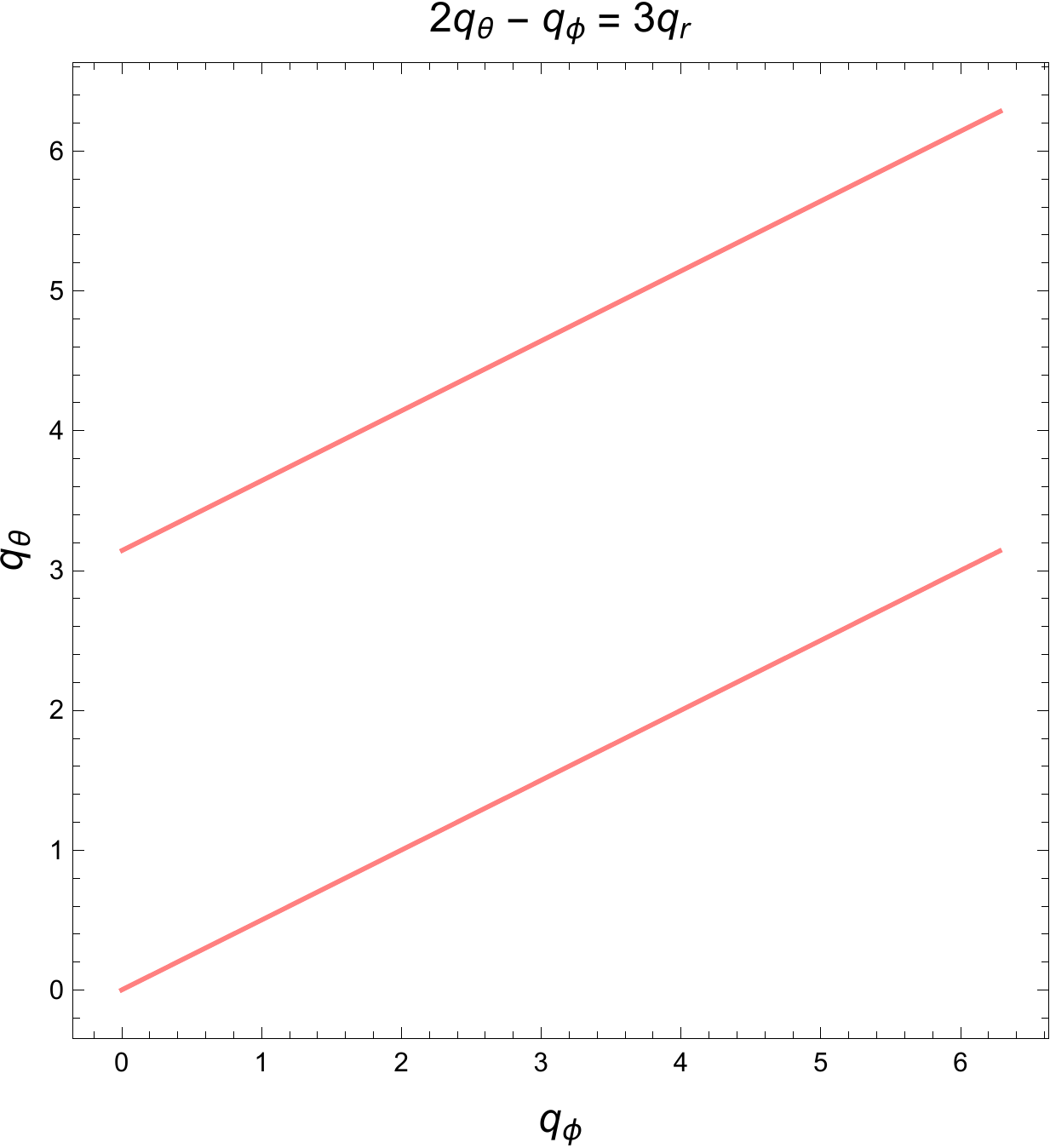}
    \hskip 0.8cm
\includegraphics[width=0.3\linewidth]{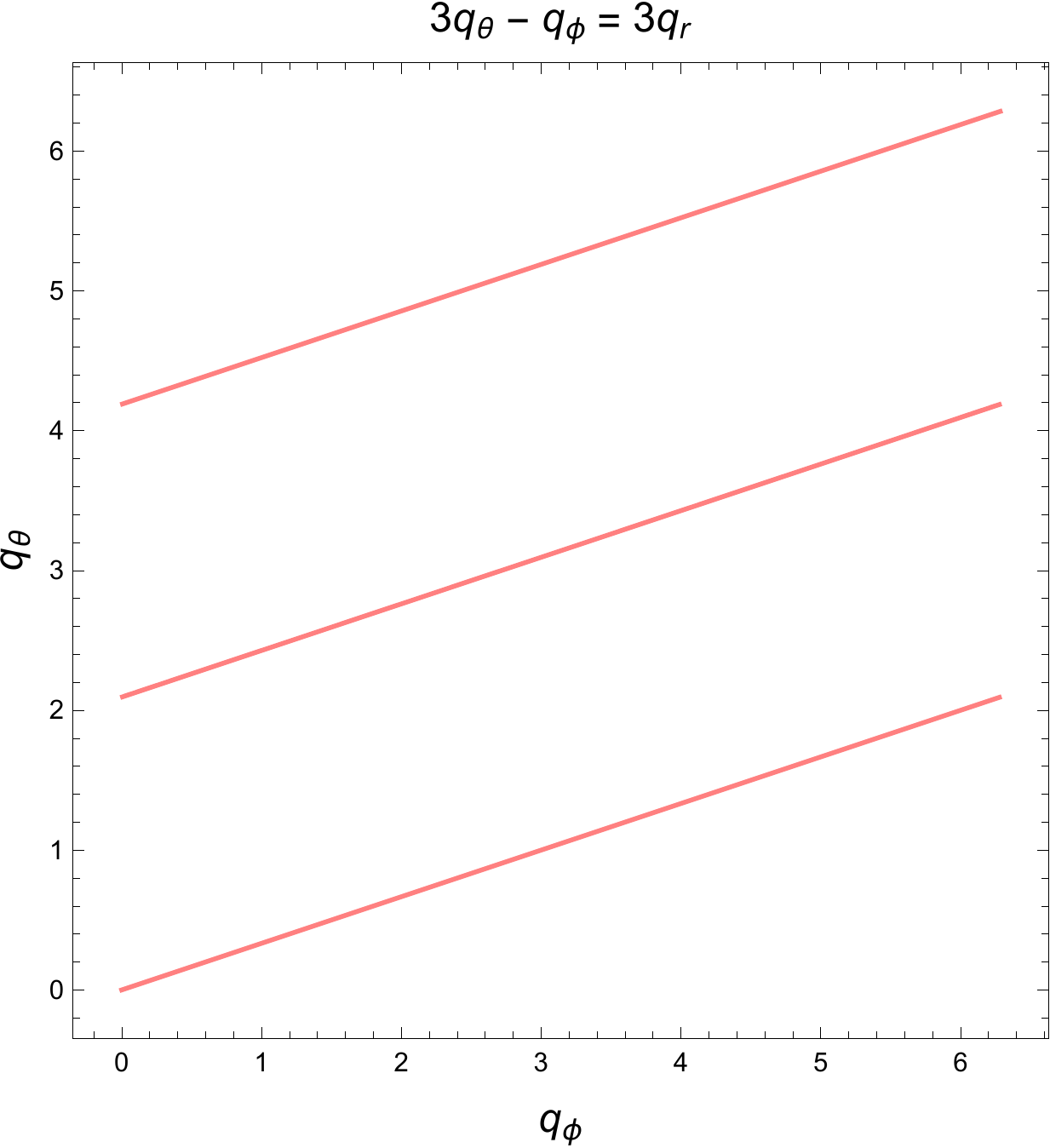}
\caption{Section of orbit in $q_\phi$ - $q_\theta$ plane for different resonance conditions.}
\label{fig:phase-cancellation}
\end{figure*}
\end{widetext}
\appendix
\section{Suppression of odd $k+m$ integer resonances}
\label{appex:A}
In Paper I, we focused our analysis on $m = \pm 2$ modes and discussed the suppression of resonances for odd $k$ integers. Here, we take a step further and investigate  $m = \pm 1$ modes. We discover that tidal resonances with odd $k+m$ integer do not give rise to a jump in the constants of motion. Hence, they do not contribute to a secular accumulation of a phase shift and are therefore not relevant for waveform modeling. On assuming a static tidal interaction, the leading order external potential at a large distance is expressed as 
$$U_{ext} \propto \mathcal{E}_{ab}\,x^ax^b,$$ 
where the symmetric tidal tensor  $\mathcal{E}_{ab}$ contains all the information about the tidal environment. For $m=\pm 1$ modes, only $\mathcal{E}_{xz}$ and $\mathcal{E}_{yz}$ contribute where $x,y,z (r,\theta,\phi)$ are standard Cartesian (spherical) coordinates. Note that transforming $q_\phi \rightarrow q_{\phi+\pi} \Rightarrow x \rightarrow -x , y \rightarrow -y $ or $q_\theta\rightarrow q_{\theta+\pi} \Rightarrow z \rightarrow -z$ leads to a sign flip of the tidal potential and hence the resulting torque. Therefore, if corresponding points (for instance, both $(q_\phi,q_{\theta})$ and $(q_\phi,q_{\theta+\pi})$) are passed by an orbit, then it results in a net cancellation of $d{L}_{z}/dt$ between the two segments of the orbit. In Fig~\ref{fig:phase-cancellation}, for illustrative purpose, we show a section of the orbit in the $q_{\phi}$ - $q_{\theta}$ plane for $k+m = 1$ (left) and $k+m = 2$ (right)  resonance combinations. In the left plot, for fixed $q_r = 0$, the distance between two lines is $\pi$. Thus, the orbit evolves in such a manner, that the net tidal force cancels out resulting in no change in $L_z$. Whereas, in the right plot, the corresponding ``cancellation" points are not crossed by the orbit. While this discussion helps understand the vanishing $d{L}_z/dt$ on crossing odd $k+m$ resonances, empirically we found that $d{Q}/dt$ also vanishes for such resonances.

\bibliography{refer}

\begin{thebibliography}{55}
\expandafter\ifx\csname natexlab\endcsname\relax\def\natexlab#1{#1}\fi
\expandafter\ifx\csname bibnamefont\endcsname\relax
  \def\bibnamefont#1{#1}\fi
\expandafter\ifx\csname bibfnamefont\endcsname\relax
  \def\bibfnamefont#1{#1}\fi
\expandafter\ifx\csname citenamefont\endcsname\relax
  \def\citenamefont#1{#1}\fi
\expandafter\ifx\csname url\endcsname\relax
  \def\url#1{\texttt{#1}}\fi
\expandafter\ifx\csname urlprefix\endcsname\relax\def\urlprefix{URL }\fi
\providecommand{\bibinfo}[2]{#2}
\providecommand{\eprint}[2][]{\url{#2}}

\bibitem[{\citenamefont{Gupta et~al.}(2021)\citenamefont{Gupta, Bonga, Chua,
  and Tanaka}}]{PaperI}
\bibinfo{author}{\bibfnamefont{P.}~\bibnamefont{Gupta}},
  \bibinfo{author}{\bibfnamefont{B.}~\bibnamefont{Bonga}},
  \bibinfo{author}{\bibfnamefont{A.~J.~K.} \bibnamefont{Chua}},
  \bibnamefont{and} \bibinfo{author}{\bibfnamefont{T.}~\bibnamefont{Tanaka}},
  \bibinfo{journal}{Phys. Rev. D} \textbf{\bibinfo{volume}{104}},
  \bibinfo{pages}{044056} (\bibinfo{year}{2021}),
  \urlprefix\url{https://link.aps.org/doi/10.1103/PhysRevD.104.044056}.

\bibitem[{\citenamefont{Abbott et~al.}(2020)\citenamefont{Abbott, Abbott,
  Abbott, Abraham, Acernese, Ackley, Adams, Adya, Affeldt, and
  et~al.}}]{Abbott_2020}
\bibinfo{author}{\bibfnamefont{B.~P.} \bibnamefont{Abbott}},
  \bibinfo{author}{\bibfnamefont{R.}~\bibnamefont{Abbott}},
  \bibinfo{author}{\bibfnamefont{T.~D.} \bibnamefont{Abbott}},
  \bibinfo{author}{\bibfnamefont{S.}~\bibnamefont{Abraham}},
  \bibinfo{author}{\bibfnamefont{F.}~\bibnamefont{Acernese}},
  \bibinfo{author}{\bibfnamefont{K.}~\bibnamefont{Ackley}},
  \bibinfo{author}{\bibfnamefont{C.}~\bibnamefont{Adams}},
  \bibinfo{author}{\bibfnamefont{V.~B.} \bibnamefont{Adya}},
  \bibinfo{author}{\bibfnamefont{C.}~\bibnamefont{Affeldt}}, \bibnamefont{and}
  \bibinfo{author}{\bibnamefont{et~al.}}, \bibinfo{journal}{Living Reviews in
  Relativity} \textbf{\bibinfo{volume}{23}} (\bibinfo{year}{2020}), ISSN
  \bibinfo{issn}{1433-8351}.

\bibitem[{\citenamefont{Abbott et~al.}(2021)\citenamefont{Abbott, Abbott,
  Abbott, Abraham, Acernese, Ackley, Adams, Adya, Affeldt, and
  et~al.}}]{Abbott_2021}
\bibinfo{author}{\bibfnamefont{B.~P.} \bibnamefont{Abbott}},
  \bibinfo{author}{\bibfnamefont{R.}~\bibnamefont{Abbott}},
  \bibinfo{author}{\bibfnamefont{T.~D.} \bibnamefont{Abbott}},
  \bibinfo{author}{\bibfnamefont{S.}~\bibnamefont{Abraham}},
  \bibinfo{author}{\bibfnamefont{F.}~\bibnamefont{Acernese}},
  \bibinfo{author}{\bibfnamefont{K.}~\bibnamefont{Ackley}},
  \bibinfo{author}{\bibfnamefont{C.}~\bibnamefont{Adams}},
  \bibinfo{author}{\bibfnamefont{V.~B.} \bibnamefont{Adya}},
  \bibinfo{author}{\bibfnamefont{C.}~\bibnamefont{Affeldt}}, \bibnamefont{and}
  \bibinfo{author}{\bibnamefont{et~al.}}, \bibinfo{journal}{SoftwareX}
  \textbf{\bibinfo{volume}{13}}, \bibinfo{pages}{100658}
  (\bibinfo{year}{2021}), ISSN \bibinfo{issn}{2352-7110}.

\bibitem[{\citenamefont{Collaboration and the
  Virgo~Collaboration}(2021)}]{ligo2021population}
\bibinfo{author}{\bibfnamefont{T.~L.~S.} \bibnamefont{Collaboration}}
  \bibnamefont{and} \bibinfo{author}{\bibnamefont{the Virgo~Collaboration}}
  (\bibinfo{year}{2021}), \eprint{2010.14533}.

\bibitem[{\citenamefont{Collaboration and the
  Virgo~Collaboration}(2020)}]{theligo2020tests}
\bibinfo{author}{\bibfnamefont{T.~L.~S.} \bibnamefont{Collaboration}}
  \bibnamefont{and} \bibinfo{author}{\bibnamefont{the Virgo~Collaboration}}
  (\bibinfo{year}{2020}), \eprint{2010.14529}.

\bibitem[{\citenamefont{Amaro-Seoane et~al.}(2017)\citenamefont{Amaro-Seoane,
  Audley, Babak, Baker, Barausse, Bender, Berti, Binetruy, Born, Bortoluzzi
  et~al.}}]{amaroseoane2017laser}
\bibinfo{author}{\bibfnamefont{P.}~\bibnamefont{Amaro-Seoane}},
  \bibinfo{author}{\bibfnamefont{H.}~\bibnamefont{Audley}},
  \bibinfo{author}{\bibfnamefont{S.}~\bibnamefont{Babak}},
  \bibinfo{author}{\bibfnamefont{J.}~\bibnamefont{Baker}},
  \bibinfo{author}{\bibfnamefont{E.}~\bibnamefont{Barausse}},
  \bibinfo{author}{\bibfnamefont{P.}~\bibnamefont{Bender}},
  \bibinfo{author}{\bibfnamefont{E.}~\bibnamefont{Berti}},
  \bibinfo{author}{\bibfnamefont{P.}~\bibnamefont{Binetruy}},
  \bibinfo{author}{\bibfnamefont{M.}~\bibnamefont{Born}},
  \bibinfo{author}{\bibfnamefont{D.}~\bibnamefont{Bortoluzzi}},
  \bibnamefont{et~al.} (\bibinfo{year}{2017}), \eprint{1702.00786}.

\bibitem[{\citenamefont{Berry et~al.}(2019)\citenamefont{Berry, Hughes,
  Sopuerta, Chua, Heffernan, Holley-Bockelmann, Mihaylov, Miller, and
  Sesana}}]{berry2019unique}
\bibinfo{author}{\bibfnamefont{C.~P.~L.} \bibnamefont{Berry}},
  \bibinfo{author}{\bibfnamefont{S.~A.} \bibnamefont{Hughes}},
  \bibinfo{author}{\bibfnamefont{C.~F.} \bibnamefont{Sopuerta}},
  \bibinfo{author}{\bibfnamefont{A.~J.~K.} \bibnamefont{Chua}},
  \bibinfo{author}{\bibfnamefont{A.}~\bibnamefont{Heffernan}},
  \bibinfo{author}{\bibfnamefont{K.}~\bibnamefont{Holley-Bockelmann}},
  \bibinfo{author}{\bibfnamefont{D.~P.} \bibnamefont{Mihaylov}},
  \bibinfo{author}{\bibfnamefont{M.~C.} \bibnamefont{Miller}},
  \bibnamefont{and} \bibinfo{author}{\bibfnamefont{A.}~\bibnamefont{Sesana}}
  (\bibinfo{year}{2019}), \eprint{1903.03686}.

\bibitem[{\citenamefont{Mei et~al.}(2020)\citenamefont{Mei, Bai, Bao, Barausse,
  Cai, Canuto, Cao, Chen, Chen, Ding et~al.}}]{Mei_2020}
\bibinfo{author}{\bibfnamefont{J.}~\bibnamefont{Mei}},
  \bibinfo{author}{\bibfnamefont{Y.-Z.} \bibnamefont{Bai}},
  \bibinfo{author}{\bibfnamefont{J.}~\bibnamefont{Bao}},
  \bibinfo{author}{\bibfnamefont{E.}~\bibnamefont{Barausse}},
  \bibinfo{author}{\bibfnamefont{L.}~\bibnamefont{Cai}},
  \bibinfo{author}{\bibfnamefont{E.}~\bibnamefont{Canuto}},
  \bibinfo{author}{\bibfnamefont{B.}~\bibnamefont{Cao}},
  \bibinfo{author}{\bibfnamefont{W.-M.} \bibnamefont{Chen}},
  \bibinfo{author}{\bibfnamefont{Y.}~\bibnamefont{Chen}},
  \bibinfo{author}{\bibfnamefont{Y.-W.} \bibnamefont{Ding}},
  \bibnamefont{et~al.}, \bibinfo{journal}{Progress of Theoretical and
  Experimental Physics}  (\bibinfo{year}{2020}), ISSN
  \bibinfo{issn}{2050-3911}.

\bibitem[{\citenamefont{Flanagan and Hinderer}(2012)}]{PhysRevLett.109.071102}
\bibinfo{author}{\bibfnamefont{E.~E.} \bibnamefont{Flanagan}} \bibnamefont{and}
  \bibinfo{author}{\bibfnamefont{T.}~\bibnamefont{Hinderer}},
  \bibinfo{journal}{Phys. Rev. Lett.} \textbf{\bibinfo{volume}{109}},
  \bibinfo{pages}{071102} (\bibinfo{year}{2012}).

\bibitem[{\citenamefont{Berry et~al.}(2016)\citenamefont{Berry, Cole,
  Cañizares, and Gair}}]{2016Berry}
\bibinfo{author}{\bibfnamefont{C.~P.} \bibnamefont{Berry}},
  \bibinfo{author}{\bibfnamefont{R.~H.} \bibnamefont{Cole}},
  \bibinfo{author}{\bibfnamefont{P.}~\bibnamefont{Cañizares}},
  \bibnamefont{and} \bibinfo{author}{\bibfnamefont{J.~R.} \bibnamefont{Gair}},
  \bibinfo{journal}{Physical Review D} \textbf{\bibinfo{volume}{94}}
  (\bibinfo{year}{2016}), ISSN \bibinfo{issn}{2470-0029},
  \urlprefix\url{http://dx.doi.org/10.1103/PhysRevD.94.124042}.

\bibitem[{\citenamefont{Speri and Gair}(2021)}]{speri2021assessing}
\bibinfo{author}{\bibfnamefont{L.}~\bibnamefont{Speri}} \bibnamefont{and}
  \bibinfo{author}{\bibfnamefont{J.~R.} \bibnamefont{Gair}},
  \bibinfo{journal}{Phys. Rev. D} \textbf{\bibinfo{volume}{103}},
  \bibinfo{pages}{124032} (\bibinfo{year}{2021}),
  \urlprefix\url{https://link.aps.org/doi/10.1103/PhysRevD.103.124032}.

\bibitem[{\citenamefont{Amaro-Seoane}(2019)}]{Amaro_Seoane_2019}
\bibinfo{author}{\bibfnamefont{P.}~\bibnamefont{Amaro-Seoane}},
  \bibinfo{journal}{Physical Review D} \textbf{\bibinfo{volume}{99}}
  (\bibinfo{year}{2019}), ISSN \bibinfo{issn}{2470-0029}.

\bibitem[{\citenamefont{Amaro-Seoane}(2020)}]{amaroseoane2020gravitational}
\bibinfo{author}{\bibfnamefont{P.}~\bibnamefont{Amaro-Seoane}}
  (\bibinfo{year}{2020}), \eprint{2011.03059}.

\bibitem[{\citenamefont{Emami and Loeb}(2020{\natexlab{a}})}]{Emami_2020}
\bibinfo{author}{\bibfnamefont{R.}~\bibnamefont{Emami}} \bibnamefont{and}
  \bibinfo{author}{\bibfnamefont{A.}~\bibnamefont{Loeb}},
  \bibinfo{journal}{Journal of Cosmology and Astroparticle Physics}
  \textbf{\bibinfo{volume}{2020}}, \bibinfo{pages}{021–021}
  (\bibinfo{year}{2020}{\natexlab{a}}), ISSN \bibinfo{issn}{1475-7516}.

\bibitem[{\citenamefont{Emami and
  Loeb}(2020{\natexlab{b}})}]{emami2020detectability}
\bibinfo{author}{\bibfnamefont{R.}~\bibnamefont{Emami}} \bibnamefont{and}
  \bibinfo{author}{\bibfnamefont{A.}~\bibnamefont{Loeb}}
  (\bibinfo{year}{2020}{\natexlab{b}}), \eprint{1903.02579}.

\bibitem[{\citenamefont{Pan and Yang}(2021)}]{pan2021formation}
\bibinfo{author}{\bibfnamefont{Z.}~\bibnamefont{Pan}} \bibnamefont{and}
  \bibinfo{author}{\bibfnamefont{H.}~\bibnamefont{Yang}},
  \bibinfo{journal}{Physical Review D} \textbf{\bibinfo{volume}{103}}
  (\bibinfo{year}{2021}).

\bibitem[{\citenamefont{Pan et~al.}(2021)\citenamefont{Pan, Lyu, and
  Yang}}]{pan2021formation2}
\bibinfo{author}{\bibfnamefont{Z.}~\bibnamefont{Pan}},
  \bibinfo{author}{\bibfnamefont{Z.}~\bibnamefont{Lyu}}, \bibnamefont{and}
  \bibinfo{author}{\bibfnamefont{H.}~\bibnamefont{Yang}},
  \bibinfo{journal}{Physical Review D} \textbf{\bibinfo{volume}{104}}
  (\bibinfo{year}{2021}).

\bibitem[{\citenamefont{Babak et~al.}(2017)\citenamefont{Babak, Gair, Sesana,
  Barausse, Sopuerta, Berry, Berti, Amaro-Seoane, Petiteau, and
  Klein}}]{Gair_2017}
\bibinfo{author}{\bibfnamefont{S.}~\bibnamefont{Babak}},
  \bibinfo{author}{\bibfnamefont{J.}~\bibnamefont{Gair}},
  \bibinfo{author}{\bibfnamefont{A.}~\bibnamefont{Sesana}},
  \bibinfo{author}{\bibfnamefont{E.}~\bibnamefont{Barausse}},
  \bibinfo{author}{\bibfnamefont{C.~F.} \bibnamefont{Sopuerta}},
  \bibinfo{author}{\bibfnamefont{C.~P.} \bibnamefont{Berry}},
  \bibinfo{author}{\bibfnamefont{E.}~\bibnamefont{Berti}},
  \bibinfo{author}{\bibfnamefont{P.}~\bibnamefont{Amaro-Seoane}},
  \bibinfo{author}{\bibfnamefont{A.}~\bibnamefont{Petiteau}}, \bibnamefont{and}
  \bibinfo{author}{\bibfnamefont{A.}~\bibnamefont{Klein}},
  \bibinfo{journal}{Physical Review D} \textbf{\bibinfo{volume}{95}}
  (\bibinfo{year}{2017}), ISSN \bibinfo{issn}{2470-0029}.

\bibitem[{\citenamefont{Fujita and Shibata}(2020)}]{Fujita_2020}
\bibinfo{author}{\bibfnamefont{R.}~\bibnamefont{Fujita}} \bibnamefont{and}
  \bibinfo{author}{\bibfnamefont{M.}~\bibnamefont{Shibata}},
  \bibinfo{journal}{Physical Review D} \textbf{\bibinfo{volume}{102}}
  (\bibinfo{year}{2020}), ISSN \bibinfo{issn}{2470-0029}.

\bibitem[{\citenamefont{Hughes et~al.}(2021)\citenamefont{Hughes, Warburton,
  Khanna, Chua, and Katz}}]{hughes2021adiabatic}
\bibinfo{author}{\bibfnamefont{S.~A.} \bibnamefont{Hughes}},
  \bibinfo{author}{\bibfnamefont{N.}~\bibnamefont{Warburton}},
  \bibinfo{author}{\bibfnamefont{G.}~\bibnamefont{Khanna}},
  \bibinfo{author}{\bibfnamefont{A.~J.~K.} \bibnamefont{Chua}},
  \bibnamefont{and} \bibinfo{author}{\bibfnamefont{M.~L.} \bibnamefont{Katz}}
  (\bibinfo{year}{2021}), \eprint{2102.02713}.

\bibitem[{\citenamefont{Chua et~al.}(2021)\citenamefont{Chua, Katz, Warburton,
  and Hughes}}]{Chua_2021}
\bibinfo{author}{\bibfnamefont{A.~J.~K.} \bibnamefont{Chua}},
  \bibinfo{author}{\bibfnamefont{M.~L.} \bibnamefont{Katz}},
  \bibinfo{author}{\bibfnamefont{N.}~\bibnamefont{Warburton}},
  \bibnamefont{and} \bibinfo{author}{\bibfnamefont{S.~A.}
  \bibnamefont{Hughes}}, \bibinfo{journal}{Physical Review Letters}
  \textbf{\bibinfo{volume}{126}} (\bibinfo{year}{2021}), ISSN
  \bibinfo{issn}{1079-7114}.

\bibitem[{\citenamefont{Katz et~al.}(2021)\citenamefont{Katz, Chua, Speri,
  Warburton, and Hughes}}]{2021Katz}
\bibinfo{author}{\bibfnamefont{M.~L.} \bibnamefont{Katz}},
  \bibinfo{author}{\bibfnamefont{A.~J.} \bibnamefont{Chua}},
  \bibinfo{author}{\bibfnamefont{L.}~\bibnamefont{Speri}},
  \bibinfo{author}{\bibfnamefont{N.}~\bibnamefont{Warburton}},
  \bibnamefont{and} \bibinfo{author}{\bibfnamefont{S.~A.}
  \bibnamefont{Hughes}}, \bibinfo{journal}{Physical Review D}
  \textbf{\bibinfo{volume}{104}} (\bibinfo{year}{2021}), ISSN
  \bibinfo{issn}{2470-0029},
  \urlprefix\url{http://dx.doi.org/10.1103/PhysRevD.104.064047}.

\bibitem[{\citenamefont{Wardell et~al.}(2021)\citenamefont{Wardell, Pound,
  Warburton, Miller, Durkan, and Le~Tiec}}]{Wardell:2021fyy}
\bibinfo{author}{\bibfnamefont{B.}~\bibnamefont{Wardell}},
  \bibinfo{author}{\bibfnamefont{A.}~\bibnamefont{Pound}},
  \bibinfo{author}{\bibfnamefont{N.}~\bibnamefont{Warburton}},
  \bibinfo{author}{\bibfnamefont{J.}~\bibnamefont{Miller}},
  \bibinfo{author}{\bibfnamefont{L.}~\bibnamefont{Durkan}}, \bibnamefont{and}
  \bibinfo{author}{\bibfnamefont{A.}~\bibnamefont{Le~Tiec}}
  (\bibinfo{year}{2021}), \eprint{2112.12265}.

\bibitem[{\citenamefont{Lynch et~al.}(2021)\citenamefont{Lynch, van~de Meent,
  and Warburton}}]{lynch2021eccentric}
\bibinfo{author}{\bibfnamefont{P.}~\bibnamefont{Lynch}},
  \bibinfo{author}{\bibfnamefont{M.}~\bibnamefont{van~de Meent}},
  \bibnamefont{and}
  \bibinfo{author}{\bibfnamefont{N.}~\bibnamefont{Warburton}},
  \emph{\bibinfo{title}{Eccentric self-forced inspirals into a rotating black
  hole}} (\bibinfo{year}{2021}), \eprint{2112.05651}.

\bibitem[{\citenamefont{Bonga et~al.}(2019)\citenamefont{Bonga, Yang, and
  Hughes}}]{byh}
\bibinfo{author}{\bibfnamefont{B.}~\bibnamefont{Bonga}},
  \bibinfo{author}{\bibfnamefont{H.}~\bibnamefont{Yang}}, \bibnamefont{and}
  \bibinfo{author}{\bibfnamefont{S.~A.} \bibnamefont{Hughes}},
  \bibinfo{journal}{Phys. Rev. Lett.} \textbf{\bibinfo{volume}{123}},
  \bibinfo{pages}{101103} (\bibinfo{year}{2019}), \eprint{1905.00030}.

\bibitem[{\citenamefont{Amaro-Seoane et~al.}(2022)}]{Amaro-Seoane:2022rxf}
\bibinfo{author}{\bibfnamefont{P.}~\bibnamefont{Amaro-Seoane}}
  \bibnamefont{et~al.} (\bibinfo{year}{2022}), \eprint{2203.06016}.

\bibitem[{\citenamefont{Flanagan et~al.}(2014)\citenamefont{Flanagan, Hughes,
  and Ruangsri}}]{Flanagan:2012kg}
\bibinfo{author}{\bibfnamefont{E.~E.} \bibnamefont{Flanagan}},
  \bibinfo{author}{\bibfnamefont{S.~A.} \bibnamefont{Hughes}},
  \bibnamefont{and} \bibinfo{author}{\bibfnamefont{U.}~\bibnamefont{Ruangsri}},
  \bibinfo{journal}{Phys. Rev. D} \textbf{\bibinfo{volume}{89}},
  \bibinfo{pages}{084028} (\bibinfo{year}{2014}), \eprint{1208.3906}.

\bibitem[{\citenamefont{Isoyama et~al.}(2013)\citenamefont{Isoyama, Fujita,
  Nakano, Sago, and Tanaka}}]{Isoyama:2013yor}
\bibinfo{author}{\bibfnamefont{S.}~\bibnamefont{Isoyama}},
  \bibinfo{author}{\bibfnamefont{R.}~\bibnamefont{Fujita}},
  \bibinfo{author}{\bibfnamefont{H.}~\bibnamefont{Nakano}},
  \bibinfo{author}{\bibfnamefont{N.}~\bibnamefont{Sago}}, \bibnamefont{and}
  \bibinfo{author}{\bibfnamefont{T.}~\bibnamefont{Tanaka}},
  \bibinfo{journal}{PTEP} \textbf{\bibinfo{volume}{2013}},
  \bibinfo{pages}{063E01} (\bibinfo{year}{2013}), \eprint{1302.4035}.

\bibitem[{\citenamefont{Isoyama et~al.}(2021)\citenamefont{Isoyama, Fujita,
  Chua, Nakano, Pound, and Sago}}]{Isoyama:2021jjd}
\bibinfo{author}{\bibfnamefont{S.}~\bibnamefont{Isoyama}},
  \bibinfo{author}{\bibfnamefont{R.}~\bibnamefont{Fujita}},
  \bibinfo{author}{\bibfnamefont{A.~J.~K.} \bibnamefont{Chua}},
  \bibinfo{author}{\bibfnamefont{H.}~\bibnamefont{Nakano}},
  \bibinfo{author}{\bibfnamefont{A.}~\bibnamefont{Pound}}, \bibnamefont{and}
  \bibinfo{author}{\bibfnamefont{N.}~\bibnamefont{Sago}}
  (\bibinfo{year}{2021}), \eprint{2111.05288}.

\bibitem[{\citenamefont{Nasipak and Evans}(2021)}]{Zachary}
\bibinfo{author}{\bibfnamefont{Z.}~\bibnamefont{Nasipak}} \bibnamefont{and}
  \bibinfo{author}{\bibfnamefont{C.~R.} \bibnamefont{Evans}},
  \bibinfo{journal}{Phys. Rev. D} \textbf{\bibinfo{volume}{104}},
  \bibinfo{pages}{084011} (\bibinfo{year}{2021}),
  \urlprefix\url{https://link.aps.org/doi/10.1103/PhysRevD.104.084011}.

\bibitem[{\citenamefont{Amaro-Seoane et~al.}(2011)\citenamefont{Amaro-Seoane,
  Brem, Cuadra, and Armitage}}]{Amaro_Seoane_2011}
\bibinfo{author}{\bibfnamefont{P.}~\bibnamefont{Amaro-Seoane}},
  \bibinfo{author}{\bibfnamefont{P.}~\bibnamefont{Brem}},
  \bibinfo{author}{\bibfnamefont{J.}~\bibnamefont{Cuadra}}, \bibnamefont{and}
  \bibinfo{author}{\bibfnamefont{P.~J.} \bibnamefont{Armitage}},
  \bibinfo{journal}{The Astrophysical Journal} \textbf{\bibinfo{volume}{744}},
  \bibinfo{pages}{L20} (\bibinfo{year}{2011}).

\bibitem[{\citenamefont{Gourgoulhon et~al.}(2019)\citenamefont{Gourgoulhon,
  Le~Tiec, Vincent, and Warburton}}]{Gourgoulhon_2019}
\bibinfo{author}{\bibfnamefont{E.}~\bibnamefont{Gourgoulhon}},
  \bibinfo{author}{\bibfnamefont{A.}~\bibnamefont{Le~Tiec}},
  \bibinfo{author}{\bibfnamefont{F.~H.} \bibnamefont{Vincent}},
  \bibnamefont{and}
  \bibinfo{author}{\bibfnamefont{N.}~\bibnamefont{Warburton}},
  \bibinfo{journal}{Astronomy \& Astrophysics} \textbf{\bibinfo{volume}{627}},
  \bibinfo{pages}{A92} (\bibinfo{year}{2019}), ISSN \bibinfo{issn}{1432-0746}.

\bibitem[{\citenamefont{Yunes and Gonzalez}(2006)}]{Yunes2006}
\bibinfo{author}{\bibfnamefont{N.}~\bibnamefont{Yunes}} \bibnamefont{and}
  \bibinfo{author}{\bibfnamefont{J.}~\bibnamefont{Gonzalez}},
  \bibinfo{journal}{Phys. Rev. D} \textbf{\bibinfo{volume}{73}},
  \bibinfo{pages}{024010} (\bibinfo{year}{2006}).

\bibitem[{\citenamefont{Gair et~al.}(2011)\citenamefont{Gair, Flanagan, Drasco,
  Hinderer, and Babak}}]{osculating-kerr}
\bibinfo{author}{\bibfnamefont{J.~R.} \bibnamefont{Gair}},
  \bibinfo{author}{\bibfnamefont{E.~E.} \bibnamefont{Flanagan}},
  \bibinfo{author}{\bibfnamefont{S.}~\bibnamefont{Drasco}},
  \bibinfo{author}{\bibfnamefont{T.}~\bibnamefont{Hinderer}}, \bibnamefont{and}
  \bibinfo{author}{\bibfnamefont{S.}~\bibnamefont{Babak}},
  \bibinfo{journal}{Phys. Rev.} \textbf{\bibinfo{volume}{D83}},
  \bibinfo{pages}{044037} (\bibinfo{year}{2011}), \eprint{1012.5111}.

\bibitem[{\citenamefont{Cutler and Vallisneri}(2007)}]{2007Curt}
\bibinfo{author}{\bibfnamefont{C.}~\bibnamefont{Cutler}} \bibnamefont{and}
  \bibinfo{author}{\bibfnamefont{M.}~\bibnamefont{Vallisneri}},
  \bibinfo{journal}{Physical Review D} \textbf{\bibinfo{volume}{76}}
  (\bibinfo{year}{2007}), ISSN \bibinfo{issn}{1550-2368},
  \urlprefix\url{http://dx.doi.org/10.1103/PhysRevD.76.104018}.

\bibitem[{\citenamefont{{Bardeen} et~al.}(1972)\citenamefont{{Bardeen},
  {Press}, and {Teukolsky}}}]{1972ApJBardeen}
\bibinfo{author}{\bibfnamefont{J.~M.} \bibnamefont{{Bardeen}}},
  \bibinfo{author}{\bibfnamefont{W.~H.} \bibnamefont{{Press}}},
  \bibnamefont{and} \bibinfo{author}{\bibfnamefont{S.~A.}
  \bibnamefont{{Teukolsky}}}, \bibinfo{journal}{\apj}
  \textbf{\bibinfo{volume}{178}}, \bibinfo{pages}{347} (\bibinfo{year}{1972}).

\bibitem[{\citenamefont{Schmidt}(2002)}]{Schmidt_2002}
\bibinfo{author}{\bibfnamefont{W.}~\bibnamefont{Schmidt}},
  \bibinfo{journal}{Classical and Quantum Gravity}
  \textbf{\bibinfo{volume}{19}}, \bibinfo{pages}{2743–2764}
  (\bibinfo{year}{2002}), ISSN \bibinfo{issn}{0264-9381}.

\bibitem[{\citenamefont{Mino}(2003)}]{Mino_2003}
\bibinfo{author}{\bibfnamefont{Y.}~\bibnamefont{Mino}},
  \bibinfo{journal}{Physical Review D} \textbf{\bibinfo{volume}{67}}
  (\bibinfo{year}{2003}), ISSN \bibinfo{issn}{1089-4918}.

\bibitem[{\citenamefont{Fujita and Hikida}(2009)}]{Fujita1_2009}
\bibinfo{author}{\bibfnamefont{R.}~\bibnamefont{Fujita}} \bibnamefont{and}
  \bibinfo{author}{\bibfnamefont{W.}~\bibnamefont{Hikida}},
  \bibinfo{journal}{Classical and Quantum Gravity}
  \textbf{\bibinfo{volume}{26}}, \bibinfo{pages}{135002}
  (\bibinfo{year}{2009}), ISSN \bibinfo{issn}{1361-6382}.

\bibitem[{\citenamefont{Poisson}(2015)}]{poisson2015tidal}
\bibinfo{author}{\bibfnamefont{E.}~\bibnamefont{Poisson}},
  \bibinfo{journal}{Physical Review D} \textbf{\bibinfo{volume}{91}},
  \bibinfo{pages}{044004} (\bibinfo{year}{2015}).

\bibitem[{\citenamefont{Misner et~al.}(2017)\citenamefont{Misner, Thorne, and
  Wheeler}}]{MTW_2017}
\bibinfo{author}{\bibfnamefont{C.~W.} \bibnamefont{Misner}},
  \bibinfo{author}{\bibfnamefont{K.~S.} \bibnamefont{Thorne}},
  \bibnamefont{and} \bibinfo{author}{\bibfnamefont{J.~A.}
  \bibnamefont{Wheeler}}, \bibinfo{journal}{The Astrophysical Journal}
  (\bibinfo{year}{2017}).

\bibitem[{\citenamefont{Mino et~al.}(1997)\citenamefont{Mino, Sasaki, and
  Tanaka}}]{Mino_1997}
\bibinfo{author}{\bibfnamefont{Y.}~\bibnamefont{Mino}},
  \bibinfo{author}{\bibfnamefont{M.}~\bibnamefont{Sasaki}}, \bibnamefont{and}
  \bibinfo{author}{\bibfnamefont{T.}~\bibnamefont{Tanaka}},
  \bibinfo{journal}{Physical Review D} \textbf{\bibinfo{volume}{55}},
  \bibinfo{pages}{3457–3476} (\bibinfo{year}{1997}), ISSN
  \bibinfo{issn}{1089-4918}.

\bibitem[{\citenamefont{Quinn and Wald}(1997)}]{Quinn_1997}
\bibinfo{author}{\bibfnamefont{T.~C.} \bibnamefont{Quinn}} \bibnamefont{and}
  \bibinfo{author}{\bibfnamefont{R.~M.} \bibnamefont{Wald}},
  \bibinfo{journal}{Physical Review D} \textbf{\bibinfo{volume}{56}},
  \bibinfo{pages}{3381–3394} (\bibinfo{year}{1997}), ISSN
  \bibinfo{issn}{1089-4918}.

\bibitem[{\citenamefont{Poisson et~al.}(2011)\citenamefont{Poisson, Pound, and
  Vega}}]{Poisson_2011}
\bibinfo{author}{\bibfnamefont{E.}~\bibnamefont{Poisson}},
  \bibinfo{author}{\bibfnamefont{A.}~\bibnamefont{Pound}}, \bibnamefont{and}
  \bibinfo{author}{\bibfnamefont{I.}~\bibnamefont{Vega}},
  \bibinfo{journal}{Living Reviews in Relativity} \textbf{\bibinfo{volume}{14}}
  (\bibinfo{year}{2011}), ISSN \bibinfo{issn}{1433-8351}.

\bibitem[{\citenamefont{Barack and Pound}(2018)}]{Barack_2018}
\bibinfo{author}{\bibfnamefont{L.}~\bibnamefont{Barack}} \bibnamefont{and}
  \bibinfo{author}{\bibfnamefont{A.}~\bibnamefont{Pound}},
  \bibinfo{journal}{Reports on Progress in Physics}
  \textbf{\bibinfo{volume}{82}}, \bibinfo{pages}{016904}
  (\bibinfo{year}{2018}), ISSN \bibinfo{issn}{1361-6633}.

\bibitem[{\citenamefont{Le~Tiec et~al.}(2020)\citenamefont{Le~Tiec, Casals, and
  Franzin}}]{LeTiec:2020bos}
\bibinfo{author}{\bibfnamefont{A.}~\bibnamefont{Le~Tiec}},
  \bibinfo{author}{\bibfnamefont{M.}~\bibnamefont{Casals}}, \bibnamefont{and}
  \bibinfo{author}{\bibfnamefont{E.}~\bibnamefont{Franzin}}
  (\bibinfo{year}{2020}), \eprint{2010.15795}.

\bibitem[{\citenamefont{Yang and Casals}(2017)}]{PhysRevD.96.083015}
\bibinfo{author}{\bibfnamefont{H.}~\bibnamefont{Yang}} \bibnamefont{and}
  \bibinfo{author}{\bibfnamefont{M.}~\bibnamefont{Casals}},
  \bibinfo{journal}{Phys. Rev. D} \textbf{\bibinfo{volume}{96}},
  \bibinfo{pages}{083015} (\bibinfo{year}{2017}).

\bibitem[{\citenamefont{Poisson and Will}(2014)}]{poisson_will_2014}
\bibinfo{author}{\bibfnamefont{E.}~\bibnamefont{Poisson}} \bibnamefont{and}
  \bibinfo{author}{\bibfnamefont{C.~M.} \bibnamefont{Will}},
  \emph{\bibinfo{title}{Gravity: Newtonian, Post-Newtonian, Relativistic}}
  (\bibinfo{publisher}{Cambridge University Press}, \bibinfo{year}{2014}).

\bibitem[{BHP({\natexlab{a}})}]{BHPC}
\emph{\bibinfo{title}{{Black Hole Perturbation Club}}},
  \bibinfo{howpublished}{(\href{https://sites.google.com/view/bhpc1996/home}{https://sites.google.com/view/bhpc1996/home})}.

\bibitem[{BHP({\natexlab{b}})}]{BHPToolkit}
\emph{\bibinfo{title}{{Black Hole Perturbation Toolkit}}},
  \bibinfo{howpublished}{(\href{http://bhptoolkit.org/}{bhptoolkit.org})}.

\bibitem[{dat(2011)}]{dataanalysis}
\emph{\bibinfo{title}{Gravitational-Wave Data Analysis}}
  (\bibinfo{year}{2011}), chap.~\bibinfo{chapter}{7}, pp.
  \bibinfo{pages}{269--347}.

\bibitem[{\citenamefont{Chua et~al.}(2017)\citenamefont{Chua, Moore, and
  Gair}}]{Chua:2017ujo}
\bibinfo{author}{\bibfnamefont{A.~J.} \bibnamefont{Chua}},
  \bibinfo{author}{\bibfnamefont{C.~J.} \bibnamefont{Moore}}, \bibnamefont{and}
  \bibinfo{author}{\bibfnamefont{J.~R.} \bibnamefont{Gair}},
  \bibinfo{journal}{Phys. Rev. D} \textbf{\bibinfo{volume}{96}},
  \bibinfo{pages}{044005} (\bibinfo{year}{2017}), \eprint{1705.04259}.

\bibitem[{\citenamefont{Katz et~al.}(2020)\citenamefont{Katz, Chua, Warburton,
  and Hughes.}}]{michael_l_katz_2020_4005001}
\bibinfo{author}{\bibfnamefont{M.~L.} \bibnamefont{Katz}},
  \bibinfo{author}{\bibfnamefont{A.~J.~K.} \bibnamefont{Chua}},
  \bibinfo{author}{\bibfnamefont{N.}~\bibnamefont{Warburton}},
  \bibnamefont{and} \bibinfo{author}{\bibfnamefont{S.~A.}
  \bibnamefont{Hughes.}},
  \emph{\bibinfo{title}{{BlackHolePerturbationToolkit/FastEMRIWaveforms:
  Official Release}}} (\bibinfo{year}{2020}),
  \urlprefix\url{https://doi.org/10.5281/zenodo.4005001}.

\bibitem[{\citenamefont{Pound and Poisson}(2008)}]{PhysRevD.77.044013}
\bibinfo{author}{\bibfnamefont{A.}~\bibnamefont{Pound}} \bibnamefont{and}
  \bibinfo{author}{\bibfnamefont{E.}~\bibnamefont{Poisson}},
  \bibinfo{journal}{Phys. Rev. D} \textbf{\bibinfo{volume}{77}},
  \bibinfo{pages}{044013} (\bibinfo{year}{2008}).

\bibitem[{\citenamefont{Bronicki et~al.}(2022)\citenamefont{Bronicki,
  C\'ardenas-Avenda\~no, and Stein}}]{Bronicki:2022eqa}
\bibinfo{author}{\bibfnamefont{D.}~\bibnamefont{Bronicki}},
  \bibinfo{author}{\bibfnamefont{A.}~\bibnamefont{C\'ardenas-Avenda\~no}},
  \bibnamefont{and} \bibinfo{author}{\bibfnamefont{L.~C.} \bibnamefont{Stein}}
  (\bibinfo{year}{2022}), \eprint{2203.08841}.

\end{thebibliography}
\end{document}